\documentclass{tufte-handout}
\usepackage[T1]{fontenc}
\usepackage[utf8]{inputenc}
\usepackage{float} % has option [H] for force placement
\usepackage{graphicx} % useful for eg. pdf or eps or svg graphics
\usepackage[space]{grffile}
\usepackage{latexsym}
\usepackage{textcomp}
\usepackage{hyperref} % see https://arxiv.org/help/faq/texlive 2017-04-25 update
\usepackage{longtable}
\usepackage{multirow,booktabs}
\usepackage{amsfonts,amsmath,amssymb}
\usepackage{natbib}
\usepackage{url}
\usepackage{wasysym}
\usepackage{lscape} % for landscape
\usepackage{pdfpages}
\usepackage{verbatim} % for \begin{comment}
\usepackage{todonotes}
\usepackage[english]{babel}
\usepackage{longtable} % For caption on head of table, long-table example.
\usepackage{afterpage}  % to put at beginning of longtables to float them
\usepackage[toc,page,title,titletoc]{appendix} % For appendices
\usepackage{tocloft}
\hypersetup{colorlinks=false,pdfborder={0 0 0}}
\newif\iflatexml\latexmlfalse

\setkeys{Gin}{width=\linewidth,totalheight=\textheight,keepaspectratio}
\graphicspath{{graphics/}}

\usepackage{units}
\usepackage{fancyvrb}
\fvset{fontsize=\normalsize}
\usepackage{multicol}
\newcommand\numberthis{\addtocounter{equation}{1}\tag{\theequation}}
\title{Planet Detection Simulations for Several Possible TESS Extended Missions}
\author{\normalsize{L. G. Bouma$^{1,2,3}$, 
		Joshua N. Winn$^{1,2,3}$, 
		Jacobi Kosiarek$^{2}$ 
		and P. R. McCullough$^{4}$}}

\date{}

\def\ltsima{$\; \buildrel < \over \sim \;$}
\def\lsim{\lower.5ex\hbox{\ltsima}}
\def\gtsima{$\; \buildrel > \over \sim \;$}
\def\gsim{\lower.5ex\hbox{\gtsima}}
\def\tess{{\it TESS}\:}
\def\kepler{{\it Kepler}\:}
\def\keplers{{\it Kepler}'s\:}

\def\jwst{{\it JWST}\:}
\def\jwsts{{\it JWST}'s\:}
\def\ktwo{{\it K2}\:}

\def\corot{{\it CoRoT}\:}

\def\cheops{{\it CHEOPS}\: }

\def\jwst{{\it JWST}\:}
\def\gaia{{\it Gaia}\:}
\def\tic{\tess Input Catalog}
\def\tesss{\textit{TESS}'s\:}

\def \npole{\texttt{pole}}
\def \nhemi{\texttt{hemi}}
\def \elong{\texttt{ecl\_long}}
\def \eshort{\texttt{ecl\_short}}
\def \shemiAvoid{\texttt{hemi+ecl}}
\def \hemis{\texttt{allsky}}

\def \moon{{\!\rightmoon}}

\defcitealias{Sullivan_2015}{S+15}
\defcitealias{campante_asteroseismic_2016}{C+16}

\begin{document}

\maketitle % print document title, author, and date

\vspace{-0.42cm}
%\vspace{0.2cm}
%\input{affiliations}
{\tiny\noindent
	$^1$Department of Physics, 77 Massachusetts Ave., Massachusetts 
	Institute of Technology, Cambridge, MA 02139\newline
	\noindent$^2$MIT Kavli Institute for Astrophysics and Space Research, 70 
	Vassar
	St., Cambridge, MA 02139\newline
	\noindent$^3$Department of Astrophysical Sciences, Princeton University, 4 
	Ivy 
	Lane, Princeton, NJ 08540, USA\newline	
	\vspace{-0.222cm}
	\noindent$^4$Department of Physics and Astronomy, Johns 
	Hopkins University, 3400 North Charles Street, Baltimore, MD 21218
}

\begin{abstract}
	\label{sec:abstract}
	
	\vspace{-0.35cm}
	%\vspace{-0.55cm}
	\noindent 
  {{Disclaimer: The views, opinions, assumptions, examples, and results expressed
  in this article are solely those of the authors and do not necessarily reflect
  the official policy or position of the \tess Science Team,  any of the authors'
  employers or affiliated institutions, NASA, or any agency of the U.S.
  government. This article has not been endorsed or reviewed by NASA or the \tess
  Science Team.}}
  \vspace{0.5cm}
		
		\noindent {\bf Executive Summary}\\
		
		The Transiting Exoplanet Survey Satellite (\textit{TESS}) will perform
		a two-year survey of nearly the entire sky, with the main goal of 
		detecting
		exoplanets smaller than Neptune around bright, nearby stars. There do
		not appear to be any fundamental obstacles to continuing science
		operations for at least several years after the two-year Primary 
		Mission.
		
		Any decisions regarding use of the \textit{TESS} spacecraft
		after the Primary Mission should be made with the broadest
		possible input and assistance from the astronomical community.
		As has recently been made clear by the NASA \textit{K2} mission, there
		are many applications of precise time-series photometry of bright 
		objects 
		besides exoplanet detection.
		
		Nevertheless, exoplanet
		detection is likely to be part of the motivation for a \textit{TESS} 
		Extended Mission.
		To provide a head start to those who are planning and proposing for 
		such a 
		mission, this white paper presents some simulations
		of exoplanet detections in a third year of \textit{TESS}
		operations. Our goal is to provide a helpful reference for the 
		exoplanet-related
		aspects of any Extended Mission, while recognizing that this will be 
		only one part of
		a larger discussion of the scientific goals of such a mission.
		
		We performed Monte Carlo simulations to try to anticipate the 
		quantities and
		types of planets that would be detected in several plausible
		scenarios for a one-year Extended Mission following the two-year
		Primary Mission. The strategies differ mainly in the schedule of 
		pointings on the sky.
		For simplicity we did not compare different choices
		for the cadence of photometric measurements, or for the target
		selection algorithm, although different choices might prove to be
		advantageous and should be studied in future work.
		
		We considered six different scenarios for Year 3
		of the \tess mission, illustrated in Figure~\ref{fig:strategies}:
		\begin{figure*}[!tb]
			\includegraphics[width=5in]{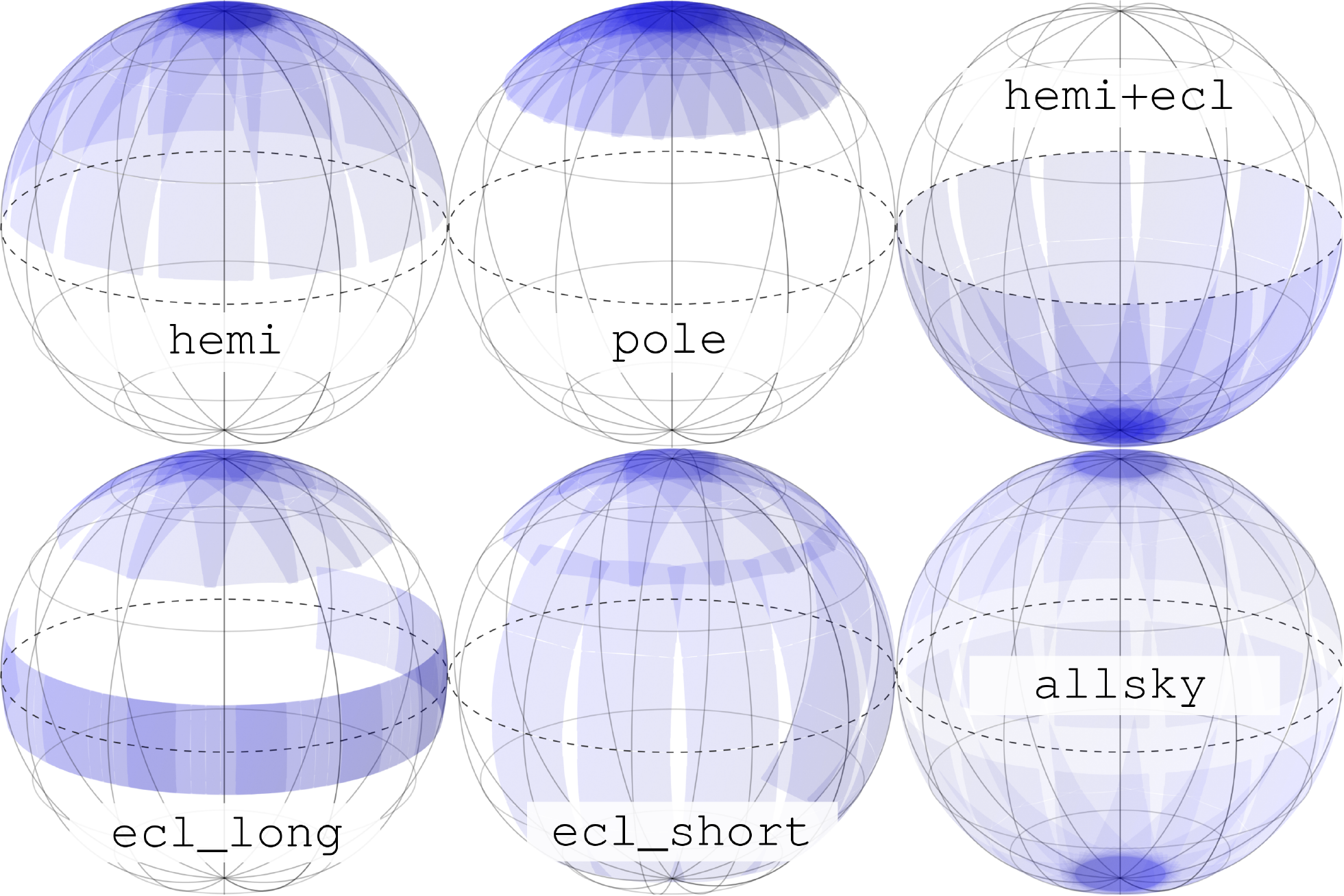}
			\caption{Six possible pointing strategies for a \tess Extended
				Mission, visualized in ecliptic coordinates.  None
				of these scenarios spend the entire year observing the
				ecliptic; we concluded that such a plan is inadvisable
				because of interruptions by the Earth and Moon (see
				Fig.~\protect\ref{fig:earth_moon_elong}).}
			\label{fig:strategies}
		\end{figure*}
		
		\begin{enumerate}
			
			\item \nhemi, which re-observes one of the ecliptic hemispheres in
			essentially the same manner as in the Primary Mission (i.e.,
			neglecting the zone within $6^\circ$ of the ecliptic);
			
			\item \npole, which focuses on one of the two ecliptic poles;
			
			\item \shemiAvoid, which re-observes an ecliptic hemisphere, but 
			moving all 
			fields $6^\circ$ closer in latitude to the ecliptic plane. This 
			scenario
			has a continuous viewing zone with angular diameter $12^\circ$ 
			rather than 
			$24^\circ$;
			
			\item \elong, which has a series of pointings with the long
			axis of the $24^\circ\times96^\circ$ field-of-view along the
			ecliptic (in combination with
			some fields near the ecliptic pole, when the Earth or Moon would
			prevent effective observations of the ecliptic);
			
			\item \eshort, which has a series of pointings with the short
			axis of the field-of-view along the ecliptic (again in combination
			with some fields near the ecliptic pole);
			
			\item \hemis, which covers nearly the entire sky with $\sim$14-day
			pointings (as opposed to the 28-day pointings of the Primary
			Mission), by alternating between northern and southern hemispheres.
			
		\end{enumerate}
		
		We numerically computed the results based on the methodology
		of~\citet{Sullivan_2015}. Some of the most important findings are:
		\begin{enumerate}
			
			\item The overall quantity of detected planets\footnote{We define 
			`detected 
				planet' to mean one with at least two observed transits, and a 
				phase-folded 
				$\mathrm{SNR} > 7.3$ (Eq.~\ref{eq:detection_criterion}). All 
				statistics are 
				quoted for $R_p<4R_\oplus$ planets.} does not depend
			strongly on the sky-scanning schedule.  Among the six scenarios
			considered here, the number of newly-detected planets with radii
			less than $4R_\oplus$ is the same to within about 30\%.
			
			\item The number of newly-detected sub-Neptune radius planets ($R_p 
			\lesssim 
			4R_\oplus$) in Year 3 is approximately the same as the number 
			detected in 
			either Year 1 or
			Year 2.  Thus, we do not expect a sharp fall-off in the planet
			discovery rate in Year 3.  This is because the Primary Mission will
			leave behind many short-period transiting planets with bright host
			stars, with a signal-to-noise ratio just below the threshold for
			detection.  These planets can be detected by collecting more data in
			Year 3.
			
			\item Apart from detecting new planets, a potentially important
			function of an Extended Mission would be to improve our ability to
			predict the times of future transits and occultations of 
			{\it TESS}-detected planets.  With data from the Primary Mission 
			alone, the uncertainty in planetary orbital periods will inhibit 
			follow-up observations after only a few years, as the transit 
			ephemerides
			become stale. By re-observing the same sky that was observed in the
			Primary Mission, \nhemi, \shemiAvoid, and \hemis\ address this 
			issue.
			
			\item Regarding newly detected sub-Neptunes, the \hemis, \npole, and
			\shemiAvoid\,strategies offer the greatest number (1350-1400, as
			compared to the 1250 during each year of the Primary Mission).
			
			\item Regarding planets with orbital periods >20 days, the 
			\hemis\:and
			\npole\:strategies discover twice as many such planets as will be
			discovered in each year of the Primary Mission.
			However, this assumes that two transits are sufficient for secure 
			detection.
			If instead we require three transits, then \npole\:detects 260 new 
			long-period planets, while the next-best scenarios, \hemis, \nhemi, 
			and 
			\shemiAvoid, all detect about 160. (The simulated Primary Mission 
			detects 145; see 
			Figs.~\ref{fig:yield_results} and~\ref{fig:Ntra_hist}).
			
			\item Regarding new planets with very bright host stars ($I_c<10$),
			the \hemis, \shemiAvoid, and \eshort\:strategies offer the greatest
			numbers ($\approx$190, about the same as are found in each year of 
			the
			Primary Mission; see Table~\ref{tab:icmag_meta}).
			
			\item Regarding planets with near-terrestrial insolation ($0.2 
			<S/S_\oplus< 2$), all the strategies considered here offer similar
			numbers (about 120, as compared to 105 in each year of the 
			simulated Primary
			Mission).
			
		\end{enumerate}
		
		The rest of this report is organized as follows.
		Sec.~\ref{sec:approach} discusses how we selected and compared 
		different pointing strategies, 
		as well as how we modeled \tesss observations.
		Sec.~\ref{sec:comparing_pointing_strategies} describes some figures of 
		merit for 
		comparing Extended Mission scenarios, including a discussion of some 
		scenarios we chose not to study.
		Sec.~\ref{sec:input_assumptions} lists the most important assumptions 
		we made for the simulations.
		Sec.~\ref{sec:newly_detected_planet_metrics} compares the 
		characteristics of 
		newly-detected planets for the six scenarios under consideration.
		Sec.~\ref{sec:gtr_1yr_horizon} discusses some considerations and 
		implications for future years of the Extended Mission, beyond the 
		one-year scenarios that were simulated in detail.
		Sec.~\ref{sec:ephemeris_times} raises the critical issue of the 
		uncertainty in 
		transit ephemerides.
		Sec.~\ref{sec:risks_caveats} discusses the reliability and limitations 
		of our methodology.
		Sec.~\ref{sec:conclusions} concludes and recommends avenues for further 
		study.
		
		The catalogs of simulated detected planets (for both the Primary and 
		Extended
		Missions) are available
		online\footnote{\url{scholar.princeton.edu/jwinn/extended-mission-simulations}\label{fn:wiki}}.
		This website also
		contains a less formal document with some ideas and questions regarding 
		the
		broader applications of a \textit{TESS} Extended Mission.  We welcome 
		any other
		ideas, comments, or corrections; please send them to:
		\url{luke@astro.princeton.edu}.
		
	\end{abstract}

\newpage
\tableofcontents
\newpage
\section{Approach}
\label{sec:approach}

\subsection{Constraints on \tesss Observing}
\label{sec:constraints_on_pointings}

\begin{figure}[!b]
	\centering
	\includegraphics{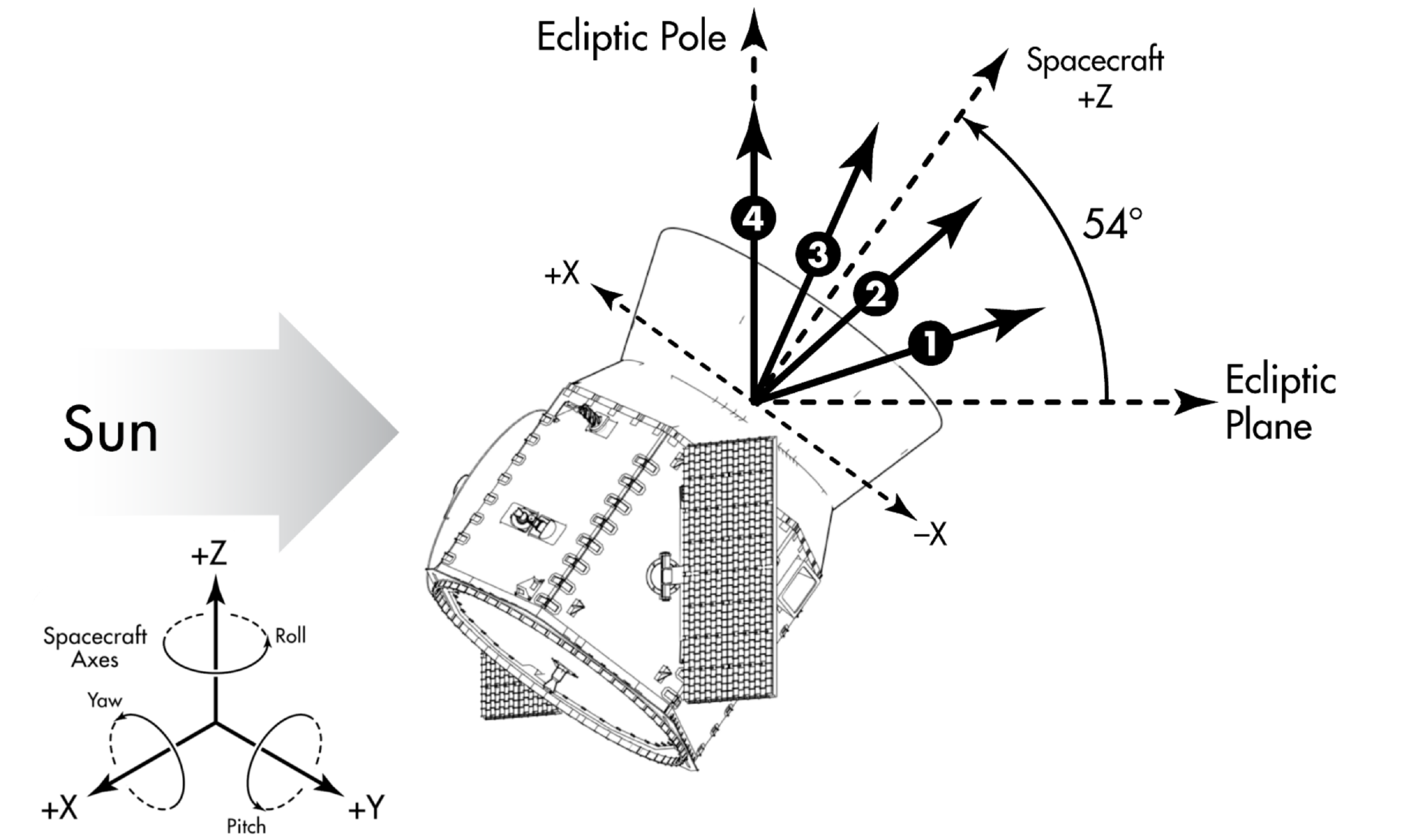}
	\caption{The spacecraft must point so that incident sunlight is collected 
		by the solar panels, and not the cameras. \tesss solar panels pitch 
		about the $+Y$ axis. (Adapted from Orbital ATK design document) }
	\label{fig:spacecraft_angles}
\end{figure}

When considering possible schedules for telescope pointings, the main
constraint is that the cameras must be directed approximately opposite
the Sun.  Specifically, the center of the combined fields-of-view is
ideally pointed within 15$^\circ$ in ecliptic longitude of the antisolar 
direction, and no more than 30$^\circ$ away.
%\todo[inline]{Roland: are these numbers still good? are they imposed by the 
%sunshade? or solar panels? Some comment appreciated.} 
This enables the solar panels to collect
sufficient sunlight to power the spacecraft.
Fig.~\ref{fig:spacecraft_angles} illustrates the geometry. The solar
panels are free to rotate about the $Y$-axis.

It is also important for the sunshade and 
spacecraft to block sunlight.
This constraint makes it difficult to remain pointed at a given field for more 
than two spacecraft orbits ($\approx$42 days). This is why \tess advances the 
fields of view by $\approx$$28^\circ$ east in ecliptic
longitude every lunar month, during the Primary Mission.  Another important 
consideration
is whether the Earth or Moon passes
through \tesss camera fields during a proposed pointing (see
Sec.~\ref{sec:earth_moon_crossings}), because these crossings are detrimental 
to high photometric precision.

In addition, the spacecraft has finite fuel reserves for necessary maneuvers.
These are expected to last at least 10 years [G. Ricker, priv.\ comm.].
Since the time horizon for this study is only 3 years, 
we do not consider the fuel reserves to be a constraint on the scenarios
we investigate.

\subsection{Proposed Pointing Strategies}
\label{sec:proposed_pointings}

For simplicity we chose to study one-year plans for an Extended
Mission, i.e., plans for Year 3 of the \tess mission. (Later in this
report we remark on some possible implications of our study for
additional years of an Extended Mission.)  Given the constraints
outlined in Sec.~\ref{sec:constraints_on_pointings}, we selected the
following scenarios for detailed study:

\begin{figure*}[!bt]
	\includegraphics{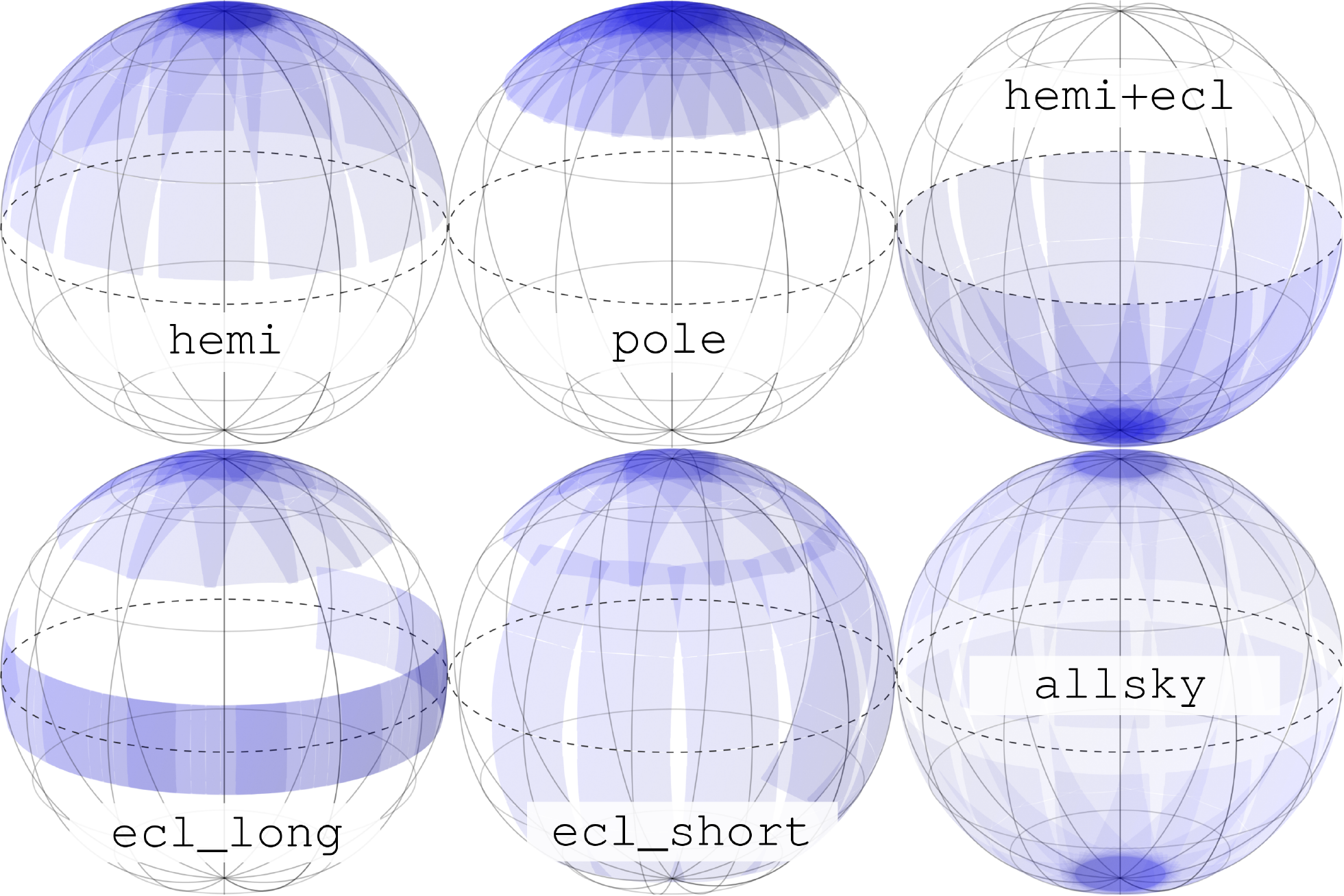}
	\caption{Proposed pointing strategies for a \tess Extended Mission, 
		visualized in ecliptic coordinates. \nhemi, \npole, \shemiAvoid, 
		\elong, 
		\eshort, and \hemis. Note for \elong\:and \eshort\:that Earth and Moon 
		crossings likely make an entire year looking at the ecliptic 
		impractical 
		(see Fig.~\protect\ref{fig:earth_moon_elong}).}
	\label{fig:proposed_pointings}
\end{figure*}

\begin{description}
	
	\item[Scenario 1.] \nhemi: Repeat observations of one of the two
	ecliptic hemispheres in a manner similar to the Primary Mission.
	For concreteness we arbitrarily chose the northern ecliptic hemisphere
	for the third year. In this scenario we could take the opportunity to 
	shift the longitudes of all sectors by an amount that would enable 
	\tess to cover the gaps that were left during the Primary Mission 
	(the ``slits'' in the sky coverage between ecliptic latitudes of 
	6--30$^\circ$).
	However for simplicity, we opted to observe the same longitudes as
	in the Primary Mission.
	\textit{Motivation:} similar to the Primary Mission.
	Long time baseline at the North Ecliptic Pole, and broad sky
	coverage. Also remeasures transit times (and substantially improves 
	ephemerides)
	of previously detected \tess planets over most of the entire
	hemisphere.
	
	\item[Scenario 2.] \npole: Focus on one of the ecliptic poles,
	arbitrarily chosen to be the north ecliptic pole for
	concreteness. \tesss sunshade and lens hoods must still
	suppress incoming sunlight in this scenario.
	A potential problem is sunlight might
	reflect off the interior of the sunshade and into a lens hood.
	Since this problem could be addressed by rotating the spacecraft to point 
	slightly anti-Sun (putting the cameras in the shadow of the sunshade), we 
	neglect any effects of extra
	scattered sunlight\footnote{Whether this minor 
		adjustment would even be necessary depends on the combined performance 
		of the
		sunshade and each camera's lens hood.
		These will be verified by measurements during commissioning.
	}.
	\textit{Motivation:} maximizes the average duration of
	observations per star; intuitively expected to provide
	greatest sensitivity to long-period planets.
	
	\item[Scenario 3.] \shemiAvoid: Repeat observations of one of the two
	ecliptic hemispheres, but in this case shifting all fields $6^\circ$
	toward the ecliptic, such that the combined fields-of-view reach all
	the way from the ecliptic to $6^\circ$ behind the ecliptic pole. We chose
	to simulate the southern ecliptic hemisphere because the northern
	version of this plan would suffer more from Earth and Moon interference
	(cf. Table~\ref{tab:dropped_fields} in
	Sec.~\ref{sec:earth_moon_crossings}).
	\textit{Motivation:} trades the long continuous viewing zone near
	the pole for greater sky coverage, and in particular, coverage of
	the ecliptic zone which was missed in the Primary Mission. Extensible
	to Year 4. Freshens ephemerides.
	
	\item[Scenario 4.] \elong: Survey the ecliptic with 7 sectors (14
	orbits) in which the long axis of the fields-of-view are oriented
	along the ecliptic.  For the other 6 sectors, during the interval
	when ecliptic observations would be interrupted by Earth and Moon
	crossings, we focus on one of the ecliptic poles.
	\textit{Motivation:} covers the ecliptic, which will not be observed
	during the Primary Mission.  Offers opportunities for follow-up
	of K2 discoveries. Minimizes Earth-moon interference.
	
	\item[Scenario 5.] \eshort: Survey the ecliptic and also cover a large
	fraction of the rest of the ecliptic hemisphere. For 7 sectors we
	observe the ecliptic but with the {\it short} axis oriented along
	the ecliptic, and the long axis reaching up to higher latitudes.
	The remaining 6 sectors are focused on the ecliptic pole, as in
	\elong.  \textit{Motivation:} similar to \elong, but with more
	overlap between this year and the Primary Mission to allow for
	improved transit ephemerides and better ability to follow-up on
	previous discoveries.
	Also covers more sky than \elong, which could improve the quantity
	of planet detections from full frame images.
	
	\item[Scenario 6.] \hemis: Cover both northern and southern ecliptic
	hemispheres in a single year, by alternating between the hemispheres
	every 13.7 days.  \textit{Motivation:} rapid coverage of the
	entire sky, allows follow-up of almost all previously detected \tess
	objects and refined ephemerides.
	
\end{description}

Although these 6 scenarios seemed like reasonable choices for further study and
direct comparison, there are many other possibilities that may be of interest
that were not studied in detail, in order to keep the scope of this report
manageable.  Among these other possibilities that were considered but not
studied are:
\begin{itemize}
	\item The \npole\:strategy applied to the south ecliptic pole rather than 
	the north (we do not expect major differences).
	\item The \nhemi\:strategy applied to the southern ecliptic hemipshere 
	rather than the northern (we do not expect major differences).
	\item The \nhemi\:strategy, but rotated about the ecliptic polar axis by
	$12^\circ$ in longitude.
	\item The Northern inversion of \shemiAvoid, which is more strongly 
	affected by Earth and Moon crossings (Sec.~\ref{sec:earth_moon_crossings}) 
	and is less able to 
	improve knowledge of mid-transit times 
	(Fig.~\ref{fig:lowering_uncertainty_tc}).
	\item A version of \nhemi\ in which all fields are $12^\circ$ closer to 
	the ecliptic, rather than $6^\circ$ as in \shemiAvoid.
	\item A full year spent observing the ecliptic.  Such a plan would 
	suffer from Earth and Moon crossings for a substantial fraction of the year.
	We show the outage as a function of time in Fig.~\ref{fig:earth_moon_elong}.
	Solar system objects (planets, asteroids) could also be an annoyance, 
	although
	we did not model their effects.
	\item Alternate between northern and southern ecliptic poles every 13.7 
	days.
	This would be similar to \hemis\:but would focus on the poles rather than 
	the entire sky.
	It would sacrifice sky coverage (and ability to refresh ephemerides over 
	the whole sky) in return for a longer mean  observation duration per star.
	\item Hybrid strategies that change from month to month.  For instance, in 
	the \nhemi\:scenario, during a month when the Earth or Moon crosses through 
	the field of a camera pointed close to the ecliptic, we could tilt all the 
	cameras away from the ecliptic as in the \npole\:scenario.        
\end{itemize}

\subsection{Metrics to Compare Pointing Strategies}
\label{sec:comparing_pointing_strategies}

We assess Extended Missions based on the risks and opportunities they
present, as well as their performance on selected technical and
science-based criteria.  These criteria are organized following an
approach originally outlined by~\citet{kepner_rational_1965}.
Summarizing them in list form:
\begin{description}
	\item[Technical musts:] Point cameras anti-sun. Allow the solar panels to 
	collect sunlight.
	\item[Technical wants:] Keep the duration of each sector $<28$ days. 
	Minimize Earth or Moon 
	crossings. Minimize zodiacal background light. Minimize scattered sunlight.
	\item[Metrics in exoplanet science:]\
	\begin{itemize}
		\item number of newly detected planets
		\begin{itemize}
			\item from stars observed with 2-minute time sampling from among 
			$\approx$200,000 subrasters, known colloquially as "postage stamps" 
			(PS),
			\item from stars observed with 30-minute time sampling in the 
			full-frame images (FFIs);
		\end{itemize}
		\item number of new long-period planets 
		\begin{itemize} 
			\item by detecting additional transits of planets for which only 
			one transit was observed in the Primary Mission,
			\item by detecting transits of long-period planets that were not 
			detected in the Primary Mission;
		\end{itemize}
		\item number of new habitable-zone planets; 
		\item number of new planets with ``characterizable'' atmospheres; 
		\item number of newly detected planets with bright host stars; 
		\item number of planet-hosting stars detected in the Primary Mission 
		for which the Extended Mission reveals an additional transiting planet;
		\item ability to improve transit ephemerides for previously detected 
		transiting planets;
		\item ability to observe more transits over a longer baseline to enable 
		searches for transit-timing variations.
	\end{itemize}
\end{description}

These metrics were chosen for their apparent importance as well as our
ability to quantify them with simulations. Of course there are other 
considerations
that may be very important but are more difficult to quantify:
\begin{itemize}
	\item Different ways to choose the stars that are observed with finer time 
	sampling.
	It will likely be advantageous to use the results of the Primary Mission to
	choose stars for which photometric variability is known to be detectable 
	and interesting.
	Examples include planet hosts, candidate planet hosts, circumbinary \&
	circumprimary planets, white dwarfs, open clusters, eclipsing binaries and 
	higher-order systems,
	pulsating stars (Cepheids, RR Lyrae, $\delta$ Scuti, slowly pulsating B 
	stars), eruptive stars, cataclysmic variables, and stars of special 
	interest for asteroseismology.
	\item Prospects for solar-system science, such as observations of main belt 
	asteroids and the brightest near-Earth asteroids.
	\item Prospects for extragalactic astronomy and high energy astrophysics; 
	for 
	instance, gathering light curves of variable active galactic nuclei, or 
	imaging
	extended low surface brightness features of galaxies.
\end{itemize}

Regarding opportunities and risks, the following need to be considered:
\begin{description}
	\item[Opportunities:]\ 
	\begin{itemize}
		\item Optimizing science beyond the 3-year horizon. For example, 
		suppose it were known in advance
		that \tess would likely continue operations for 5-10 years: how would 
		this knowledge affect
		the decision on what to do in the year following the Primary Mission?
		\item Ability to promote targets that were detected in FFIs to PSs in 
		the Extended Mission.
		\item Alter the number of PSs, the time sampling of the PS, and the 
		time sampling of the FFIs, under the constraint of fixed data volume. 
		An extreme case is eliminating the PSs and returning only FFIs.
		\item \textit{\tess as follow-up mission:} ability to
		observe \corot objects; ability to observe the \kepler
		field;
		ability to observe \ktwo fields (follow-up \ktwo
		few-transit objects); ability to observe targets
		previously monitored by ground-based surveys.
		\item \textit{Prospects for follow-up with other resources:}
		potential for \jwst follow-up, 
		potential for \cheops follow-up,
		ability to obtain \tess photometry contemporaneously with ground-based 
		observations,
		ability to follow up with resources in both hemispheres.
		\item Impact on Guest Investigator program.
	\end{itemize}
	
	\item[Risks:] 
	Risk of spacecraft damage. 
	Risk of not meeting threshold science (however it is defined for the 
	Extended Mission).
	Risk of poorer photometric precision than desired, e.g., from confusion in 
	crowded fields.
	Would partial instrument failure in Primary Mission make this scenario 
	infeasible? 
	Would reduced precision (from aged CCDs, worse pointing accuracy, or other 
	mechanical sources) invalidate this scenario? 
	Risk of planet detection simulation over- or under-estimating planet yield.
\end{description}

\subsection{Description of Planet Detection Model}
\label{sec:planet_detection_model}

\citet{Sullivan_2015} (hereafter, \citetalias{Sullivan_2015})
developed a simulation of \tesss planet and false positive detections
based on the spacecraft and payload design specified by
\citet{ricker_transiting_2014}.  We adapt this simulation for Extended
Mission planning.  With our additions, we can change where \tess looks
in additional years of observing while holding fixed all other
mission-defining parameters.  Our approach is to run our planet
detection simulation for each plausible pointing strategy, and to
compare the relative yields of detected planets.  This lets us compare
Extended Mission scenarios with one another and with the Primary
Mission.

\paragraph{Background on synthetic catalogs:}

\tess will be sensitive to sub-Neptune sized transiting planets orbiting M
dwarfs out to $\sim\!200\,\text{pc}$ and G dwarfs out to
$\sim\!1\,\text{kpc}$~(\citetalias{Sullivan_2015}, Sec. 2.3).  It will be
sensitive to giant planets and eclipsing binaries across a significant
fraction of the galactic disk.  With this sensitivity in mind, the
stellar catalog we `observe' in our planet detection simulation is
drawn from the output of TRILEGAL, a population synthesis code for the
Milky Way~\citep{girardi_star_2005}.  ~\citetalias{Sullivan_2015} made
some modifications to the catalog, notably in the M dwarf
radius-luminosity relation, to better approximate interferometric
stellar radii measurements.  We retain these modifications; the modified
TRILEGAL stellar catalog shows acceptable agreement with
observations\footnote{Looking closely at the radius-luminosity
	relations, we do see non-physical interpolation artifacts. These
	outliers are visible in Figs.~\ref{fig:fig17_replica}
	and~\ref{fig:fig17_radius_on_x} below, but are a small enough subset
	of the population that they can be ignored on statistical grounds in this 
	work.}, specifically
the Hipparcos
sample~\citep{perryman_hipparcos_1997,van_leeuwen_validation_2007} and
the $10\text{pc}$ RECONS sample~\citep{henry_solar_2006}.

With a stellar catalog defined, we populate the stars in the catalog
with planets based on occurrence rates derived from the
\textit{Kepler} sample.  We use rates~\citet{fressin_false_2013} found
for planets orbiting stars with $T_\text{eff} > 4000\text{K}$ and
those that~\citet{dressing_occurrence_2015} found for the remaining M
and late K dwarfs.

\paragraph{Detection process:}
We then simulate transits of these planets.  Assuming the transit
depth and number of transits are known, we use a model of \tesss point
spread function (PSF) to determine optimal photometric aperture sizes
for each postage stamp star (\textit{i.e.,} we compute the noise for
all plausible aperture sizes, and find the number of pixels that
minimizes this noise).  With the aperture sizes and noise
corresponding to a given integration time known, we compute a signal
to noise ratio for each transiting object.  Our model for planet
detectability is a simple step function in SNR: if we have two or more
transits and $\text{SNR} > 7.3$, the planet is `detected', otherwise
it is not detected\footnote{The value of this threshold is chosen to
	ensure that no more than one statistical false positive is present in
	a pipeline search of $2\times10^5$ target stars. When observing a
	greater number of stars, for instance in the full frame images, 
	a higher threshold value should be imposed to maintain the same condition. 
	We
	discuss this further in Sec.~\protect\ref{sec:risks_caveats}.}. 
Cast as an equation,
\begin{equation}
	\mathrm{Prob(detection)}=\begin{cases}
		1 \quad \mathrm{if\ SNR_{phase-folded}} > 7.3\ \mathrm{and}\ 
		N_\mathrm{tra} \geq 2 \\
		0 \quad \mathrm{otherwise},
	\end{cases}
	\label{eq:detection_criterion}
\end{equation}
for $N_\mathrm{tra}$ the number of observed transits and 
$\mathrm{SNR_{phase-folded}}$ the phase-folded signal to noise ratio, defined 
in Eq.~\ref{eq:snr} below.
Our model for \tesss photometric precision is described 
by~\citetalias{Sullivan_2015} and also shown in
Fig.~\ref{fig:noise_with_moon}.

\paragraph{Assumptions of SNR calculation:}
A realistic simulation of the signal detection process
would begin with a realistic noise model for each 2-second CCD readout, taking 
into account both
instrumental and astrophysical variations. These 2-second data would be stacked 
into 2-minute
averages for the PSs and 30-minute averages for the FFIs. The time series would 
then be prepared and searched for transit signals using the same code that is 
used on real data.

Our calculation is much simpler. We calculate the phase-folded SNR based on an 
analytic approximation.
We implicitly assume that the correct period can be identified and that the 
noise follows
the simple model of \citetalias{Sullivan_2015}, which includes instrumental 
noise
and an empirically-determined level of stellar variability.
We also assume the transit signal to be constant in amplitude, although it may 
be reduced
in amplitude (``diluted'') below $(R_p/R_\star)^2$ due to the constant light 
from other stars
that fall within the same photometric aperture.
We then simply tally the number of \tess fields a
given host star falls in, which corresponds to a known total
observing baseline.  Knowing a planet's orbital period, and assuming the
planet is randomly phased in a circular orbit when \tess begins observing, 
we can count the number of transits \tess observes for the planet over the 
total baseline.
Using a model PSF, we determine ideal aperture sizes and obtain 
an estimated noise per transit.
Summarizing the relevant terms in an equation,
\begin{align}
	\mathrm{SNR}_\mathrm{phase-folded} &\approx
	\sqrt{N_\mathrm{tra}} \times \mathrm{SNR}_\mathrm{per-transit}\nonumber \\
	&= \sqrt{N_\mathrm{tra}} \times 
	\frac{\delta \cdot D}{\left(\frac{\sigma_\mathrm{1hr}^2}{T_\mathrm{dur}} 
		+ \sigma_v^2 \right)^{1/2}}, 
	\label{eq:snr} 
\end{align}
for $\delta$ the undiluted transit depth; $D$ the dilution factor
computed from background and binary contamination (Eq.~\ref{eq:dilution});
$\sigma_\mathrm{1hr}$ the summed noise contribution from CCD read noise, 
photon-counting noise from the star, a systematic 
$60\,\mathrm{ppm\cdot hr^{1/2}}$ noise floor, and zodiacal noise;
$T_\mathrm{dur}$ the transit duration in hours, and $\sigma_v$ the intrinsic
stellar variability (cf.~\citetalias{Sullivan_2015} Sec 3.5).
Note that our only noise contribution which is not white comes from stellar 
variability, 
which following~\citetalias{Sullivan_2015} we assume to be independent over 
transits and also independent of the duration of a given transit.
Finally, we multiply the SNR per transit by the square root of the number of 
transits
observed.

We have changed some aspects of this simulation
since the work by~\citetalias{Sullivan_2015} was published. These
changes are described in Sec.~\ref{sec:changes_from_S15}.

\subsection{Selecting Target Stars (and Modelling Full Frame Images)}
\label{sec:selection_criteria}
For the Primary Mission, \tesss short cadence (2 min) targets will be drawn 
from a subset of the \tic. 
The exact manner in which these targets will be chosen has not yet been 
determined.

We know that for \tess to detect small transiting planets
it should observe stars that are small and bright.  For this work, we
define a simple statistic, \texttt{Merit}, proportional to the SNR we
should expect from a given planet orbiting the star:
\begin{align}
	\texttt{Merit} &\equiv 
	\frac{1/R_\star^2}{\sigma_\text{1-hr}(I_c)/\sqrt{N_\text{obs}}}\ ,
	\label{eq:merit}
\end{align}
where $R_\star$ is the radius of the star in question,
$\sigma_\text{1-hr}$ is the relative precision in flux measurements
over one hour of integration time, taken from the solid black line of
Fig~\ref{fig:noise_with_moon}, $I_c$ is the Cousins band $I$ magnitude
\tess observes for the star (or more precisely, the star system) and
$N_\text{obs}$ is the number of observations the star receives over
the course of the mission.  For multiple star systems, $R_\star$ is
taken to be the radius of
the planet-hosting star, and the $I_c$ magnitude is computed from the 
combined flux of all the stars.

\begin{figure*}[!tb] %[!thb]
	\centering
	\includegraphics{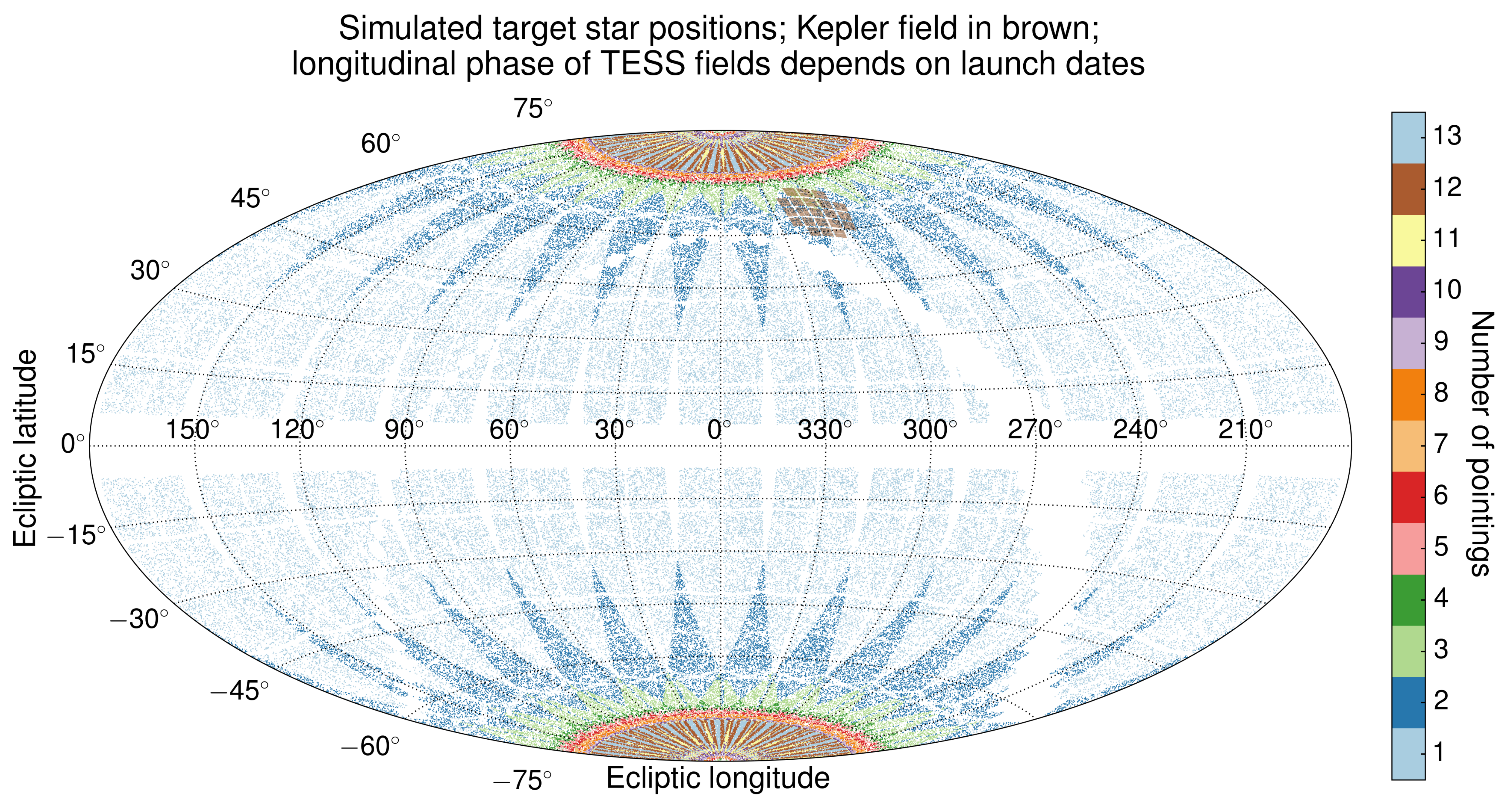}
	\caption{Selected target stars in the Primary Mission. Their surface 
		density increases as $\sqrt{N_\text{obs}}$ towards the poles because of 
		the 
		weight used	in Eq.~\protect\ref{eq:merit}. Gaps in the focal plane 
		array 
		between each of a 
		given camera's four CCDs leads to the slight deviations from 
		continuous observing at the ecliptic pole.}
	\label{fig:positions_pointings}
\end{figure*}
\begin{figure}[!tb] %[!thb]
	\centering
	\includegraphics{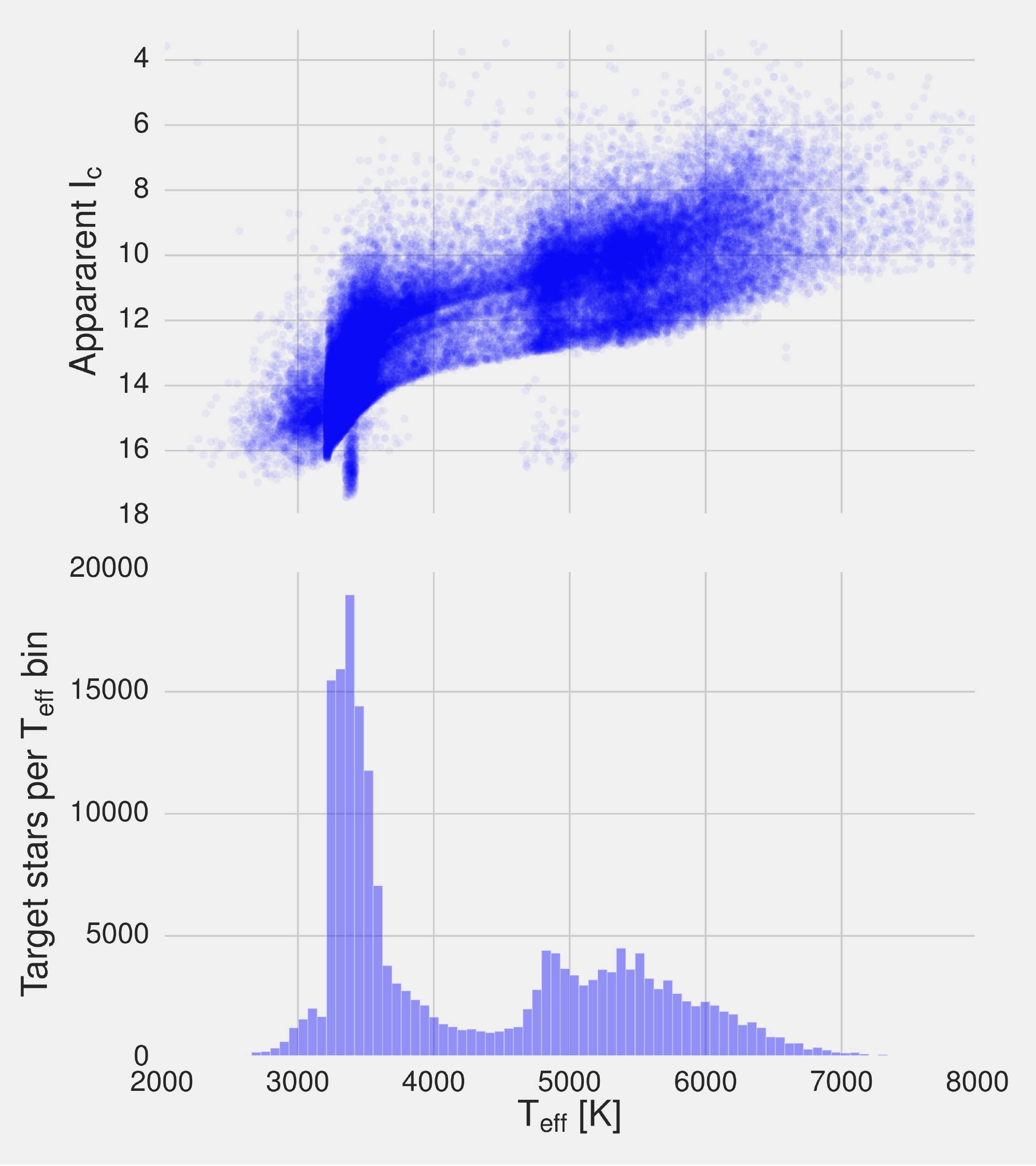}
	\label{fig:fig17_replica}
	\caption{Target star Cousins I magnitude against effective temperature 
		(replicates Figure 17 of~\protect\citetalias{Sullivan_2015}). 
		Target stars are selected as the best $2\times10^5$ stars according to 
		$\texttt{Merit}\equiv 
		\sqrt{N_\text{obs}}(1/R_\star^2)/\sigma_\text{1-hr}(I_c)$. 
		The top subplot shows 1 in 10 stars. This simple model could inform the 
		target selection to be performed on the \tic. 
		The lower histogram is bimodal, selecting heavily for M
		dwarfs, and selecting more F and G dwarfs than K dwarfs. This
		shape arises from the combined $1/R_\star^2$ and
		$1/\sigma_\text{1-hr}(I_c)$ weights: the fact that the minimum
		falls between FG and M stars is because there are many more FG stars 
		than K stars in our
		catalog, and because M stars offer the largest signals. Note also the
		dip in the TRILEGAL (\& observed) $V$-band luminosity functions for K 
		stars
		(see~\citetalias{Sullivan_2015} Figure 5).
		% N.b. Winn showed this bias ANALYTICALLY in the searchable stars memo 
		%(he showed there are ~4* more stars e.g. in 0.75<R/Rsun<0.85 than 
		%0.5<R/Rsun<0.6)
		
		The outliers visible in the upper scatter plot represent nonphysical 
		"stars",
		possibly artifacts of the Padova-to-Dartmouth conversion performed 
		by~\citetalias{Sullivan_2015}.
		These comprise fewer than 1\% of the target stars, and are ignored in 
		what follows.
	}
\end{figure}
\begin{figure}[!tb]
	\includegraphics{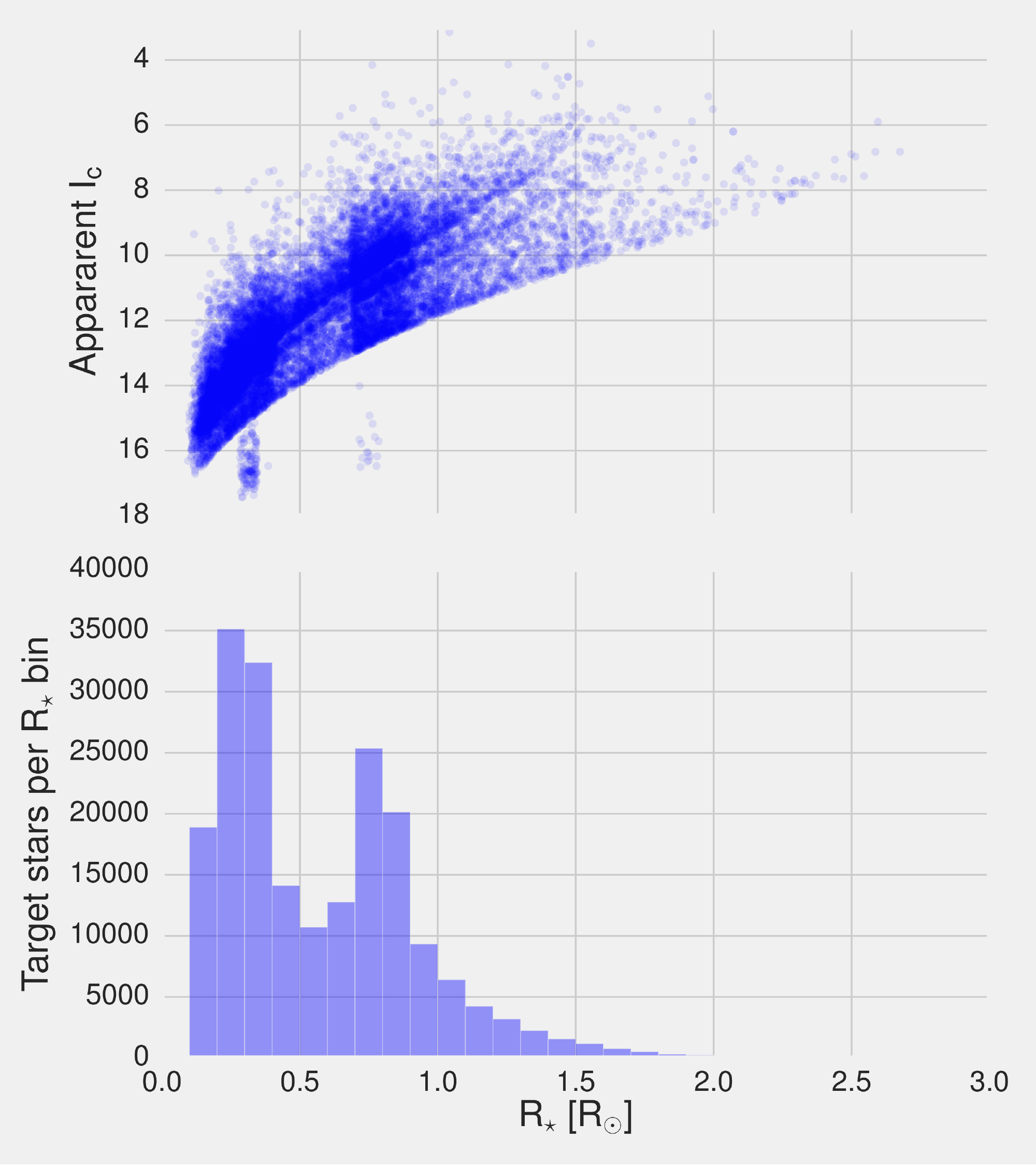}
	\label{fig:fig17_radius_on_x}
	\caption{Same as Fig.~\protect\ref{fig:fig17_replica}, but as a function of 
		stellar radius. $1/R_\star^2$ selection weight is visible as the 
		envelope 
		of the upper subplot, and the same outliers are present.
	}
\end{figure}

We evaluate \texttt{Merit} for all the star systems in our modified
TRILEGAL catalog, and then choose the highest-\texttt{Merit} 
$2\times10^5$ as target
stars to be observed at 2 minute cadence.  Target stars selected in
this manner are shown in Fig.~\ref{fig:positions_pointings}. 
This statistic is simpler than the procedure outlined in Section 6.7
of~\citetalias{Sullivan_2015} and it produces a nearly identical
population of target stars (shown in Fig~\ref{fig:fig17_replica}).
Our approach for full frame image simulation is different from that
of~\citetalias{Sullivan_2015}, and we describe it below.

We generalize our \texttt{Merit} statistic to Extended Missions as 
follows: over an
entire mission, the total number of observations a star receives is
the sum of its observations in the Primary and Extended Missions:
$N_\text{obs}=N_\text{primary}+N_\text{extended}$.  If
$N_\text{extended}=0$ for a given star, then do not select that star as a
target star in the Extended Mission.  Else, compute its \texttt{Merit}
(Eq.~\ref{eq:merit}) substituting
$N_\text{obs}=N_\text{primary}+N_\text{extended}$.  In this manner
stars that are observed more during the Primary Mission are more
likely to be selected during the Extended Mission.

\paragraph{Alternative prioritization approaches:}

It is worth emphasizing that our scheme for selecting target stars for
an Extended Mission does not make use of any information on whether
candidate transit events were detected during the Primary Mission.  If
a star were observed at short cadence for an entire year, and no
candidate events were found, it might be sensible to disregard that
star in the Extended Mission in favor of stars that were never
observed at short cadence -- particularly those with candidate events
that were detected in the Primary Mission full-frame images.  These
and related concerns are discussed further in the accompanying
document\textsuperscript{\ref{fn:wiki}}.

More abstractly, the procedure of simply applying Eq.~\ref{eq:merit}
attempts to select a stellar sample that will yield the most small
transiting planets around the brightest stars.  An alternative
approach would be to select stars that will give the most 
\textit{relative benefit} in 2 minute postage stamps over 30 
minute cadence observations since all stars will be present in the
30 minute images.  This `relative benefit' could be based on
improvement in transit detectability, or perhaps improved 
capacity to resolve the partial transit phases.

For purposes of transit detection, the difference between 2 and 30
minute cadence matters most when transits have short durations -- in
other words for small stars, and for close-in planets.  Switching to
this alternative approach would consequently bias us even more
strongly toward selecting M dwarfs.  We already select almost every M
dwarf with $I_c < 14$.  The limiting $I_c$ magnitude for detecting
$R_p > 4R_\oplus$ planets with \tess is $\sim\!16$, which is where we
see the dimmest stars in Fig.~\ref{fig:fig17_replica}.

Additionally, the procedure of applying Eq.~\ref{eq:merit} and
assuming that it will maximize the number of small planets that \tess
will detect about bright stars ignores the dependence of
planet occurrence rates on stellar properties, or on the properties of
planets already known to exist around those stars. If the goal is literally to 
maximize the number of planet
detections (a goal which we do not advocate) then \tesss target selection might 
take these dependences into
account. For example the planet occurrence rates measured with {\it Kepler} 
data could be used
to upweight those types of stars most likely to have planets.
Likewise~\protect\citet{kipping_transit_2016} note that the
probability for a star with short-period transiting planets to have additional 
transiting
planets is dependent upon the radii and periods of the
short-period planets.  They navigate the optimization problem using artificial 
neural
networks (ANNs) trained to select for features that improve the
probability of detecting transiting outer companions.  \tess might
benefit from a similar approach in target selection.

\paragraph{Alternative prioritization approaches in Extended Missions:}
Our \texttt{Merit} statistic also neglects the possibility of an Extended
Mission which only observes stars with known planets or planet
candidates (\tesss objects of interest, or those from other transit
and RV surveys) at short cadence.  This approach would free up a
considerable portion of \tesss data mass for full frame images at
\textit{e.g.}, 15 minutes rather than the current nominal 30 minutes.

\paragraph{Approach to full-frame images:}
\label{sec:FFI_simulation}
We want to simulate the full frame image detections in a
computationally tractable manner.  While~\citetalias{Sullivan_2015}
evaluated the phase-folded SNR for every potentially transiting object
about each of the $\sim\!1.6\times10^8$ stars in our synthetic catalog,
we consider only the stars for which \tess could plausibly detect a
sub-Neptune planet during a 3-year mission.  Most stars that
\tess sees are too dim or too large to detect $R_p<4R_\oplus$ planets
-- while we expect many giant planet detections towards the galactic
plane (\citetalias{Sullivan_2015} Fig 19), small-planet detections are
more nearly isotropic, since practically all occur for stars at
$<1\rm\ kpc$.  While thousands of transiting giant planets will be revealed by 
\tess,
we assume here that the prospects for detecting smaller planets are more likely
to help discriminate between different scenarios for the Extended Mission.

In this vein, we only simulate full frame image detections for the
$3.8\times10^6$ highest \texttt{Merit} stars following the
$2\times10^5$ highest \texttt{Merit} stars that are selected for 2-minute 
observations.
This number ($3.8\times10^6$) was initially estimated based
on the number of searchable stars about which we expect \tess to be
able to detect sub-Neptune radius
planets~\citep{winn_searchable_2013}. Our model for the detection process for 
the FFI data is 
identical to that of the PS data, except that the effects of time-smearing on
the apparent transit duration and depth are taken into account (which is
important for transit durations $\lesssim 1$~hour).

To check that the $4\times10^6$ highest \texttt{Merit} stars includes 
all stars about which \tess might detect sub-Neptune radius planets, 
we repeated this process for the Primary Mission using $5.8\times10^6$,
$9.8\times10^6$ and $19.8\times10^6$ `full frame image' stars, and
confirmed that there was no significant difference in the planet
yields at $R_p\le4R_\oplus$ between any of the cases.
%see ext_sim_notes/160817_on_the_FFI_assumtion. (10 Monte Carlo realizations 
%each)
With an increased number of FFI stars, the number of detected giant planets 
also increased,
particularly near the galactic
disk.  Meanwhile the number of sub-Neptune planets remained
fixed. This convinced us that $4\times10^6$ is a sufficient
number of target stars to compute detection statistics for sub-Neptune radius 
planets.

\subsection{Earth and Moon Crossings}
\label{sec:earth_moon_crossings}
\begin{figure}[!tb]
	\centering
	\includegraphics{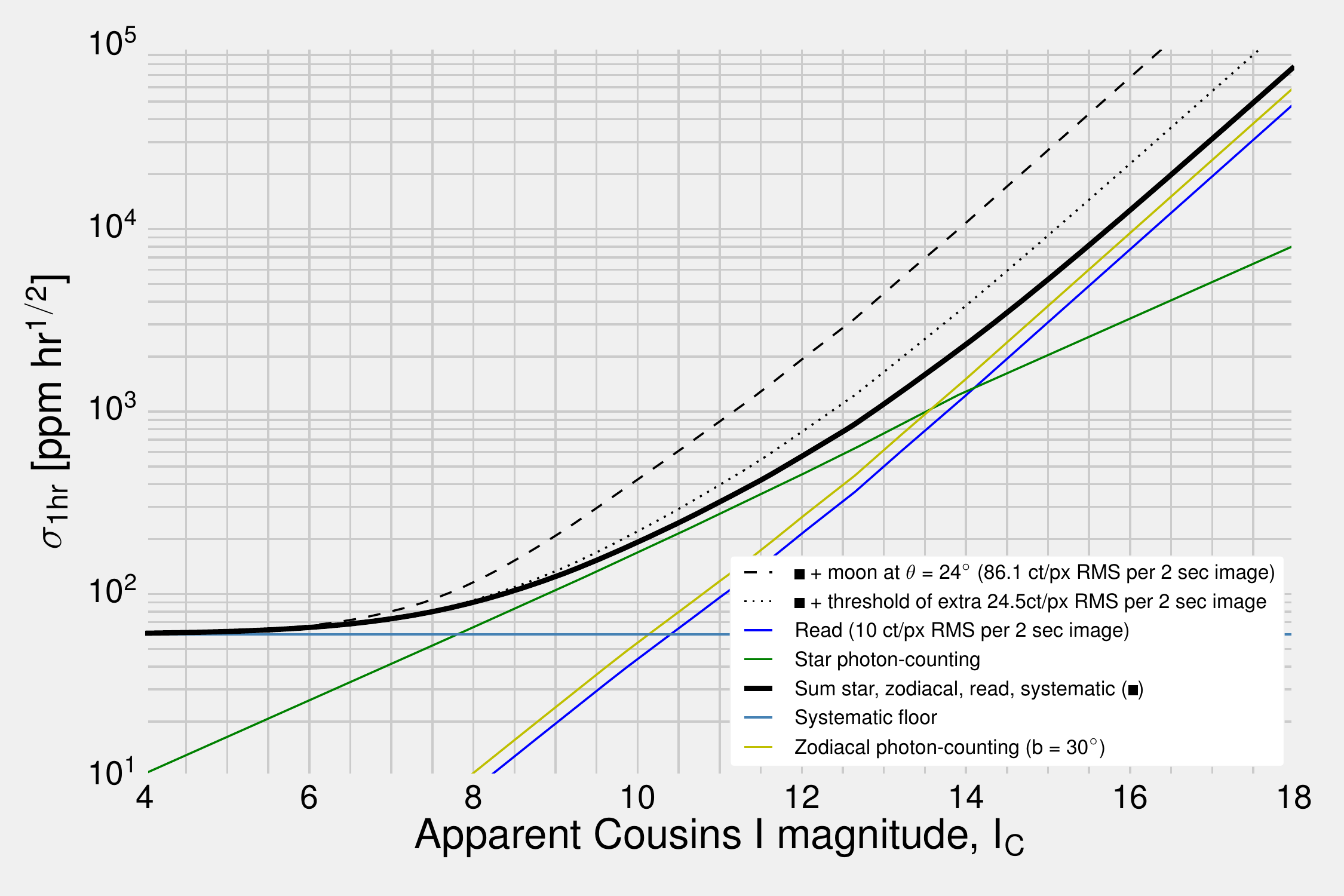}
	\caption{Relative precision in measured flux over a one hour integration 
		time. The noise sources described by~\protect\citet{Sullivan_2015} are 
		solid thin lines.
		Additional scatter from contaminating background stars is not shown.
		The dotted line corresponds to the predicted noise budget with an 
		additional background flux from the Moon 
		at $24^\circ$ from a camera's boresight.	
		The dashed line corresponds to the predicted noise budget including an 
		additional threshold flux, $F_\mathrm{thresh}\equiv 
		300\,\mathrm{ct/px/s}$, that we use to flag Moon and Earth 
		contamination in 
		\tesss field of view (see text).
		Zodiacal background is plotted for an ecliptic latitude of $30^\circ$ 
		using 
		Eq. 21 of~\protect\citet{Sullivan_2015}.
		The appended 
		Figs.~\protect\ref{fig:sun_scat}-\protect\ref{fig:earth_scat} 
		show this plot with scattered sunlight, moonlight, and Earthlight at 
		arbitrary angles.
	}
	\label{fig:noise_with_moon}
\end{figure}

When the Earth or Moon passes through \tesss camera fields, the CCD
pixels are flooded to their full well capacity ($\sim2\times10^5$
photoelectrons).  Precision differential photometry becomes impossible
in any pixels that are directly hit during these crossings.  Even when
the Earth or Moon is not directly in the camera's field of
view, light may scatter off the interior of a camera's lens hood
and act as a source of contaminating flux. The Poisson noise in the number of 
photons arriving from the Earth or the Moon in such a scenario degrades \tesss
photometric performance, although a detailed model for the scattered light is 
not yet available.

We first illustrate with an example.
When the Moon is $24^\circ$ from a camera boresight the lens hood 
is expected to suppress $1/100^\mathrm{th}$ of a mean 
$3\times10^6\,\mathrm{ct/px}$ (per 2 second image) from incoming 
moonlight\footnote{Our model for suppression as a function of angle is 
	appended in Fig.~\ref{fig:lens_hood_suppression}. For estimates in this 
	section 
	we assume 
	100\% quantum efficiency, so photons, electrons, and counts are 
	interchangeable; 
	$1\,\mathrm{ph} = 1\,\mathrm{e^{-1}} = 1\,\mathrm{ct}$.}.
Thus a mean flux of $3\times10^4\,\mathrm{ct/px}$ reaches the cameras per 
image. 
Assuming the arrival of these photons is Poissonian, the additional RMS is 
then $173\,\mathrm{ct/px}$ per 2 second image.
Including this additional variance in the quadrature sum of variances of all 
noise sources, Fig.~\ref{fig:noise_with_moon} shows that this background 
is the dominant noise source for stars with $I_c \gtrsim 8$ -- nearly all 
target stars.
While the effect's magnitude is highly dependent on the angle between a 
camera's
boresight and the Moon or Earth, the general picture is that for field angles 
$\theta \lesssim25^{\circ}$, the impact can be severe.

Separately from our planet detection simulation, we study these 
crossings in a dynamical simulation based on JPL NAIF's standard SPICE toolkit.
Given a nominal launch date (we assume Dec 20, 2017), this code determines 
\tesss orbital phasing throughout its entire mission. 
At every time step of the three-body orbit, we calculate the distance between 
\tess and the other two bodies of interest, and the separation angles among 
each of the four cameras and each of the two bodies (eight angles in total). 
The spacecraft's inclination oscillates in the simulation as it will in 
reality.

Taking the Earth and Moon's integrated disk brightnesses as fixed 
values\footnote{The full moon's apparent magnitude is $I_\moon \approx -13.5$. 
	Scaling from photon fluxes tabulated in~\citet{winn_photonflux_2013}, this 
	gives $2.5\times10^{13}\ \mathrm{ct/s}$. Averaging over the number of 
	pixels 
	in the focal plane array, 
	this gives a mean additional flux per image of $3\times10^6\ 
	\mathrm{ct/px}$.
	The Earth will contribute a mean flux $\sim {80}$ times greater.
	This substantiates the claim that the Earth and Moon flood the cameras to 
	their full well capacity. 
	The effect on observing precision is appended in 
	Fig.~\ref{fig:outage_vs_background}.} we use a model for scattered light 
suppression from 
the \tess lens hoods (Fig.~\ref{fig:lens_hood_suppression}) to compute the 
mean photon flux from each of these bodies onto each of the cameras throughout 
the orbit. We then compute the corresponding variance in incoming flux, and 
compare it to \tesss noise budget, similar to the example in 
Fig.~\ref{fig:noise_with_moon}.

To evaluate the cumulative impact of Earth and Moon crossings on \tesss 
Primary and Extended Missions we ask: for each camera, what fraction of the 
total observing time is \tess unable to operate at desired photometric 
precision because of Earth and Moon crossings?
An upper limit for what we mean by `unable to operate at desired photometric 
precision' is when terrestrial or lunar flux make it impossible to observe a 
sizable portion of the stars in the \tess target star catalog with reasonable 
precision.
Following our example of when the Moon is $24^\circ$ from the camera boresight, 
the appended Fig.~\ref{fig:outage_vs_background} shows that $\sim$50\% of the 
stars that could be observed at sub-mmag precision over an hour no longer can.
This of course depends on the target star catalog's apparent magnitude 
distribution (Fig.~\ref{fig:fig17_replica}), and the cutoff of `sub-mmag 
precision over an hour' is arbitrarily selected.

\begin{figure}[!t]
	\centering
	\includegraphics[angle=90,width=1.05\textwidth]{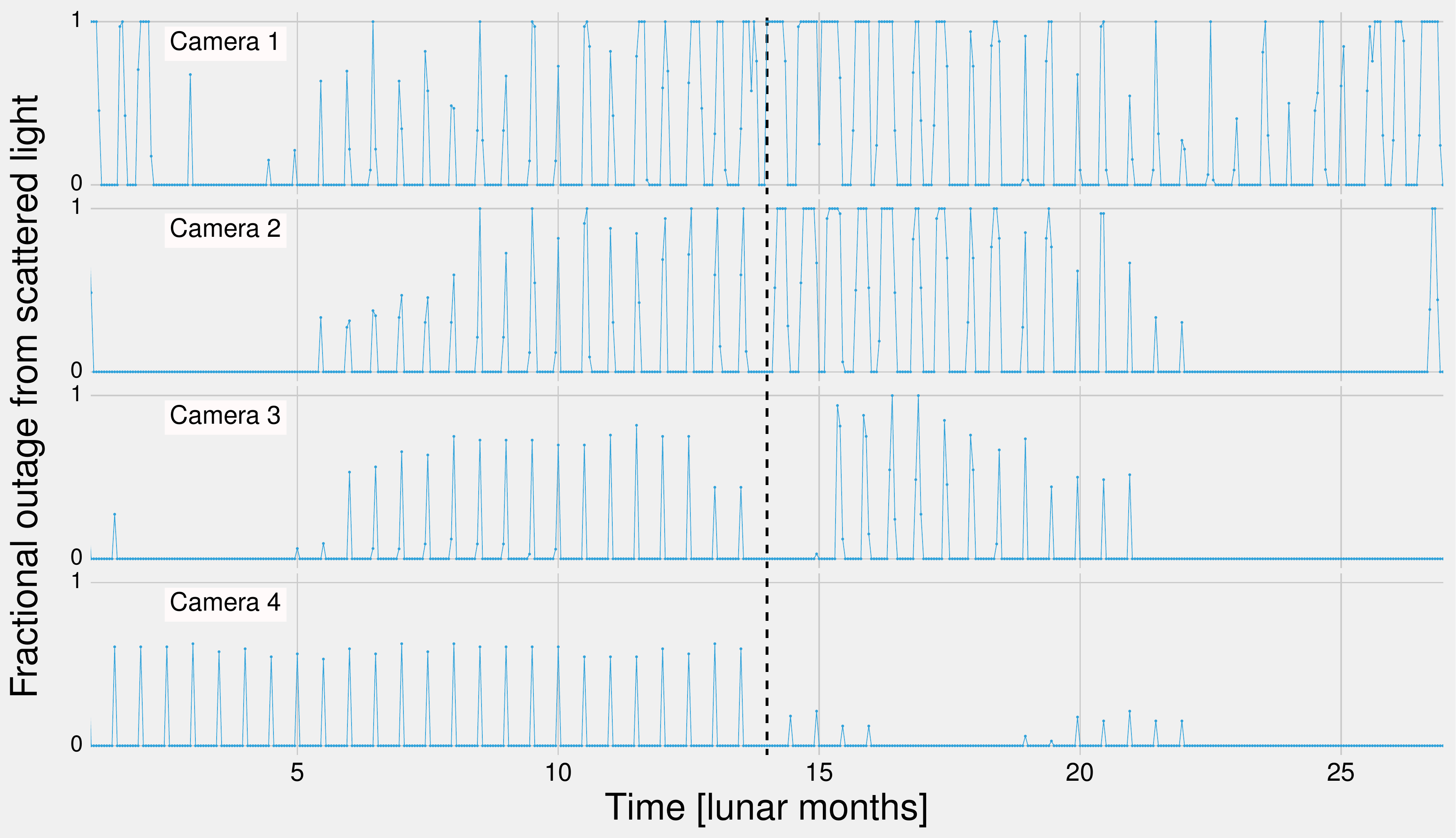}
	\caption{`Outage' caused by direct and scattered Earthlight and Moonlight 
		as a function of time in \tesss orbit.
		Specifically, the $y$-axis shows the probability of having a polluted 
		frame 
		with mean incident flux $>300\,\mathrm{ct\,px^{-1}\,s^{-1}}$ in a 
		time-step 
		bin of 1/20$^\mathrm{th}$ of an 
		orbit (where the probability is the number of polluted frames divided 
		by 
		the total number of frames in that bin).
		The mean of this outage across any given year is used to compute the 
		number of dropped fields in Table~\protect\ref{tab:dropped_fields}, 
		which should then be an upper bound to the impact of Earth and Moon 
		crossings.
		In this plot, the first year of observations are in the southern 
		ecliptic 
		hemisphere. The dashed line marks the beginning of `Year 2' (northern 
		hemisphere). 
		\tess completes two orbits per lunar month.
		The worst fractional outage per orbit is in Camera 1, which 
		points towards the ecliptic, over the first $\sim\!5$ orbits of the 
		second 
		year.
		The plot has `spikes' because outages only occur over a small 
		fraction of \tesss orbit when the relevant orbits align.}
	\label{fig:earth_moon_primary}
\end{figure}
%\clearpage
\begin{figure}[!t]
	\centering
	\includegraphics[angle=90,width=1.05\textwidth]{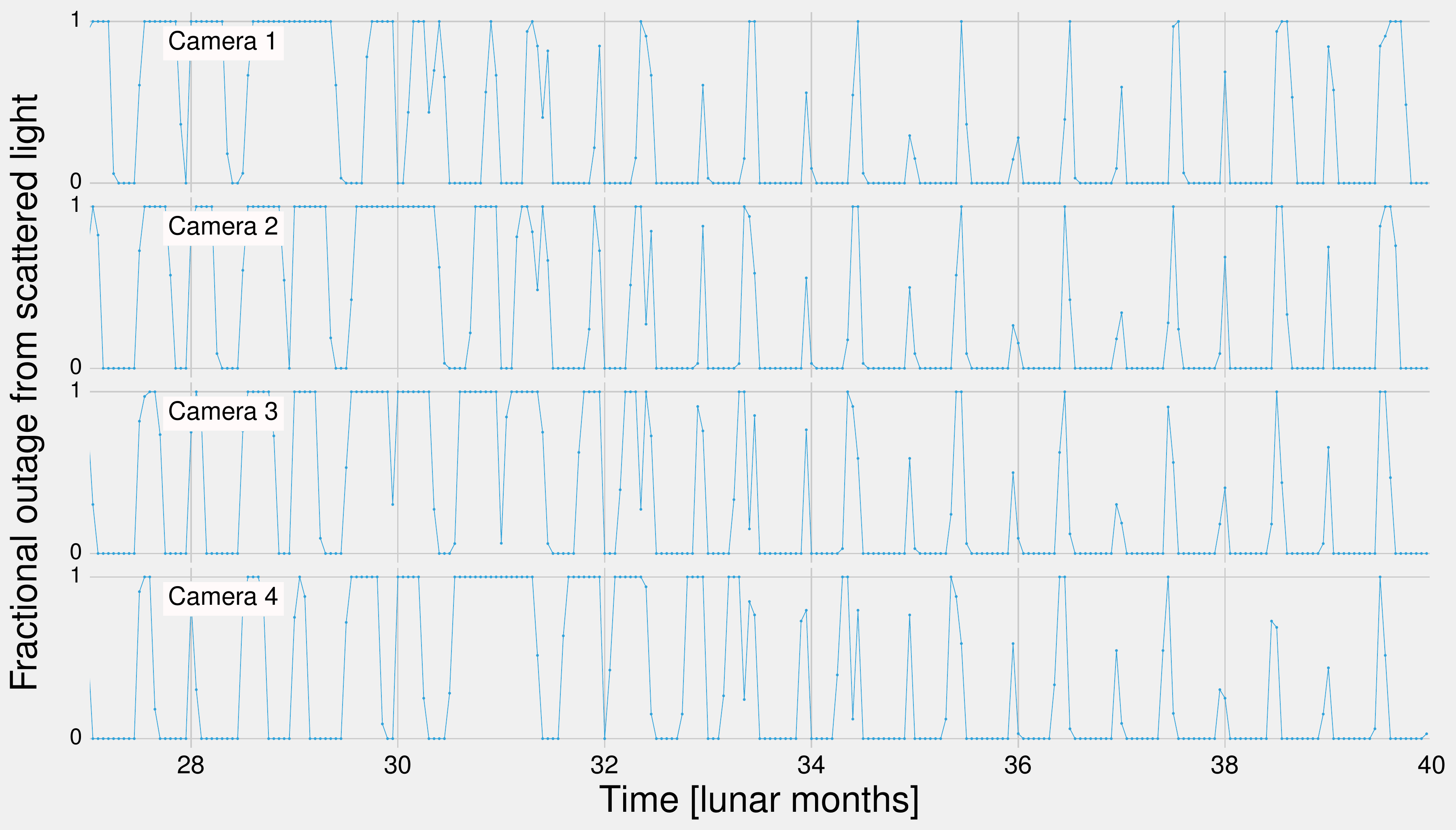}
	\caption{Same as Fig.~\protect\ref{fig:earth_moon_primary}, except showing 
		a hypothetical third year in which \tess observes the ecliptic with the 
		cameras' long axis along the ecliptic plane for the entire year. The 
		latter 
		half of the year experiences far less Earth and Moon interference than 
		the 
		first half. Considering it implausible that we would opt to sacrifice 
		such 
		a large fraction of our observing time, we study the \elong\:and 
		\eshort\:scenarios instead, as their observations avoid this effect 
		during 
		the first $\sim5$ months.}
	\label{fig:earth_moon_elong}
\end{figure}
\begin{table}[!tb]
	\centering
	\begin{tabular}{ | l | l | l | l | l | }
		\hline
		\ & Camera 1 & Camera 2 & Camera 3 & Camera 4 \\ \hline
		\multicolumn{1}{|c|}{Year 3 selected} & \  & \  & \  & \  \\ \hline
		\npole & 0 & 0 & 0 & 0 \\ \hline
		\nhemi & 2 & 1 & 0 & 0 \\ \hline
		\shemiAvoid & 2 & 0 & 0 & 0 \\ \hline
		\elong & 1 & 1 & 1 & 1 \\ \hline
		\eshort & 0 & 1 & 1 & 0 \\ \hline
		\hemis & 1 & 1 & 0 & 0 \\ \hline
		\multicolumn{1}{|c|}{Year 3 omitted} & \  & \  & \  & \  \\ \hline
		\texttt{pole\,(south)}  & 0 & 0 & 0 & 0 \\ \hline
		\texttt{hemi\,(south)} & 1 & 0 & 0 & 0 \\ \hline
		\texttt{hemi+ecl\,(north)} & 4 & 3 & 0 & 0 \\ \hline
		\texttt{ecl\_long\,(1yr)} & 4 & 4 & 4 & 4 \\ \hline
		\texttt{ecl\_short\,(1yr)} & 1 & 3 & 4 & 2 \\ \hline
		\texttt{allsky\,(poles)} & 0 & 0 & 0 & 0 \\ \hline
		\multicolumn{1}{|c|}{Primary Mission} & \  & \  & \  & \  \\ \hline
		\texttt{hemi\,(south year 1)} & 2 & 1 & 0 & 0 \\ \hline
		\texttt{hemi\,(north year 2)} & 4 & 2 & 0 & 0 \\ \hline
	\end{tabular}
	\caption{Number of sectors (of 13 per year) `dropped' due to the Earth and 
		Moon crossings in both selected and omitted Extended Missions, with 
		those 
		of the Primary Mission for reference. The method of `dropping' fields 
		(which neglects the temporal nature of the crossings, discussed in the 
		text) gives an approximate sense of the cumulative impact of Earth and 
		Moon 
		crossings. 
		Scenarios with \texttt{(south)} or \texttt{(north)} appended refer to 
		their
		counterparts in that hemisphere.
		\texttt{ecl\_long\,(1yr)} corresponds to a full year with the long axis 
		of \tesss field of view along the ecliptic, while 
		\texttt{ecl\_short\,(1yr)} corresponds to the same, but with the short 
		axis along (long axis perpendicular to) the ecliptic. These scenarios 
		are neglected because their outages are time-correlated (see 
		Fig.~\protect\ref{fig:earth_moon_elong}).
		\texttt{allsky\,(poles)} would be a scenario that observes in the 
		manner of \hemis, but only on the ecliptic poles as in \npole.}
	\label{tab:dropped_fields}
\end{table}

Developing a detailed model of how these crossings impact \tesss photometry is 
beyond the scope of this work.
To account at least qualitatively for their effect in our simulation we adopt a 
simple approximation: we impose that a camera has an `outage' if there are over 
$F_\text{thresh}\equiv300\ \text{ct/px/s}$.
This is roughly twice the maximal zodiacal background noise, which in our model 
ranges from $48$-$135\,\mathrm{ct/px/s}$ (see \citetalias{Sullivan_2015} Sec 
6.4.1). At an ecliptic latitude of $30^\circ$, Eq. 21 of 
\citetalias{Sullivan_2015} gives $75\,\mathrm{ct/px/s}$, or about 
$F_\mathrm{thresh}/4$.

We then compute the time-averaged fraction of each observing sector for which 
\tesss cameras remain above this threshold. 
A given sector has $\sim660$ hours of observing time over two spacecraft orbits.
As an example, suppose that $220$ of these hours had the Earth, the Moon, or 
both shining with a background flux $F > F_\text{thresh}$. 
This would correspond to a fractional outage of $1/3$. 
We then compute the mean of this fractional outage across all 13 annual sectors 
to derive a `mean camera outage' for each proposed pointing scenario.
We then approximate the effect of Earth and Moon crossings by selectively 
omitting the closest integer number of observing sectors corresponding to the 
`mean camera outage' described above.
For instance, if the `mean camera outage' was 17\% of \tesss observing time 
over a given year, we would omit the 2 (of 13) observing sectors that suffer 
the greatest number of lost hours, for that given camera.
The relevant number of omitted sectors is shown in 
Table~\ref{tab:dropped_fields}; an example of the resulting spatial 
distribution of detected planets is shown in Fig.~\ref{fig:skymap_nhemi}.

As mentioned in Sec.~\ref{sec:planet_detection_model}, our planet detection 
simulation is not explicitly time-resolved; it takes the ecliptic coordinates 
of camera fields for each orbit as input to compute the observed baselines for 
each star in our galactic model.
While our procedure ignores the temporal nature of the `outages' (which is 
shown resolved over time-steps of 1/20$^\mathrm{th}$ of an orbit in 
Figs.~\ref{fig:earth_moon_primary} and~\ref{fig:earth_moon_elong}), it gives a 
sense of the cumulative impact of Earth and Moon crossings over 
the course of a year, which is sufficient for the purposes of this study.

%Our approximation is `conservative' in the sense that 
The additional background corresponding to $F_\mathrm{thresh}=300\ 
\mathrm{ct/px/s}$ becomes a noticeable noise source ($\gtrsim 1.5\times$ lower
precision) for targets with $I_c \gtrsim 12$ (cf.
Fig.~\ref{fig:noise_with_moon}).
This is roughly half of the target stars, with the greatest impact on the M 
dwarf sub-sample.
By approximating all of the relevant time as an `outage' lost across all stars 
(even for the bright ones for which the additional scatter is subsumed by 
other noise sources), our simulation results might be more sensitive to 
scattered light than \tess will be in reality.
However, scattered light might have a non-Poisson component, and is   expected 
to 
be non-uniformly distributed across each focal plane as the Moon or Earth 
moves relative to \tesss pointing.
%would make removing scattered light in post-processing more difficult, and 
This could make the effect worse than simulated.

The impact of this approximation on the planet yields of the Primary and 
Extended Missions are discussed in 
Secs.~\ref{sec:results_from_primary_missions} 
and~\ref{sec:results_from_all_extended_missions}
respectively.
The summarized result is that this model of Earth and Moon crossings leads to 
$\lesssim 10\%$ fewer $R_p<4R_\oplus$ planet detections in the Primary Mission 
compared to the case of 
not accounting for the crossings at all.
%There is thus little reason to raise our low threshold, given that the 
%magnitude of the effect is so small.
%Moreover, the Earth and Moon crossings typically last for a small fraction of 
%an orbit (Fig.~\ref{fig:earth_moon_primary}). If the timescales required for 
%the cameras to `re-settle' after the crossings are small compared to orbital 
%timescales, our approach may over-estimate the effect's importance even 
%further.
Detailed models of this process remain to be developed during the spacecraft's 
commissioning period.

% I think this might be a bit better than it sounds because of how the 
%earth/moon crossings are phased in the orbits. They're only happening for at 
%worst ~1/2 of an orbit (see 
%https://drive.google.com/folderview?id=0B2941jxrlPq0Z0tQZnRpd3pVOHc&usp=drive_web),
% but at roughly the same half across adjacent sectors. I.e. it's the same 
%patches of sky, roughly, that aren't being observed across adjacent sectors, 
%so 
%just dropping the entire sector isn't as bad as you might think on first 
%guess. 
%It's admittedly somewhat sketch.

%A detailed model of this effect would take this dynamical simulation and 
%incorporate it with the extant \tess instrument model to produce full 
%simulated 
%images of the \tess fields every 2 minutes over the course of a specified 
%mission.
%Reduction to postage stamps and full frame images could be modeled in a `true' 
%sense from these simulated images.
%The main apparatus of the planet detection simulation -- the process of 
%creating planets, selecting transiting planets, and producing small images 
%with 
%the transiting planet signal injected, and computing a SNR for these  -- this 
%is a somewhat long process. Maybe more like a masters thesis, and a phd thesis 
%if you did it right.

%\input{assumptions_to_planet_detection_sim}
\subsection{Summary of Key Assumptions and Attributes of Planet Detection 
	Simulations}
\label{sec:input_assumptions}

\begin{itemize}
	
	\item We focus almost exclusively on planets with $R_p < 4R_\oplus$.
	
	\item We assume the TRILEGAL catalog (modified to match
	interferometric radii, as described
	by~\citetalias{Sullivan_2015}) is an accurate representation
	of the stellar neighborhood to $\lesssim2\text{kpc}$.
	
	\item We omit the 5\% of the sky closest to the galactic disk
	(see Fig.~\ref{fig:positions_pointings}). We expect that
	\tesss large pixel size ($21\times21''$) combined with
	crowding near the galactic disk will cause substantial
	source confusion and a large astrophysical false positive
	rate in this area.  On a practical note, TRILEGAL cannot be
	queried within its run-time limit for some of these fields
	(cf.~\citetalias{Sullivan_2015} Sec 3.1).
	
	\item We assume prior knowledge of the radii and apparent
	magnitudes of TRILEGAL's synthetic stars, so we can
	prioritize the stars with a simple planet-detectability statistic
	(Eq.~\ref{eq:merit})\footnote{Although it is difficult to determine
		accurate radii based on currently available data, we expect that by the 
		time
		\tess launches \gaia will provide parallaxes and proper
		motions for many potential \tess targets, which will enable more 
		reliable giant/dwarf discrimination and radius estimates.}.
	
	\item In evaluating a star's \texttt{Merit}
	(Eq.~\ref{eq:merit}), we compute the apparent magnitude of each "star"
	based on the sum of the flux from the star
	itself as well as any companion or background stars
	(whose presence will not be known {\it a priori}).
	
	\item We use occurrence rates of planets as a function of radius
	and orbital period as calculated by~\citet{fressin_false_2013}
	and~\citet{dressing_occurrence_2015}.
	We assume these are accurate for the 
	$P \lesssim 180\ \text{day}$ planets to which \tess is 
	sensitive. We note in passing that~\citet{burke_terrestrial_2015} 
	find slightly higher occurrence rates for planets orbiting GK dwarfs 
	than~\citet{fressin_false_2013}.
	
	\item We assume the occurrence statistics of multiple-planet systems 	
	can be approximated with repeated independent draws from the 
	single-planet distributions. We further assume the orbits of planets 
	in multiple planet systems to be 
	coplanar and stable (with period ratios of at least 1.2 between 
	adjacent planetary orbits).
	
	\item For our instrument and noise models, we assume:
	\begin{itemize}
		
		\item A point-spread function (PSF) derived from ray-tracing
		simulations.
		Compared with the PSF described by~\citetalias{Sullivan_2015} in 
		their Sec 6.1, ours incorporates lower charge diffusion (based on 
		laboratory measurements) and as-built (rather than ideal) optics.
		The net result is a slightly wider PSF, leading to $10\%$ fewer 
		$R_p<4R_\oplus$ planets than
		the same PSF model used by \citetalias{Sullivan_2015}.
		
		\item All stars are observed at the \textit{center} of the
		\tess CCDs. This ignores off-axis and chromatic
		aberrations within the \tess optics, and consequently
		ignores the angular dependence of the pixel response
		function (the fraction of light from a star that is
		collected by a given pixel).
		While~\citetalias{Sullivan_2015} attempted to model the
		field-angle dependence, we found some problems with this approach (see 
		appendix
		Sec.~\ref{sec:changes_from_S15}).  
		Ignoring the field-angle dependence is a simplification that may 
		lead to loss of absolute accuracy, but since this applies uniformly 
		to all 
		the scenarios under consideration, it is still be possible
		to {\it compare} the results of different scenarios
		without much loss of accuracy.
		
		\item The time/frequency structure of all noise (except for
		stellar variability, see Eq.~\ref{eq:snr}) is `white'.  
		This means that we ignore time-correlated instrumental effects 
		such as spacecraft jitter, thermal fluctuations, and mechanical
		flexure, which we expect will be at least partly mitigated
		by the mission's data reduction pipeline.
		
		\item We assume the instruments perform equally well in year 3
		as in years 1 and 2.
		
		\item The assumed contributors to white noise include: CCD
		read noise, shot noise from stars, a systematic noise floor
		of 60 $\mathrm{ppm}\cdot\mathrm{hr}^{1/2}$, and zodiacal
		background. See Fig.~\ref{fig:noise_with_moon} for the
		relative contributions of these terms as a function of
		apparent magnitude.
		
		\item The noise contributions from stellar intrinsic
		variability are assumed to be identical to those described
		by~\citetalias{Sullivan_2015} Sec3.5, which uses variability
		statistics from the \textit{Kepler} data computed
		by~\citet{basri_comparison_2013}.  Unlike all previously
		mentioned noise sources, we do not scale noise from stellar
		variability as $t_\mathrm{obs}^{-1/2}$, since the photon
		flux from stars may vary over time-scales similar to typical
		transit durations.  Instead, we assume the noise
		contribution from stellar variability is independent across
		transits, and thus scales as $N_\mathrm{tra}^{-1/2}$, for
		$N_\mathrm{tra}$ the number of observed transits.
	\end{itemize}
	
	\item Our detection model is specified by Eq.~\ref{eq:detection_criterion}.
	Specifically,
	
	\begin{itemize}
		
		\item We require $\geq$2 transits for detection.  We assume
		the period can be recovered without ambiguity and likewise
		there is no ambiguity in identifying which target star is
		exhibiting a given transit signal.
		We also assume a step-function detection threshold: for 
		$\mathrm{SNR_\mathrm{phase-folded}} > 7.3$ and $N_\mathrm{tra} \geq 2$, 
		we rule transiting 
		planets as detected, else they are not.
		
		\item The top $2\times 10^5$ $\mathtt{Merit}$-ranked targets
		(Eq.~\ref{eq:merit}) are observed at two-minute cadence, and
		the next $3.8\times10^6$ stars are observed at thirty-minute
		cadence.  We use~\citetalias{Sullivan_2015} Sec. 6.8
		approach to `blurring' transits with durations $\lesssim
		1\mathrm{hr}$, so that for longer cadence images shorter
		transits get shallower depths and longer apparent durations.
		As described in Sec.~\ref{sec:FFI_simulation}, we verify
		that under this assumption, our selection of PS stars is
		nearly complete for all detectable planets with
		$R_p<4R_\oplus$, and it is highly incomplete for Jupiter-sized planets.

	\end{itemize}
	
	\item We do not assume any prior knowledge of previous
	observations that may have been performed on our stars.  For
	instance, observing the ecliptic, we do not account for
	\tess\!-\ktwo overlap.
	
	\item For Earth and Moon crossings, we assume we can drop a
	fixed number of orbits of observing time for the cameras
	that suffer most from the Earth, the Moon, or both being in
	\tesss camera fields. We summarize this effect in
	Table~\ref{tab:dropped_fields}. Although this ignores the
	time-correlated nature of the outages shown in
	Figs.~\ref{fig:earth_moon_primary}
	and~\ref{fig:earth_moon_elong}, it is sufficient for
	comparing detected planet yields across missions, and
	leads to $\lesssim$10\% fewer $R_p<4R_\oplus$ planet 
	detections compared to the case of not accounting for the 
	crossings at all.
	
	\item We assume that we can (eventually) discriminate between
	astrophysical false positives (for instance background
	eclipsing binaries or hierarchical eclipsing binaries) and
	planet candidates.
	
	\item We can compute SNR with effective transit depth
	$\delta_\text{dil} = \delta \times D$ for
	$\delta=(R_p/R_\star)^2$ the transit depth, and
	\begin{equation}
		D = \frac{\Gamma_T}{\Gamma_N + \Gamma_T}
		\label{eq:dilution}
	\end{equation}
	where $D$ is the dilution factor of a target star with
	incident photon flux $\Gamma_T$ from the target star and
	incident photon flux $\Gamma_N$ from neighboring stars 
	(\textit{i.e.}, background stars or binary companions).
	
	The noise is computed by creating a synthetic image for every
	host star with a planet that transits above a `pre-dilution'
	SNR threshold (this threshold is imposed for the sake of
	lowering our computational cost).  This $16\times16$ pixel
	image is of the number of photoelectrons \tess sees from the
	star and its companions/background stars at each pixel of each
	CCD.  We produce it through our PSF model, which in turn
	requires the host star's $T_\mathrm{eff}$ and apparent $I_c$.
	Over each image, the noise is computed for a range of possible
	aperture sizes about the brightest pixel
	(see~\citetalias{Sullivan_2015} Secs. 6.2 and 6.3), and then
	finally a single `noise' for each transit is selected by
	choosing the aperture size that minimizes the noise.
	
\end{itemize}

\section{Planet Detection Statistics}  
\label{sec:newly_detected_planet_metrics}

Using the planet detection model described in 
Sec.~\ref{sec:planet_detection_model} and the target selection procedure of 
Sec.~\ref{sec:selection_criteria}, we simulate three years of \tess observing, 
for six different possibilities for the third year:
\nhemi, \npole, \shemiAvoid, \elong, \eshort, and \hemis.
How many new planets do we detect, and how do their properties differ between 
Extended Missions?

\subsection{Planet Yield from the Primary Mission}
\label{sec:results_from_primary_missions}
\begin{marginfigure} %[t]
	\centering
	\includegraphics[width=\textwidth]{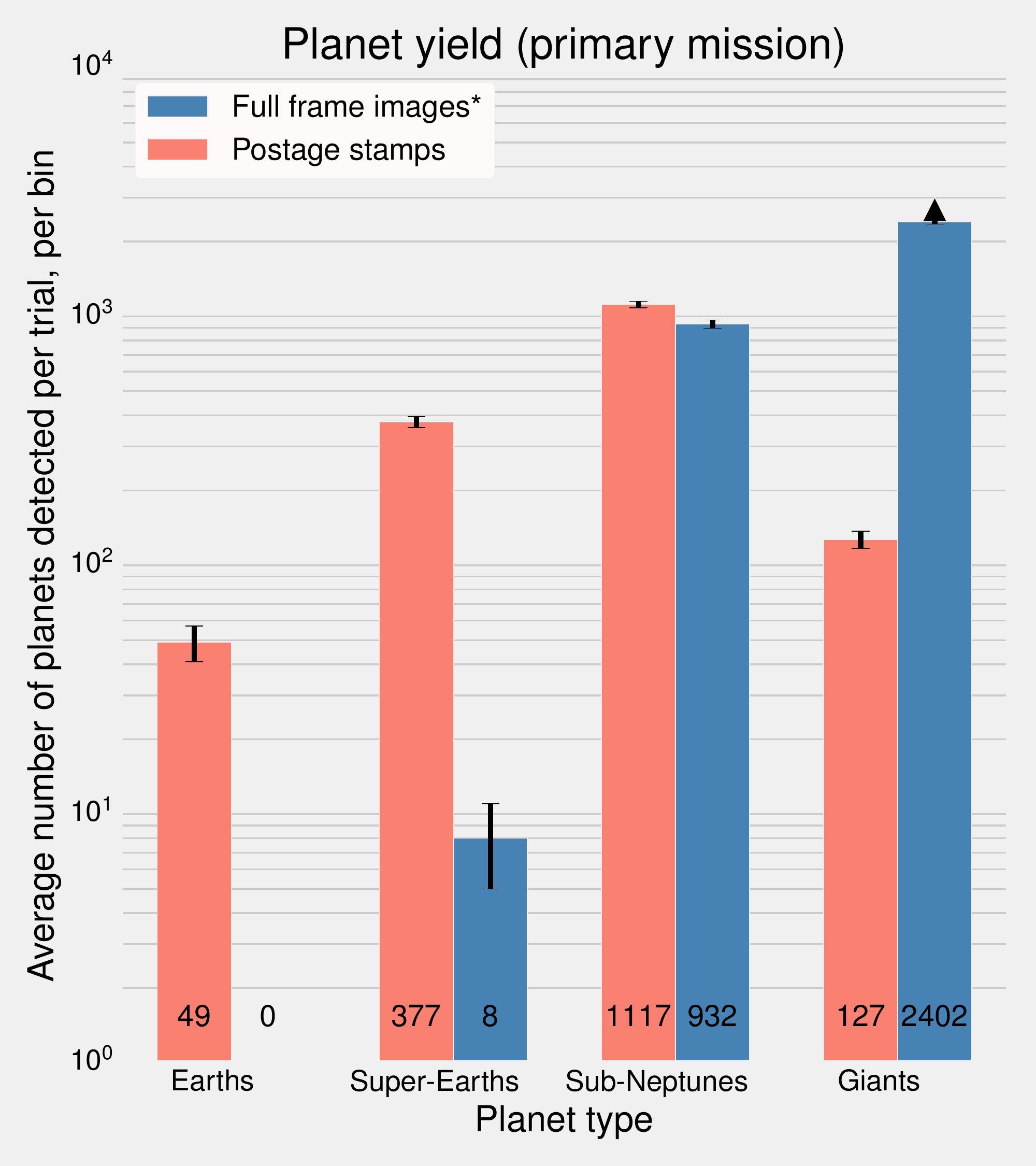}
	\caption{Mean numbers of planets detected in \tesss Primary Mission.
		The number of Earths ($R_p < 1.25R_\oplus$), super-Earths 
		($1.25R_\oplus \le R_p < 2R_\oplus$), sub-Neptunes ($2R_\oplus \le R_p 
		< 4R_\oplus$) and giants agree with the respective values quoted in 
		\protect\citet{Sullivan_2015} to $\lesssim 50\%$. 
		%despite modifications to our target selection procedure 
		%(Sec.~\protect\ref{sec:selection_criteria}).
		Our full frame images detections are complete for $R_p < 4R_\oplus$, 
		and 
		incomplete for giant ($R_p > 4R_\oplus$) planets, shown with a lower 
		limit 
		(see text for discussion). 
		Error bars are from only Poisson fluctuations and do not account for 
		systematic uncertainty.}
	\label{fig:primary_planet_yield}
\end{marginfigure}
We first examine our results for just the Primary Mission -- the first two 
years of \tesss observing. 
We follow with an analysis of our detected planet populations from a single 
Year-3 Extended Mission (Sec.~\ref{sec:results_from_nhemi_extended_mission}), 
and then all six of our proposed Extended Missions 
(Sec.~\ref{sec:results_from_all_extended_missions}).
Here we highlight commonalities and differences 
between~\citetalias{Sullivan_2015} and this work.

\paragraph{Detected planet yield}
The first point of consideration is the detected planet yield, shown in 
Fig.~\ref{fig:primary_planet_yield}.
The number of Earths, super-Earths, and sub-Neptunes we detect agrees with the 
numbers quoted by~\citetalias{Sullivan_2015} to within $50\%$, despite the 
modifications described in Sec.~\ref{sec:selection_criteria} to the target 
selection procedure.
Other changes to our simulation's assumptions, for instance using an as-built 
model of \tesss PSF informed by laboratory tests (courtesy Deborah Woods) 
rather than the idealized PSF described in Sec 6.1 
of~\citetalias{Sullivan_2015}, had only minor impact on this final result 
($10\%$ change in yield).

In the preparation of this report, a
discrepancy emerged between our predicted $\approx 400$ super-Earth detections 
and those shown in Fig.~18 of~\citetalias{Sullivan_2015}, which 
displayed $\approx1400$ planets. The subsequent investigation led to the 
discovery of a bug in the plotting script used to create 
\citetalias{Sullivan_2015}'s Fig.~18 (an Erratum has been published). The 
error did not
affect any of the results described in~\citetalias{Sullivan_2015}'s text, or 
the simulation results that were tabulated in the paper and sent electronically 
to interested parties.
The corrected version 
of~\citetalias{Sullivan_2015}'s Fig.~18 shows $\approx 500$ expected 
super-Earths, which closely agrees with our work when also accounting for the 
dilution error described below.

Another modification was the correction of a bug in the dilution calculations
of~\citetalias{Sullivan_2015}. A single missing symbol\footnote{An $\texttt{=}$ 
	rather than 
	a $\texttt{+=}$} led~\citetalias{Sullivan_2015} to under-account for this 
effect. After correcting this error, the simulations yielded
about 30\% fewer Earth-sized and super-Earth planets than reported by 
\citetalias{Sullivan_2015}.

\paragraph{Properties of planets detected in Primary Mission} 

We show the population properties of planets detected in postage
stamps and full frame images during the Primary Mission in
Figs.~\ref{fig:radius_vs_period_nhemi}
and~\ref{fig:imag_vs_teff_nhemi}.  In terms of the apparent planet
radii $R_p$, orbital periods $P$, host star brightness, and host star
$T_\mathrm{eff}$, we qualitatively agree with the results
of~\citetalias{Sullivan_2015} for the planets detected
in postage stamps. 
For instance, the dearth of $P<5$ day
Neptune-radius ($R_p \lesssim 4R_\oplus$) planets in 
Fig.~\ref{fig:radius_vs_period_nhemi} was
observed by \textit{Kepler}~\citep{mazeh_dearth_2016}, and thus it is
present in our input occurrence rates, rather than being an
observational bias.  It was also seen by~\citetalias{Sullivan_2015}.

The differences between planets detected in postage stamps vs. in full frame 
images follow our expectation from our \texttt{Merit} statistic. 
Namely, Fig.~\ref{fig:imag_vs_teff_nhemi} shows that at a fixed brightness, 
full frame image detections tend to occur at larger stellar effective 
temperature (and thus stellar radius).
At a fixed host star radius, postage stamp detections occur around brighter 
stars.

\paragraph{Impact of earth and moon crossings on Primary Mission's detected 
	planet yield}
During the Primary Mission, of the four cameras, Camera 1 (closest to the 
ecliptic) suffers the most from Earth and Moon crossings.
As noted in Table~\ref{tab:dropped_fields}, we remove 4 of its 13 
`observing sectors' from that year.
This reduces the number of planet detections near the ecliptic, and is visible 
in the orange points of Fig.~\ref{fig:skymap_nhemi}.
In the Primary Mission \tess detects $\sim20$ planets with $R_p<4R_\oplus$ 
(PS+FFI) in each $24^\circ\times24^\circ$ camera field nearest to the ecliptic.
As implemented in our simulation, Earth and Moon crossings result in some 
fields simply not being observed, so in these cases planets orbiting stars in 
these fields are never detected.
Considering only the Primary Mission, we would naively expect that dropping a 
total of 9 fields over the two years (again, see 
Table~\ref{tab:dropped_fields}) would result in a loss of $\sim9\times20=180$ 
planets.
This agrees with what our simulations actually give: running them without 
accounting for Earth and Moon losses returns a mean of 2678 detected planets 
with $R_p<4R_\oplus$, while running them with Earth and Moon crossings gives a 
mean of 2482 such planets (a loss of 196 planets; $7\%$ of the $R_p<4R_\oplus$ 
planet yield).
% data from 160708-t50 and 160729-t20. Slightly apples to oranges because of 
%number of trials difference, but makes the point. 

%\input{planet_yield_from_example_extended_mission_nhemi}
\subsection{Planet Yield from Example Extended Mission: {\rm \nhemi}}
\label{sec:results_from_nhemi_extended_mission}

\begin{figure*}[t]
	\centering
	\includegraphics[]{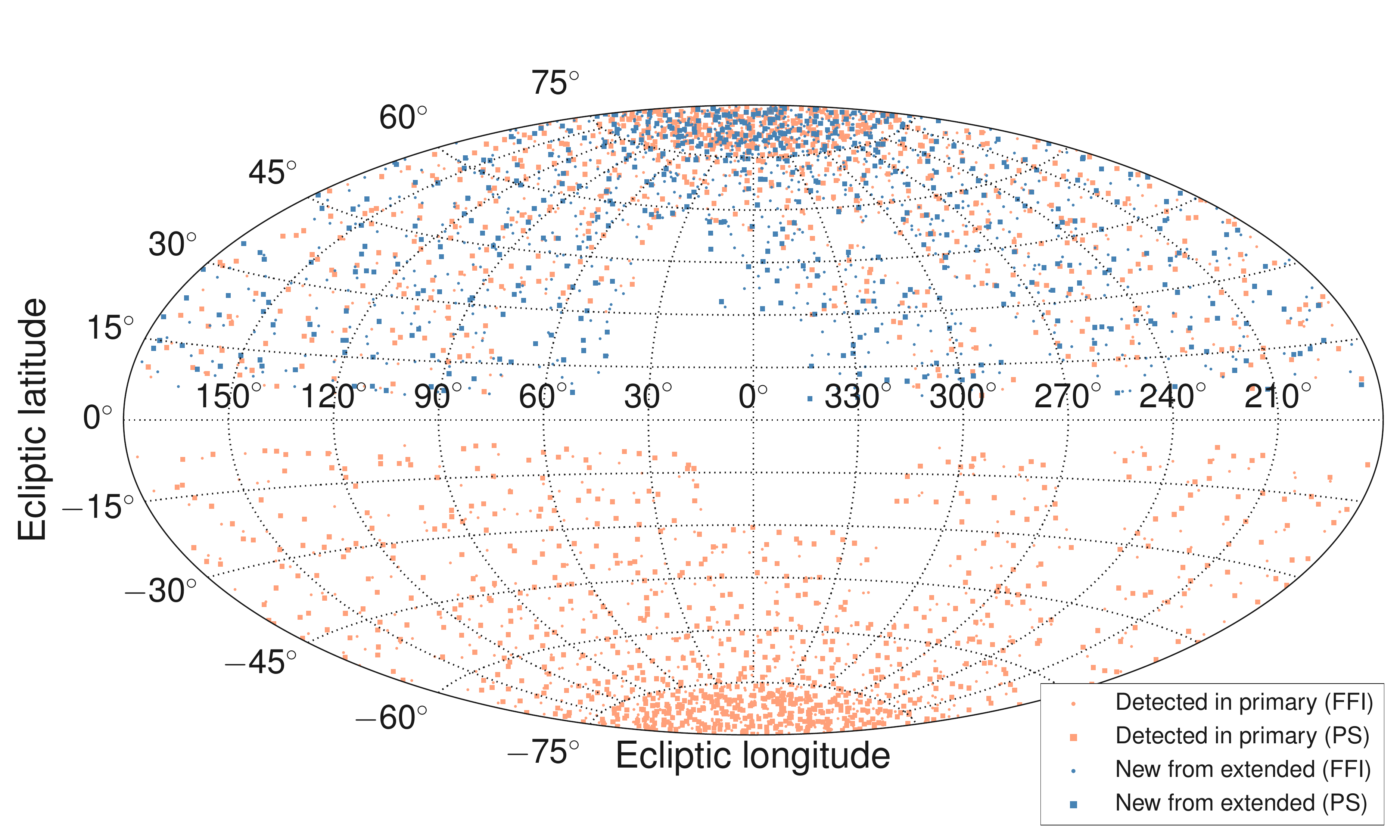}
	\caption{Positions of $R_p<4R_\oplus$ planets detected in the 
		\nhemi\:scenario. Squares (postage stamps) and dots (full frame images) 
		are 
		observed at 2 and 30 minute cadence respectively. Orange denotes 
		detection 
		over the first two years of observing (the Primary Mission), and blue 
		denotes newly detected planets from the extra third year. The `gaps' in 
		fields due to Earth and Moon crossings during the Primary and Extended 
		Missions are noted in Table~\protect\ref{tab:dropped_fields}. For 
		instance, 
		the field centered at $(\lambda=330^\circ,\beta=18^\circ)$ is observed 
		in 
		the Extended but not the Primary Mission. }
	\label{fig:skymap_nhemi}
\end{figure*}
\begin{figure*}[t]
	\centering
	\includegraphics[]{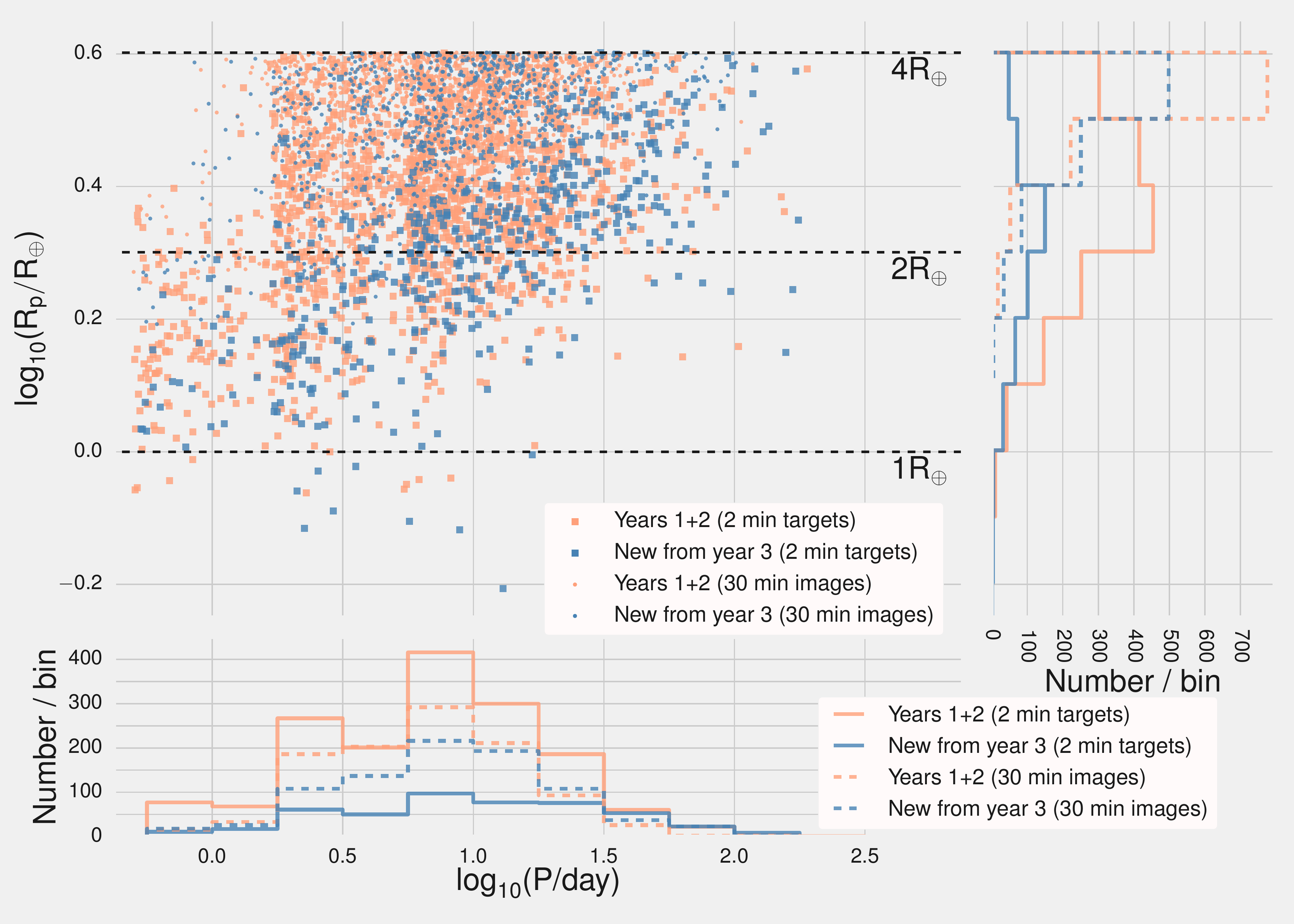}
	\caption{Radius vs period of detected $R_p<4R_\oplus$ planets from one 
		Monte Carlo realization of the \nhemi\:scenario.
		At a fixed period, Extended Missions help us detect smaller planets; at 
		a fixed radius, they let us probe out to longer periods.
		The radius histogram, and the location of all dots (rather than 
		squares) on the scatter plot show that almost all $R_p<2R_\oplus$ 
		planets are detected in postage stamps, not full frame images (also 
		shown in Fig.~\protect\ref{fig:primary_planet_yield}).}
	\label{fig:radius_vs_period_nhemi}
\end{figure*}
\begin{figure*}[t]
	\centering
	\includegraphics[]{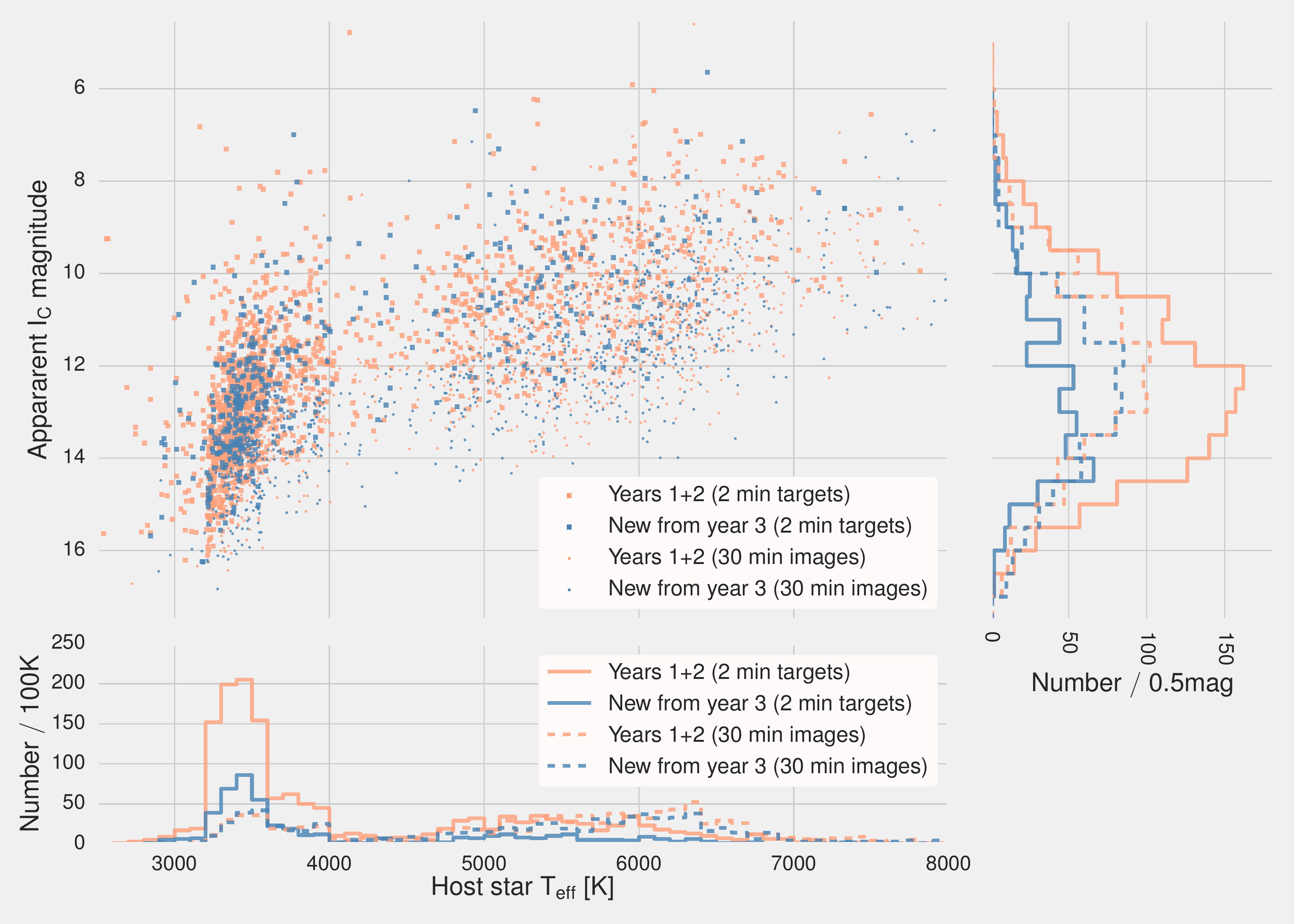}
	%from 160729_pm0_shemi_nhemi_nhemi_t20-diffReport.pdf
	\caption{Apparent Cousins I magnitude plotted against effective temperature 
		for $R_p<4R_\oplus$ planets detected from one Monte Carlo realization 
		of 
		the \nhemi\:scenario. 
		Postage stamp (PS) detections are biased towards M dwarfs in part 
		because of our selection procedure.
		For a given effective temperature, full frame images (FFIs) are taken 
		of dimmer stars.}
	\label{fig:imag_vs_teff_nhemi}
\end{figure*}
%all from 160729 runs. In this case for nhemi.

Before comparing our six selected Extended Mission scenarios simultaneously 
(Sec.~\ref{sec:results_from_all_extended_missions}), we describe the detected 
planet populations from a single realization of an Extended Mission.
As an example case, we choose the \nhemi\:scenario.

A sky map showing the positions of detected planets for this mission is drawn 
in Fig.~\ref{fig:skymap_nhemi}.
Commenting on this map, we note that:
\begin{itemize}
	\item Any planet detected following this scenario's Primary Mission is also 
	detected following its Extended Mission (using the combination of Primary 
	\& Extended datasets).
	We consequently color the detected planets depending on if they are 
	discovered in the Primary Mission, or whether they are detected only by 
	virtue of the extra data collected in the Extended Mission.
	In our simulation, these extra observations will lead to new detections (a) 
	because of an increased number of observed transits leading to a higher 
	phase-folded SNR, which causes the transiting object's SNR to clear our 
	threshold of 7.3, and/or (b) because raising the number of observed 
	transits clears the minimum number of transits we require for detections 
	($N_\mathrm{tra} \geq 2$).
	\item The `dropped' fields described in Sec.~\ref{sec:earth_moon_crossings} 
	owing to Earth and Moon crossings are visible for both the Primary and 
	Extended Missions in the $\lambda=(30^\circ, 0^\circ, 330^\circ, 
	300^\circ)$ fields.
\end{itemize}

In addition to examining the positions of the detected planets, we select and 
plot some of their key properties:
planet radius $R_p$, orbital period $P$, apparent magnitude $I_c$, and 
effective temperature $T_\mathrm{eff}$.
See Figs.~\ref{fig:radius_vs_period_nhemi} and~\ref{fig:imag_vs_teff_nhemi}.
Both of these figures are visualizations from a single Monte Carlo realization 
of the \nhemi\:scenario, and only show planets with $R_p < 4R_\oplus$.
These plots clarify a few points:
\begin{itemize}
	\item At a fixed period, Extended Missions help us detect smaller planets. 
	At a fixed radius, Extended Missions let us probe out to longer periods. 
	These are two important reasons to continue \tesss observations.
	\item Nearly all $R_p<2R_\oplus$ planets are detected in postage stamps, 
	not full frame images. This is an indication that the top $2\times10^5$ 
	\texttt{Merit} stars are a sufficient sample to detect nearly all of the 
	$R_p<2R_\oplus$ planets that \tess can detect over its first three years.
	\item Postage stamp detections are biased towards M dwarfs. Per 
	Fig.~\ref{fig:fig17_replica}, this is largely because our selection 
	procedure chooses many M dwarfs.
	\item For a given effective temperature, the stars for which planets are 
	detected in the full frame images are systematically fainter, compared to 
	PS stars. Projecting the FFI detections onto apparent $I_c$ magnitude 
	(Fig.~\ref{fig:imag_vs_teff_nhemi}, right panel), the median brightness of 
	stars with planets detected from FFIs is actually greater than the median 
	brightness of planets detected from PSs. This is because our FFI sample is 
	has more bright large stars than the target star sample, since our 
	selection 
	procedure assigns higher \texttt{Merit} to dim M dwarfs than to dim K, G, 
	or 
	F dwarfs. 
\end{itemize}

There are a few other statistics that help quantify the value of this 
Extended Mission -- how many new 
planets do we detect? How many are at long orbital periods? How many 
are in habitable zones? We respond to these questions in 
Sec.~\ref{sec:results_from_all_extended_missions}, in particular 
showing our detected planet yields in Fig.~\ref{fig:yield_results}.

\subsection{Comparing Planet Yields from all Extended Missions}
\label{sec:results_from_all_extended_missions}
To compare Extended Missions in terms of planet detection statistics, we focus 
on the subset of all detected planets that are \textit{newly} detected from 
each Extended Mission.
These may come from stars that were not observed at all in the Primary Mission 
(notably for scenarios such as \elong), or they may also come from transiting 
planets that were observed in the Primary Mission with $\mathrm{SNR}<7.3$, or 
from planets that were single-transiters in the Primary Mission (we require 
$N_\mathrm{tra}\geq2$) for detection.
With this in mind, for each Extended Mission scenario we ask the following 
questions:

\begin{enumerate}
	\item $N_\mathrm{new}$: How many new planets do we detect?
	\item $N_\mathrm{new,P>20d}$: How many of these new planets have periods 
	$>20$ days?
	\item $N_\mathrm{new,HZ}$: How many are in the habitable zone?
	\item $N_\mathrm{sys,extra\ planets}$: In how many systems in which at 
	least one planet was detected during the Primary Mission do we find extra 
	planets in the Extended Mission?
	\item $N_\mathrm{new,atm}$: How many new planets do we find that are 
	amenable to atmospheric characterization (defined below by 
	Eq.~\ref{eq:atmosphere_Deming})?
	\item $N_\mathrm{new,new\ stars}$: How many of the new planets come from 
	stars that were not observed in the Primary Mission? 
	\item $N_\mathrm{new,SNR\lor N_{tra}}$: How many of the new planets come 
	from candidates that after the Primary Mission had either a) 
	$\mathrm{SNR}<7.3$ and/or b) less than 2 transits?
\end{enumerate}
\begin{figure}[!t]
	\centering
	\includegraphics[scale=2.]{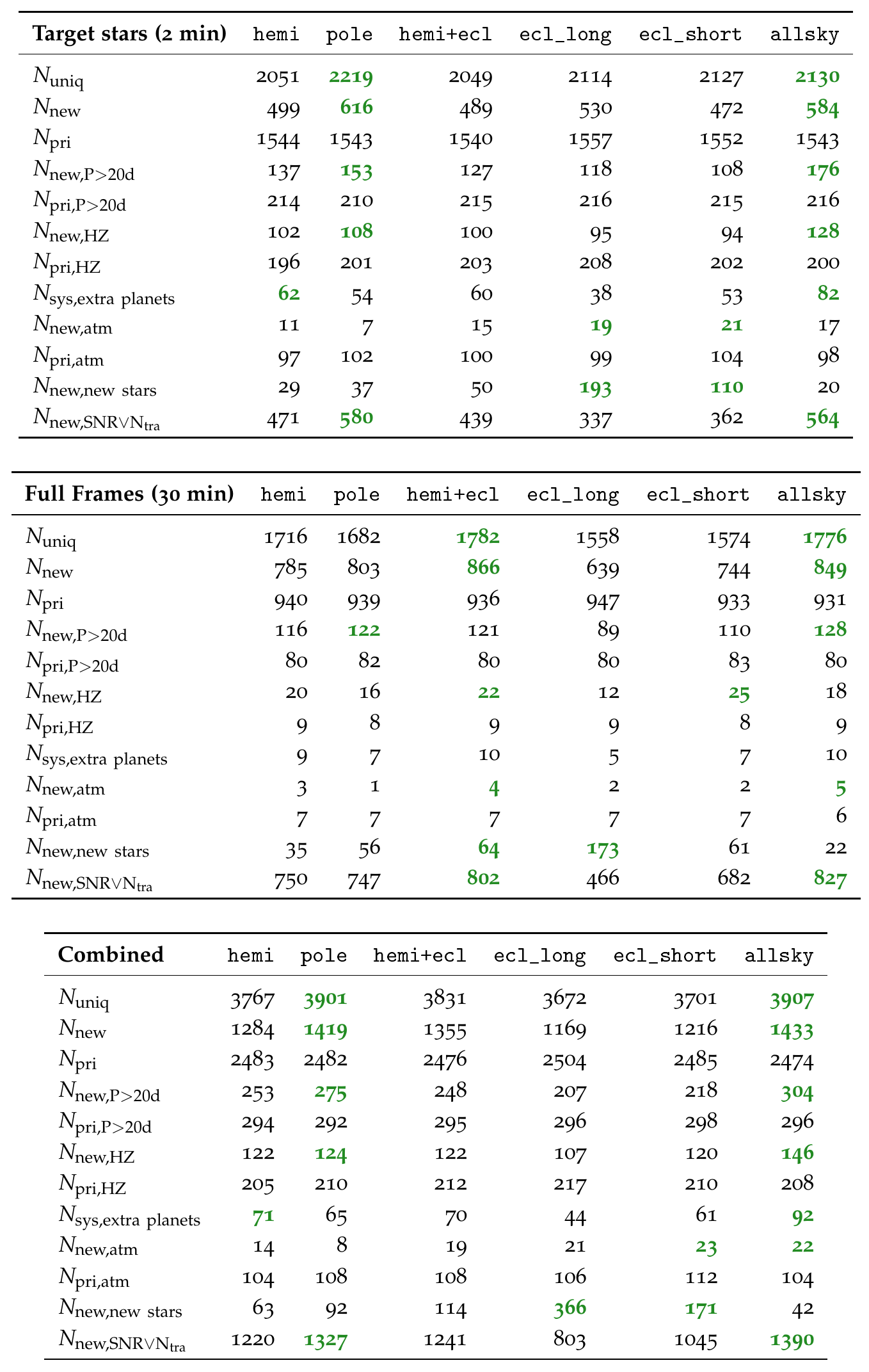}
	%from 160729_t50 data. generate with tables_vis.tex
	\caption{Detected planet metrics for six possible Extended Missions (values 
		are means of 50 Monte Carlo realizations of our calculation, all for 
		$R_p<4R_\oplus$).
		\textit{Top:} postage stamp detections, \textit{Middle:} full frame 
		image only detections, \textit{Bottom:} sum of both.
		The best two scenarios for select statistics are bolded and highlighted 
		in 
		green.
		\newline
		$N_\mathrm{uniq}$: number of unique planets detected over all 3 years.
		$N_\mathrm{new}$: number of planets detected in year 3 that were not 
		detected in years 1\&2 (newly detected planets).
		$N_\mathrm{pri}$: number of planets detected in the Primary Mission 
		(years 1\&2).
		$N_\mathrm{new,P>20d}$: number of newly detected planets with orbital 
		periods greater than 20 days.
		$N_\mathrm{pri,P>20d}$: same as previous, but from the Primary Mission.
		$N_\mathrm{new,HZ}$: number of newly detected planets satisfying 
		$0.2<S/S_\oplus<2.0$ (approximate habitable zone).
		$N_\mathrm{pri,HZ}$: same as previous, from the Primary Mission.
		$N_\mathrm{sys,extra\ planets}$: number of systems in which extra 
		planets are detected.
		$N_\mathrm{new,atm}$: number of newly detected planets with SNR in 
		transmission greater than (that of GJ 1214b)/2, as measured by 
		\jwst\,-- see text.
		$N_\mathrm{pri,atm}$: same as previous, from Primary Mission.
		$N_\mathrm{new,new\ stars}$: number of newly detected planets that 
		orbit 
		stars that were not observed during the Primary Mission.
		$N_\mathrm{new,SNR\lor N_{tra}}$: number of newly detected planets that 
		were 
		observed during the Primary Mission, but either (a) had 
		$\mathrm{SNR}<7.3$ and/or b) had $N_\mathrm{tra}<2$, and so would not 
		be 
		`detected'.}
	\label{fig:yield_results}
\end{figure}

For each Year-3 scenario, we compare these numbers to the
corresponding numbers from the Primary Mission as well as to the other
5 scenarios for Year 3. We show the results of our simulations in
Fig.~\ref{fig:yield_results}.  The first point to notice is that for
all but one of the new planet detection metrics ($N_\mathrm{new}$,
$N_\mathrm{new,P>20d}$, $N_\mathrm{new,HZ}$,
$N_\mathrm{sys,extra\ planets}$, $N_\mathrm{new,atm}$,
$N_\mathrm{new,SNR\lor N_{tra}}$) the yields between Extended Missions
vary by less than a factor of two.  The exception is in
$N_\mathrm{new,new\ stars}$, in which \elong\:detects roughly twice as
many planets orbiting never-before-observed stars as any other
proposed mission.

The second point is on the absolute yields of new planets: postage stamp 
observations find $\mathcal{O}(500)$ new planets, relative to the Primary 
Mission's $\mathcal{O}(1500)$.
Full frame image observations find $\mathcal{O}(800)$ new planets, relative to 
the Primary Mission's $\mathcal{O}(900)$.
All Extended Missions find $\mathcal{O}(1300)$ new planets, relative to the 
Primary Mission's $\mathcal{O}(2500)$.
We discuss these results -- the rough invariance of the number of new planets 
to different pointing scenarios, and the essential contribution of FFIs -- 
further in point \#1 below.

Skimming the bottom panel for which missions are highlighted in green when 
accounting for both PSs and FFIs, we see that \npole\:and \hemis\:are the 
`superlative-winning' missions in terms of detected planet statistics: 
considering both PSs and FFIs, \npole\:places top-2 in 5 of 8 relevant 
categories, and \hemis\:does the same in 6 of 8.
They both do well at maximizing the number of newly detected planets, while 
also performing well at detecting long period planets, and thus planets in 
their stars' habitable zones.
\hemis\:also has the largest number of systems in which extra planets are 
detected.

We now discuss each metric of Fig.~\ref{fig:yield_results} in more depth:
\begin{figure}[!t]
	\centering
	\includegraphics[scale=1.]{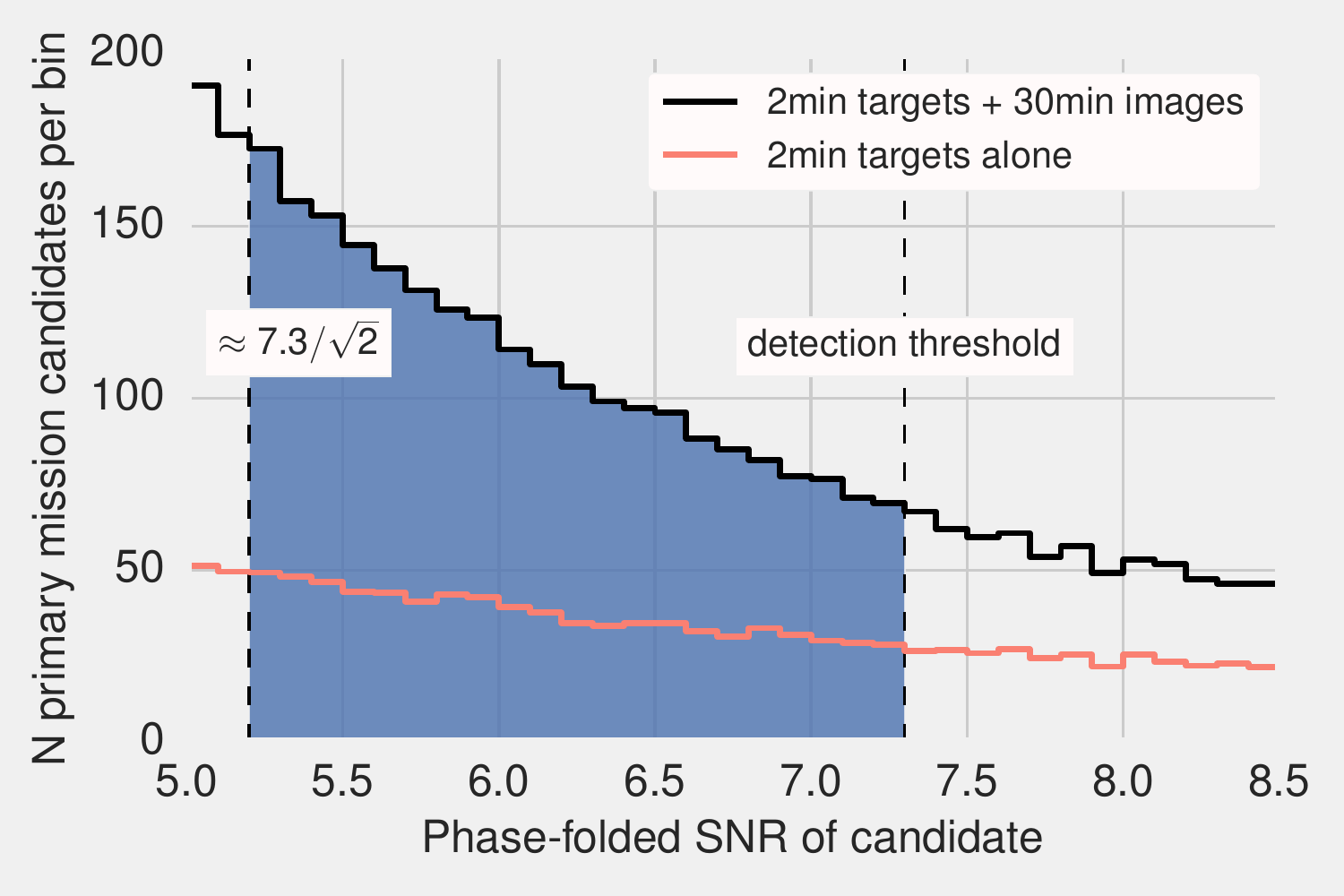}
	\caption{Histogram of phase-folded SNRs for candidate $R_p<4R_\oplus$ 
		planets following the Primary Mission (from both PS and FFI 
		observations; values are means of 20 Monte Carlo trials; 
		$N_\mathrm{tra}\geq2$).
		If an Extended Mission observes half of the sky, it roughly doubles 
		the number of observed transits for half of the planets observed in 
		the Primary Mission, enabling detection of $\approx 2316/2 = 1158$ 
		planets (half of the blue integrated area in the plot). This coarse 
		estimate is a similar result to our detailed calculations, and 
		shows the value of continuing \tesss observations 
		\textit{irrespective of where we observe}.}
	\label{fig:snrf_histogram}
\end{figure}
\begin{enumerate}
	\item $N_\mathrm{new}$: we detect about as many new planets in Year 3 as we 
	detect planets in either Years 1 or 2: roughly $1250$.
	The worst and the best scenarios (\elong\:and \hemis, respectively) differ 
	only by a factor of 1.2.
	The fact that there are so many new planets to be detected from extended 
	observations, particularly from full frame images, and that the absolute 
	number of new planets does not depend strongly on the schedule of pointings,
	can be justified (post-facto) with Fig.~\ref{fig:snrf_histogram}.
	This figure illustrates a point that was originally noted 
	by~\citetalias{Sullivan_2015}:
	\tesss Primary Mission will miss many short-period planets around bright 
	stars, and is therefore
	incomplete even in its intended hunting ground.
	For stars with $I_c<11$, there are $\sim\! 10\times$ more transiting 
	$2<R_p<4R_\oplus$ planets on $P<20\,\mathrm{day}$ orbital periods than 
	\tess will detect in its first two years\footnote{Fig 22
		of~\citetalias{Sullivan_2015}; only considers $R_\star<1.5R_\odot$ 
		stars}.
	\citetalias{Sullivan_2015} similarly note that there are $\sim\! 
	20\times$ more transiting $R_p<2R_\oplus$ ($P<20\,\mathrm{day}$, 
	$I_c<11$ host star) planets in the sky than \tess will discover in 
	its Primary Mission.
	Our findings agree: there are a substantial number of planets just below 
	the detection threshold, predominantly with $2R_\oplus < R_p <4R_\oplus$ 
	(Fig.~\ref{fig:primary_planet_yield}).
	An Extended Mission will probe and detect this population, in any of the 
	scenarios we investigate.
	This result should hold equally well for realistic detection efficiency 
	thresholds, and it demonstrates that extended observations will be valuable 
	because \tess will not yet have detected all the planets in the parameter 
	space of $R_p<4R_\oplus$ planets orbiting bright stars on short-period 
	orbits;
	there will still be small, short-period planets orbiting bright stars to 
	be discovered after \tesss Primary Mission.

	\item $N_\mathrm{new,P>20d}$: it will be possible to detect as many new 
	$P>20$ day planets in one year of \tesss Extended Mission as in both years 
	of the Primary Mission.
	The Primary Mission detects about 295 such planets; \hemis\:and 
	\npole\:scenarios detect similar numbers.
	These two scenarios are achieving the goal of long-period planet detection 
	in slightly different ways: % (see discussion in Sec.~\ref{sec:discussion}):
	\npole\:maximizes the average observing baseline per star, while 
	\hemis\:observes the greatest possible number of stars for longer than 40 
	days.
	The latter approach could succeed at detecting many planets (our result is 
	that \hemis\:detects the most $P>20$ day planets), but it relies heavily 
	upon the assumption that we can detect planets from only two transits over 
	the course of the entire mission, even if this means only one transit in 
	the Primary, and one transit in the Extended.
	\begin{figure}[!t]
		\centering
		\includegraphics{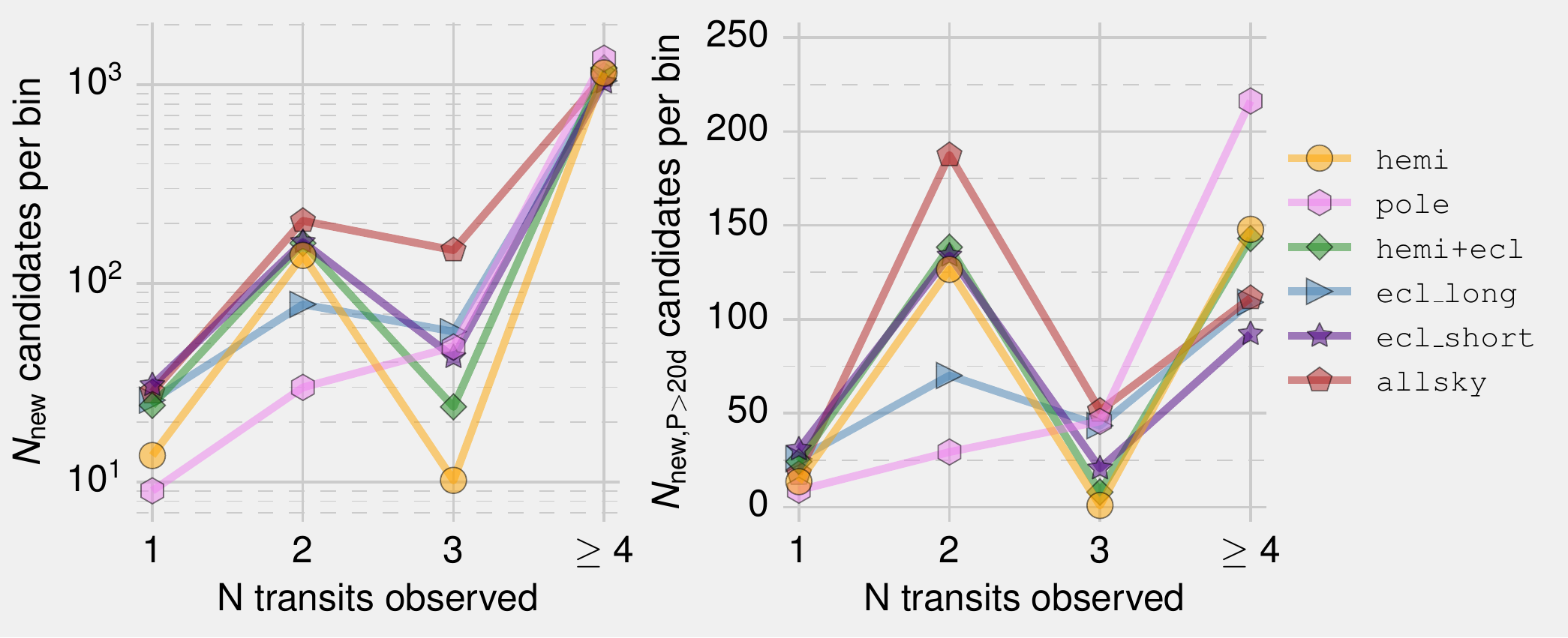}
		%this figure uses 160802 data, which should be the same as 
		%160728_t50, but has ntra_min set to 1 in yieldCode. Read my notes 
		%from ext_sim_notes on this -- something's a bit off with the 
		%normalization. That said, rather than redo the previous figures 
		%with a code that might have a minor bug, my point here -- which is 
		%the hemis14d has this issue with few-transit detections -- isn't 
		%really any different.
		\caption{ \textit{Left:} Histogram of new $R_p<4R_\oplus$ planet 
			candidates from each Extended Mission as a function of the 
			total number of 
			observed transits over all 3 years.
			Planet candidates are `detected' if $N_\mathrm{tra}\geq2$ 
			(see Eq.~\protect\ref{eq:detection_criterion}).
			\textit{Right:} Same as left, restricted to $P>20$ day planets.
			If any given scenario has a `bump' at 2 observed transits, then 
			that scenario depends more heavily on our assumption of being 
			able to make detections based on only two transits.
			Lines joining points have no physical meaning; they are 
			intended to improve readability.}
		\label{fig:Ntra_hist}
	\end{figure}
	\begin{marginfigure}[-1in]
		\centering
		\includegraphics[scale=1.]{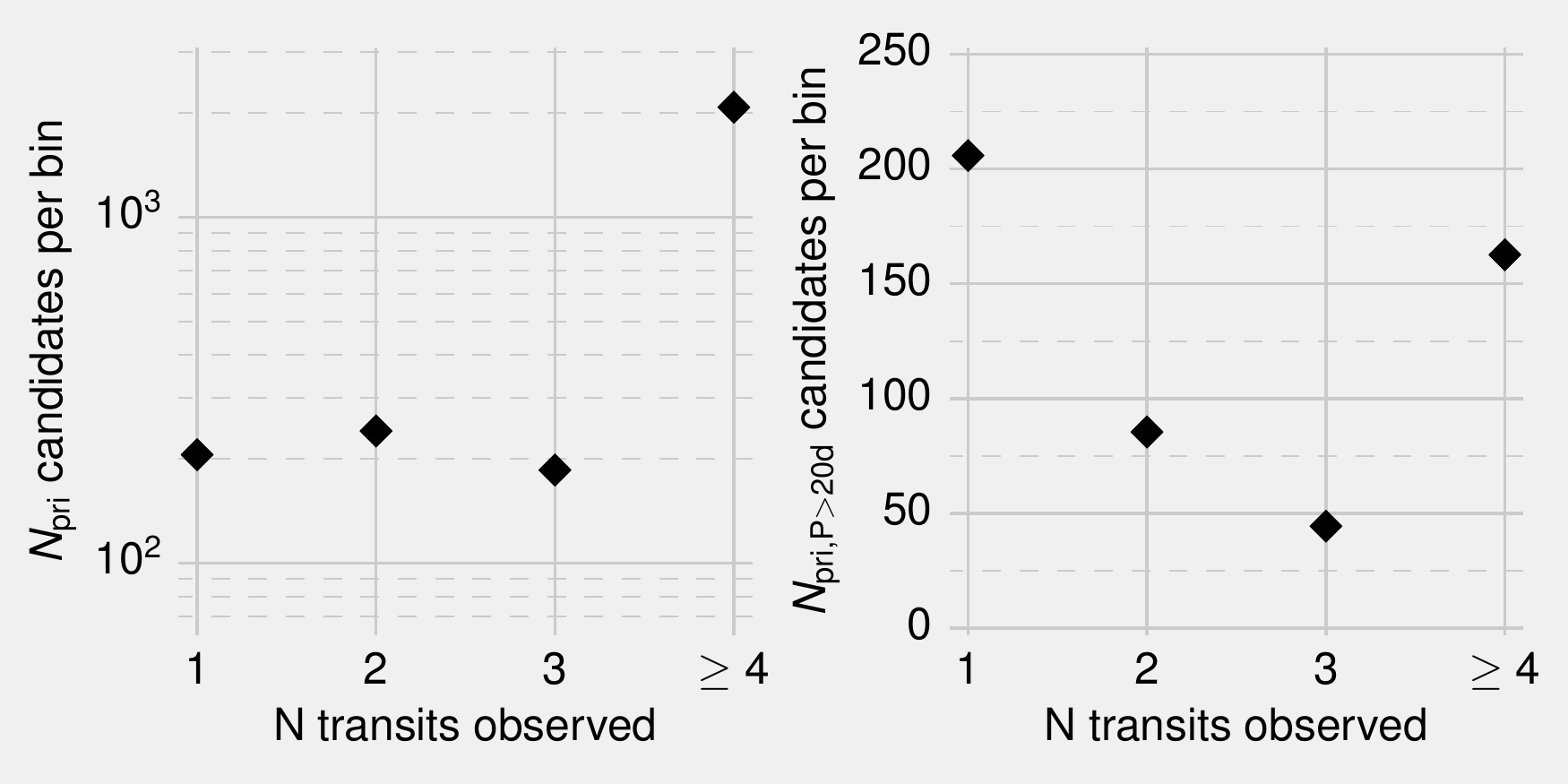}
		%\missingfigure{histogram of Primary Mission number of transits for 
		%detected planets}
		% this figure uses BLAH data
		\caption{ 
			Similar to Fig.~\protect\ref{fig:Ntra_hist}, but for the Primary 
			Mission.
		}
		\label{fig:Ntra_hist_primary}
	\end{marginfigure}
	%\vspace{-0.47cm}

	This point -- that the ability of the \hemis\:scenario to detect many long 
	period planets is grounded on the assumption that two transits at high 
	enough SNR are sufficient for detection -- is made explicit in the right 
	panel of Fig.~\ref{fig:Ntra_hist}.
	About half of the long period planets that \hemis\:finds are detected with 
	only two transits.
	By way of comparison, \npole\:detects most of its long-period planets with 
	$\ge 4$ transits.
	This means that the \npole\:detections are more secure.
	For two-transit detections, especially those separated by a gap of a year 
	or more in the \tess data, it will be
	difficult to be confident in either the detection or in the derived 
	orbital period.
	Experience from the \kepler mission showed that requiring 3 or more 
	self-consistent transits substantially lowers the fraction of false 
	signals~\citep{burke_Q1Q8_2014}.
	If we restrict `detections' to planets with both $N_\mathrm{tra}\geq 3$ and 
	$\mathrm{SNR_{phase-folded} \geq 7.3}$, we find that \npole\:detects 
	$\sim\!260$ new long-period planets, while the next-best scenarios, \hemis,
	\nhemi, and \shemiAvoid, all detect about 160.

	\item $N_\mathrm{new,HZ}$: we approximate the habitable zone as the 
	geometric shell around a host star in which a planet's insolation satisfies 
	$0.2>S/S_\oplus>2.0$.
	With this approximation, the \hemis\:scenario finds the most new habitable 
	zone planets: 146 (which is subject to the same caveats discussed above for 
	long period planet detections). 
	The next-best scenarios, \npole, \npole\:and \shemiAvoid, all detect around 
	120.
	Relative to the Primary Mission's 210 detections, this means Extended 
	Missions boost the number of detected habitable zone planets by a factor of 
	$\sim1.6$.
	For purposes of weighing the value of habitable-zone detections in deciding 
	between missions, the result that these scenarios all detect a similar 
	number of HZ planets indicates that this metric will likely not `tip the 
	scales' in any direction.
	
	We note in passing that $\sim\!80\%$ of the habitable zone planets that 
	\tess detects orbit M dwarfs with spectral classes ranging from M4 - M0, 
	and $\sim\!15\%$ of them orbit M dwarfs later than M4.
	We show the relevant cumulative distribution in 
	Fig.~\ref{fig:CDF_habitable_zone}.
	% see hz_primary_george/CDF_most_HZ_planets_orbit_early_M_dwarfs.png
	Additionally, our values for the number of $0.2<S/S_\oplus<2$ planets from 
	the Primary Mission are slightly revised from those 
	of~\citetalias{Sullivan_2015}: while~\citetalias{Sullivan_2015} found 
	$48\pm7$ planets with $0.2<S/S_\oplus<2$ and $R_p<2R_\oplus$, we find $34 
	\pm 5$ such planets.
	Adopting the habitable zone 
	of~\citet{kopparapu_habitable_2013},~\citetalias{Sullivan_2015} found 
	$14\pm4$ planets with $R_p<2R_\oplus$. We find $11 \pm 3$ such planets.
	The rule of thumb that Extended Missions give roughly $1.6\times$ the 
	number of new $0.2<S/S_\oplus<2$ habitable zone planets applies to 
	the~\citet{kopparapu_habitable_2013} habitable zone as well as 
	$0.2<S/S_\oplus<2.0$.
	Another point raised by Fig.~\ref{fig:scatter_habitable_zone} is that 
	the~\citet{kopparapu_habitable_2013} habitable zone, which is physically 
	motivated by 1-D radiative-convective cloud-free climate models with 
	accurate absorption coefficients, results in roughly 3 times fewer 
	`habitable zone' planet detections than our ad-hoc criterion of 
	$0.2<S/S_\oplus<2$.
	\begin{marginfigure}[-5in]
		\centering
		\includegraphics[scale=1.]{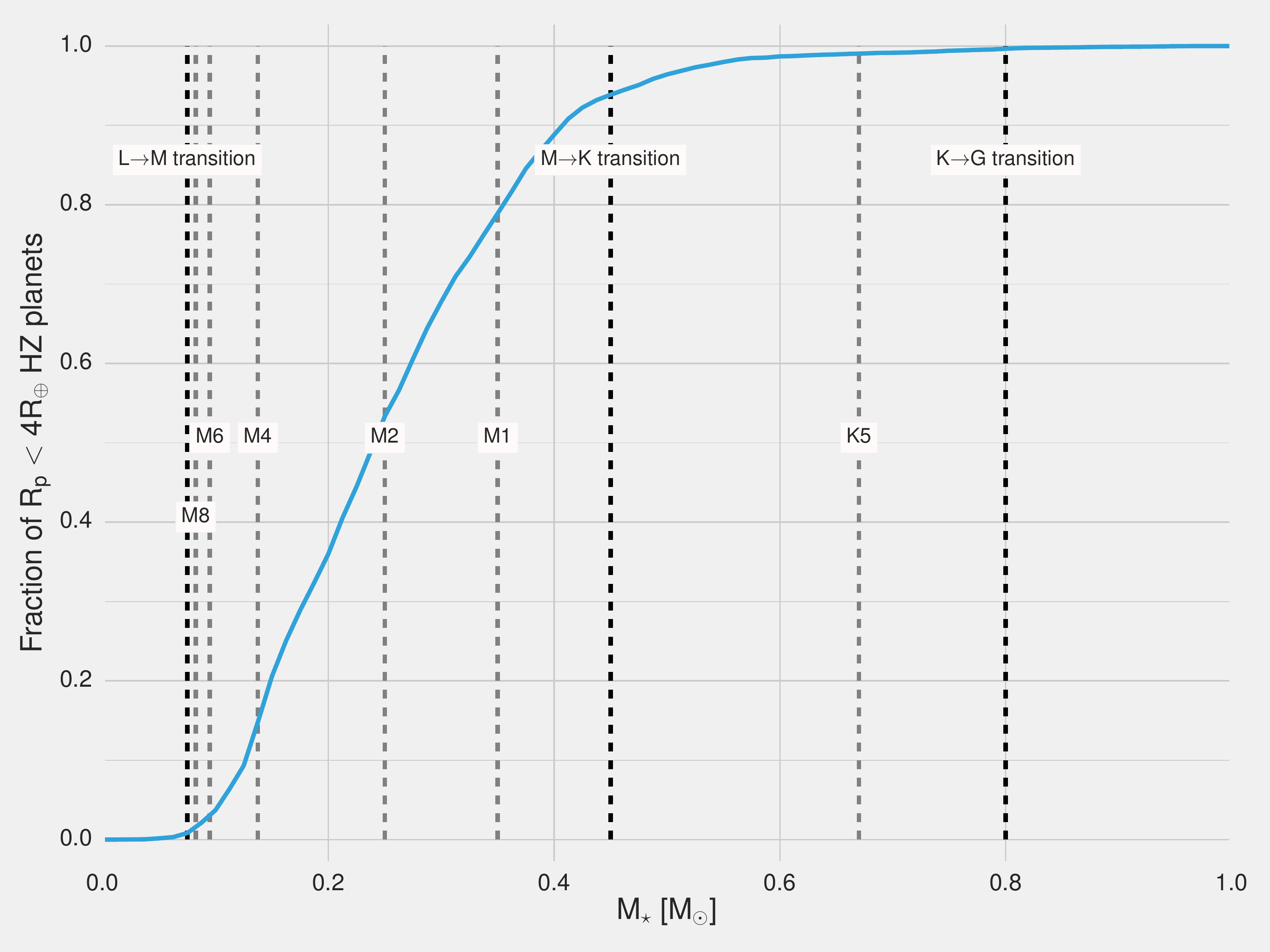}
		\caption{Cumulative distribution of $R_p<4R_\oplus$ and 
			$0.2<S/S_\oplus<2$ planet candidates from the Primary Mission (a 
			proxy 
			for the habitable zone). Boundaries of spectral classes are highly 
			approximate, and taken from 
			from~\protect\citet{habets_empirical_1981} 
			and~\protect\citet{baraffe_massspectral_1996}.}
		\label{fig:CDF_habitable_zone}
	\end{marginfigure}
	\begin{marginfigure}[-1in]
		\centering
		\includegraphics[scale=1.]{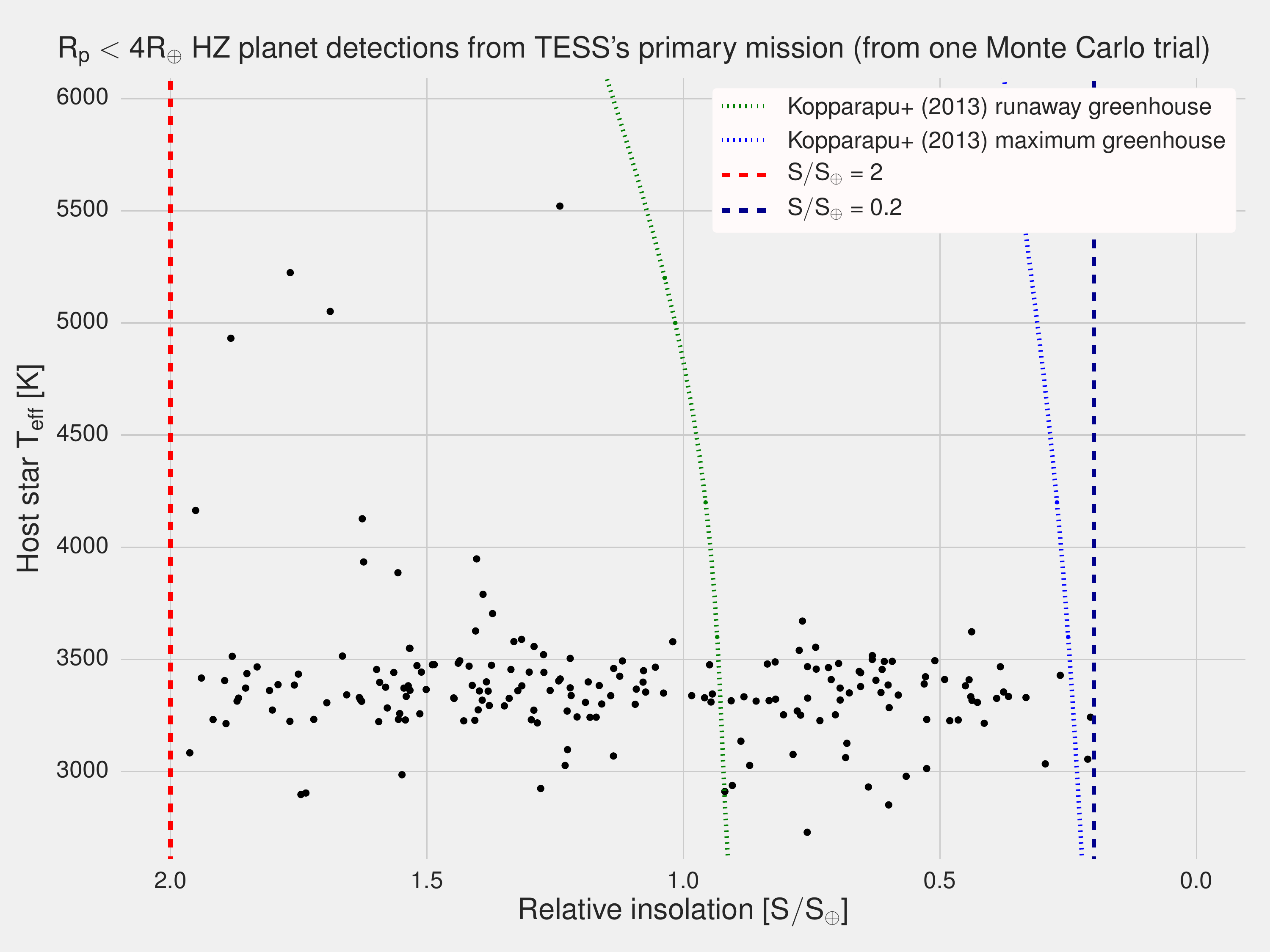}
		\caption{Scatter plot of $R_p<4R_\oplus$ planet candidates falling 
			in the~\protect\citetalias{Sullivan_2015} or 
			\protect\citet{kopparapu_habitable_2013} habitable zones. }
		\label{fig:scatter_habitable_zone}
	\end{marginfigure}	
	
	\item $N_\mathrm{sys,extra\ planets}$: for how many systems do we detect 
	extra planets?
	Our assumptions about multiple planet system distributions are crude -- we 
	assume independent probability draws from single planet occurrence 
	distributions.
	Thus our simulated planet population may not have systems of tightly packed 
	inner planets in realistic numbers.
	That said, we expect this statistic to be some indication of the 
	information that we are not explicitly modeling, but which can be obtained 
	from extended observations of planetary systems post-planet detection. 
	This additional information includes improved precision on physical and 
	dynamical parameters of the system.
	It also includes transit timing variations, which could be used to discover 
	non-transiting planets as well as transiting outer companions.
	TTVs can also give dynamical hints for the formation history of planetary 
	systems, for instance, discriminating between \textit{in situ} formation 
	and inward migration as~\citet{mills_resonant_2016} argued for the Kepler 
	223 system.

	The most prominent feature in the results for this metric is that 
	\elong\:detects the fewest systems with extra planets (44, which is $39\%$ 
	worse than the next-best). 
	This is reasonable because \elong\:spends the most time looking at new sky, 
	and in the process observes fewer systems that were detected in the Primary 
	Mission.
	\nhemi, \shemiAvoid, \npole, and \eshort\:all perform similarly, detecting 
	$\sim65$ such planets.
	\hemis\:detects the most, at 92. While this is still subject to the 
	assumption of two-transit recoverability, in this case the requirement is 
	not too strong: only 10 of \hemis's systems with newly detected planets 
	come from the case where the extra detected planet comes from two transits.
	% makeReport: search for "ntrahemis14d". same for commit.
	
	\item $N_\mathrm{new,atm}$:
	We define `planets that are amenable to atmospheric characterization' to 
	mean planets whose SNR in transmission is at least half as large as that of 
	GJ~1214b. We chose ``at least half as large'' rather than
	``equal to'' in order to give a sufficiently large sample to prevent 
	Poisson fluctuations from hindering our comparisons.
	The relevant signal in transmission spectroscopy is the ratio of the areas 
	of atmosphere's annulus to the star's disk on the sky plane, 
	$\delta_\mathrm{atm} = 2\pi R_p h_\mathrm{eff}/(\pi R_\star^2)$, where the 
	effective scale height of the atmosphere $h_\mathrm{eff}$ is proportional 
	to 
	the actual scale height.
	Assuming that the planet is in thermal equilibrium with incident radiation 
	from the host star, and that its atmosphere has known mean molecular weight 
	and Bond albedo, we can compute a representative signal.
	The noise performance depends on the observing instrument, and could be 
	complex if not simply dominated by shot-noise from IR photons.
	We circumvent such complexities via an empirical formula provided to us by 
	Drake Deming, based on a multi-variate regression fit to detailed 
	simulations performed by Dana Louie.
	This formula estimates the SNR in transmission from 4 transits observed 
	with \jwsts NIRISS instrument:
	\begin{align*}
		\log_{10} \mathrm{SNR} =\ 
		&2.98\log_{10}\left(\frac{R_p}{R_\oplus}\right)
		- 1.019\log_{10}\left(\frac{M_p}{M_\oplus}\right) \\
		&- 1.459\log_{10}\left(\frac{R_\star}{R_\odot}\right)
		- 0.249\log_{10}\left(\frac{a}{\mathrm{AU}}\right) \\
		&- 0.147\left(V - 5.0\right) + 0.193  \numberthis
		\label{eq:atmosphere_Deming}
	\end{align*}
	for $V$ the host star's apparent $V$-band magnitude (calibrated for 
	$3>V>22$), $R_p$ the planet radius, $M_p$ the planet mass, $R_\star$ the 
	star radius, and $a$ the planet's semi-major axis.
	The coefficients are physically sensible: the 2.98 coefficient of $R_p$ 
	minus the 1.019 coefficient of $M_p$ implies that the SNR depends inversely 
	as bulk density, with puffier planets giving higher SNR for transit 
	spectroscopy.
	Although this formula uses a $V$ band magnitude ($\approx$0.5--0.6$\mu$m), 
	while NIRISS's SOSS mode covers 0.6--2.8$\mu$m, the only difference if we 
	were to use $J$ band magnitudes would be in the coefficients preceding the 
	stellar radius and the semi-major axis terms (and thus implicitly, in the 
	stellar mass).
	Focusing our analysis to a SNR measured by \jwst is sensible given \tesss 
	role as a `\jwst finder scope'~\citep{deming_jwst_tess_2009}.
	We focus specifically on NIRISS but our analysis will be broadly applicable 
	to other JWST instruments for transmission 
	spectroscopy (see review by~\citet{beichman_observations_2014}).
	\begin{figure*}[!t]
		\centering
		\includegraphics[]{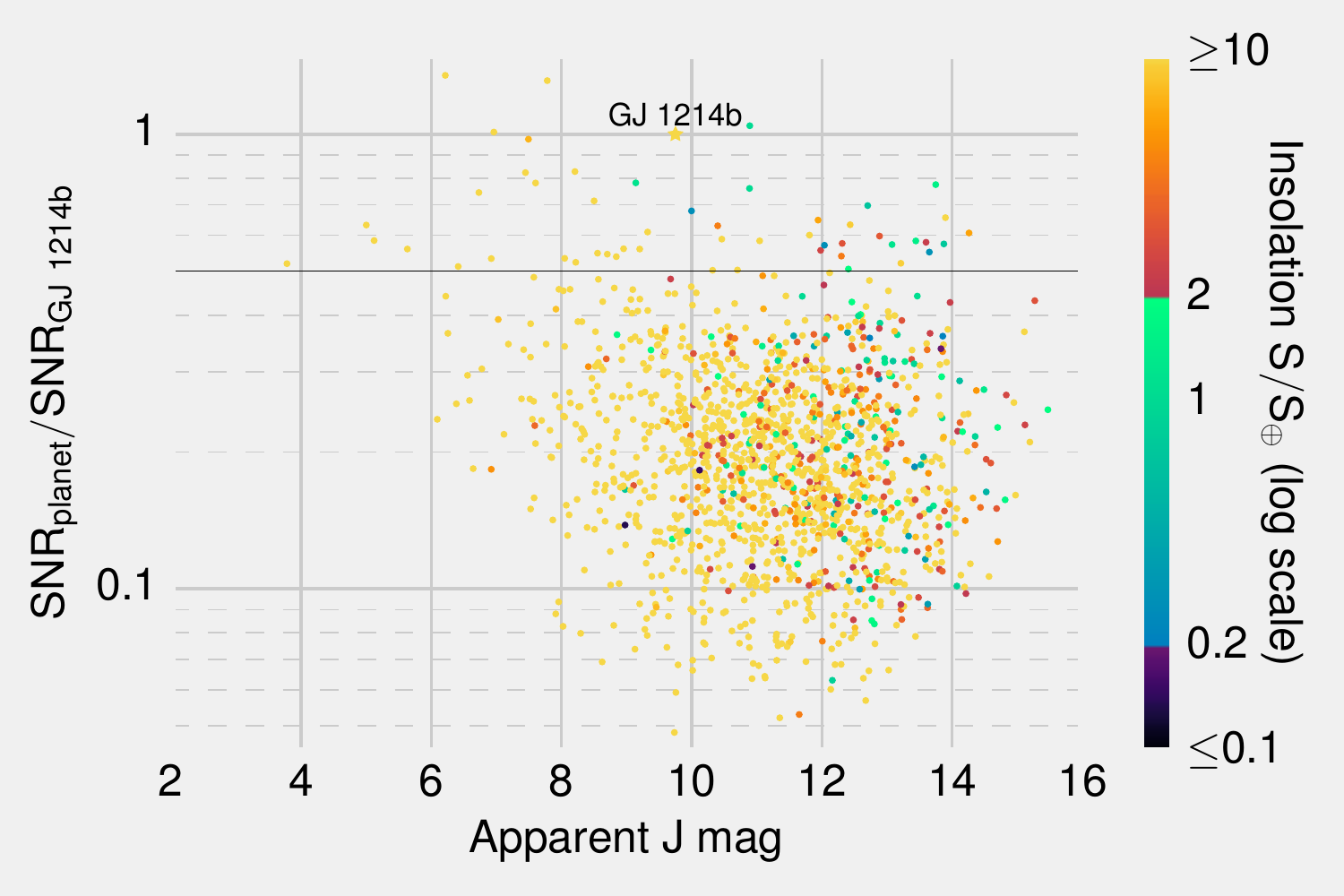}
		\caption{ Scatter plot showing the SNR in transmission of detected 
			planets with $R_p<4R_\oplus$ from one Monte Carlo realization of 
			all 3 
			years of the \npole\:scenario.
			The SNR is computed from Eq.~\protect\ref{eq:atmosphere_Deming}.
			Planets above the horizontal black line 
			($\mathrm{SNR_{planet}/SNR_{GJ\ 1214b}} = 0.5$) are counted for 
			Fig.~\protect\ref{fig:yield_results}'s metric of planets with 
			`good' atmospheres for transmission spectroscopy.
			GJ 1214b is marked with a star.
			The coloring of planets indicates their relative insolation, as 
			well as 
			whether they are in our approximate habitable zone 
			($0.2<S/S_\oplus<2)$.
		}
		\label{fig:atmosphere_scatter}
	\end{figure*}
	
	The system values for GJ 1214b are those found
	by~\citet{charbonneau_gj1214b_2009}: $R_p = 2.678R_\oplus$, $M_p = 
	6.55M_\oplus$,
	$R_\star = 0.211R_\odot$, $a = 0.0144\mathrm{AU}$, $V = 15.1$.
	Using Eq.~\ref{eq:atmosphere_Deming}, we compute the SNR in transmission 
	for 
	all detected planets, for all Extended Mission scenarios.
	We normalize them to the equivalent SNR for GJ 1214b (3.94, per 
	Eq.~\ref{eq:atmosphere_Deming}).
	Fig.~\ref{fig:atmosphere_scatter} shows one realization of the resulting 
	distribution for planets detected in all three years of the 
	\npole\:scenario.
	
	\tess mostly detects strongly irradiated planets (most points on 
	Fig.~\ref{fig:atmosphere_scatter} are yellow).
	A very small number, $\lesssim 10$, are both in the approximate habitable 
	zone and also `favorable for atmospheric characterization'.
	Of course, a highly compelling target with lower SNR in transmission per 
	transit might merit a more ambitious \jwst observing program.
	We note that all of these planets are assumed to have identical mean 
	molecular weights and cloud properties.
	%\todo[inline]{what cloud model/opacity does this use?}
	
	More importantly, Fig.~\ref{fig:yield_results} shows that most of the 
	planets with atmospheres that are best for transmission spectroscopy are 
	already discovered after two years.
	The best Extended Missions (\shemiAvoid, \elong, \eshort\:and \hemis) boost 
	the yield of such planets from $\sim\!100$ ($N_\mathrm{pri,atm}$) to 
	$\sim\!120$ ($N_\mathrm{pri,atm} + N_\mathrm{new,atm}$).
	The worst, \npole, finds about an additional 10.
	This best-case boost of $1.25\times$ more `good' planets for atmospheric 
	characterization is less than the relative boost of $1.6\times$ more newly 
	detected long period planets.
	Put differently, among the various possibilities for the Extended Mission,
	there is more variation in $N_\mathrm{new,P>20d}$ than in 
	$N_\mathrm{new,atm}$.

	\item $N_\mathrm{new,new\ stars}$:
	Intuitively we expect that to maximize the number of planets detected 
	around ``new'' stars (those which were not observed during the Primary 
	Mission) we should collect as many photons as possible from new stars.
	And the region on the sky with the greatest number of new stars is the 
	ecliptic.
	It is not surprising, then, that the \elong\:scenario finds the largest 
	number of planets around new stars.
	The \elong\:scenario dedicates 7 of a single year's 13 observing sectors to 
	the ecliptic (where the other 6 are spent centered at the North Ecliptic 
	Pole due to excessive Earth and Moon crossings).
	It consequently detects twice as many new planets about newly observed 
	stars as the next-best scenarios: \eshort\:and \shemiAvoid\:(366 vs 171 and 
	114, respectively).
	These latter two scenarios also spend time observing the ecliptic, but with 
	only one camera, rather than with all four cameras simultaneously.
	We note that even though \elong\:is the scenario most successful in 
	detecting planets about new stars, the new stars represent only 
	$\sim\!30\%$ of the total number of new detections.\footnote{Here, we 
		remind the reader that ``new'' refers only to \tess observations. Some 
		of 
		these stars will
		have been observed by K2 or other projects (see discussion in 
		Sec.~\ref{sec:discussion}).}
	
	\item $N_\mathrm{new,SNR\lor N_{tra}}$:
	This statistic is the number of newly detected planets that are detected 
	either (a) due to their final SNR clearing our threshold (logical) or (b) 
	their number of observed transits being greater than or equal to 2.
	It is the complement to $N_\mathrm{new,new\ stars}$: scenarios like \nhemi, 
	\npole, and \hemis\:that do not observe many new stars will detect all of 
	their planets from a boosted SNR and/or clearing the minimum transit 
	threshold.
	
\end{enumerate}

\paragraph{Comment on meaning of `detected in postage stamps' vs `detected in 
	FFIs'}
The invested reader may inquire ``what about the cross-over case of planets 
that are observed as PSs during the Primary (Extended) Mission, but as FFIs in 
the Extended (Primary) Mission? These are not explicitly listed in 
Fig.~\ref{fig:yield_results}''.
%This is what is done in Fig.~\ref{fig:yield_results}.
When describing the entire unique planet population detected from Years 1-3, 
for simplicity of language we use `postage stamp detections' to refer to 
planets that are observed at any time (Primary or Extended Missions) at 2 
minute cadence.
In these cases, the dominant contribution to the final signal to noise ratio 
tends to come from the PS observations.
When describing new planet detections, we use `postage stamp detections' to 
mean planets that were newly detected due to being observed as postage stamps 
in the Extended Mission.
In other words, this `cross-over' point only matters in discussions of the 
unique planet population from an entire mission, Years 1-3.
Considering just the newly detected planet population, we can unambiguously 
specify whether the new detections came from full frame images or postage 
stamps, irrespective of their observations from the Primary Mission.

\subsection{On the Brightness of Stars with Detected Planets}
\begin{figure}[!t]
	\centering
	\includegraphics[]{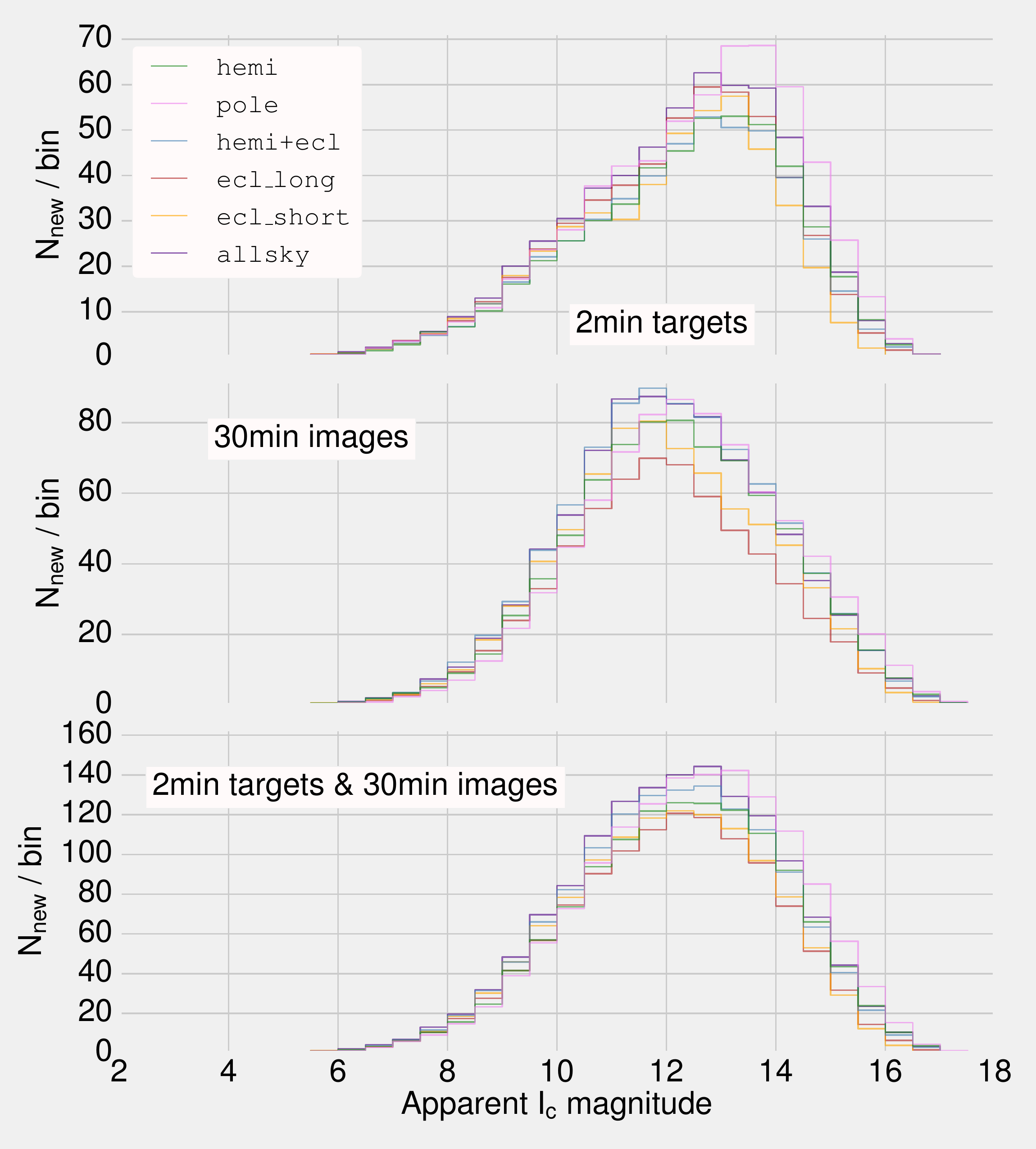}
	\caption{Histogram of apparent $I_c$ magnitude of host star for newly 
		detected $R_p<4R_\oplus$ planets from all Extended Mission scenarios, 
		for 
		\textit{top:} postage stamp detections, \textit{middle:} full frame 
		image 
		detections, \textit{bottom:} the sums thereof.
		While \npole\:does well by most metrics, a larger proportion of the new 
		planets it detects orbit dim stars compared to alternatives like 
		\hemis\:or \shemiAvoid.}
	\label{fig:icmag_meta}
\end{figure}
While \npole\:does well by most metrics, a larger proportion of its newly 
detected planets orbit dimmer stars than the planet populations detected from 
alternatives like \hemis\:or \shemiAvoid.
We demonstrate this in Fig.~\ref{fig:icmag_meta}.
The main point here is that if our only priority were to maximize the number of 
new detections around bright host stars, then \npole\:would be the worst among 
the scenarios considered here.
For instance, arbitrarily setting the bound at $I_c<10$ and numerically 
integrating from Fig.~\ref{fig:icmag_meta}'s data, we see in 
Table~\ref{tab:icmag_meta} that there is a $\sim30\%$ difference between the 
missions.
For point of reference, the Primary Mission detects 386 such planets -- so a 
single year of Extended Mission detects roughly as many planets orbiting bright 
hosts as a single year of Primary Mission.
%cf bright_star_comparison_Ic_lt_10.ipynb
%or just run /postProcessing/ext_mission_comparisons/memo_processing.py 
%interactively, and then look at the length of the relevant data
\begin{table}[!ht]
	\centering
	\caption{Number of new, $I_c<10$, $R_p<4R_\oplus$ planets from each 
		Extended Mission (average of 50 Monte Carlo realizations of our code; 
		showing sum of PSs \& FFIs). \npole\:detects the fewest new planets 
		orbiting bright stars.}
	\label{tab:icmag_meta}
	\begin{tabular}{|c|c|c|c|c|c|}
		\hline
		\nhemi & \npole & \shemiAvoid & \elong & \eshort & \hemis \\ \hline
		162    & 154    & 188         & 167    & 183     & 198    \\ \hline
	\end{tabular}
	%I put together this table by running 
	%/ext_mission_comparisons/memo_processing with the 160729_t50 data, and 
	%then 
	%just directly printing the numbers.
\end{table}

\section{Discussion} 
\label{sec:discussion}

\subsection{Planning Year 3 with Years 4--$N$ in mind}
\label{sec:gtr_1yr_horizon}

% Contextualize importance of thinking beyond just 1 yr Extended Mission
\tesss orbit is stable, in principle, for more than 1000 
years~\citep{gangestad_high_2013}.
While minor mechanical failures should be expected on the timescale of a few 
years, it is plausible that the spacecraft could be operable for up to a decade.
Therefore, when choosing any particular plan for a one-year Extended Mission, 
it makes
sense to also consider the implications of an even longer Extended Mission.

% Preliminary discussion of immediate extensibility
A simple point is that \nhemi, \npole, and \shemiAvoid\:can all be inverted to 
their southern or northern complements for a fourth year of observing.
This will yield a comparable number of new planets to what they find in year 3 
($\mathcal{O}(1300)$ with $R_p<4R_\oplus$).
This would continue \tesss role as a planet-discovery machine, while also 
addressing the matter of refining ephemerides in order to enable detailed 
characterization with suitable instruments.

While our simulations show that observing opposite hemispheres in Years 3 and 4 
leads to many planet detections, we have not compared this strategy with 
observing the same hemisphere in both years (\textit{e.g.,} performing 
\nhemi\:in both Years 3 and 4). The latter approach might improve planet 
detection statistics, in particular with small and long-period planets.
An argument against such a strategy would be that it postpones refining 
the \tess planet ephemerides until Year 5.
We leave the detailed comparison for future work. 

In terms of the other strategies, \elong, \eshort, and \hemis\:are less 
obviously extensible to multiple years.
The main reasons to return to the ecliptic after performing \elong\:or 
\eshort\:would be to make \tesss survey truly `all-sky', and to perform \ktwo 
follow-up observations (see discussion below).
Of course, this would need to happen during intervals in which the Moon and 
Earth were not in the way.
\shemiAvoid\ would also achieve this goal.
%Continuous observations have serious value. 

%Leave the detailed characterization for \jwst, Spitzer, HST, ground-based 
%ELTs, and posterity.

% Move on to very-long term comments.
Any of our proposed scenarios could simply be repeated indefinitely,
as could possible two-year Extended Missions in which our scenarios 
would be followed by their respective complement hemispheres.
However, qualitatively novel trade-offs may arise when comparing two-year
Extended Missions.
For instance, after completing the Primary Mission, compare the 2-year
extension scenarios of \#1) two years of \hemis\ vs. \#2) repeating the 
Primary Mission for two years.
If we repeat the Primary Mission over 2 years, the northern
and southern CVZs get 1 additional year each of continuous observation.  
If we execute \hemis\:for 2 years, the northern and southern `long viewing
zones' each get 2 years of 14-day windowed observations. 
While each CVZ star receives the same total observing time, CVZ stars in \#1 
have a 1-year baseline, continuously sampled, while those in \#2 have a 2-year 
baseline, half-sampled.
Thus \#2 allows 2-transit detections of $P\lesssim1$ year planets.  \#1 allows 
2-transit detections of $P\lesssim6$ month planets, with less risk of 
ambiguities in derived orbital periods.  
In the competition between better time-sampling and longer baselines,
it is not clear which strategy is superior.

Another longer-term question is ``when will \tess hit the point of
diminishing returns?''  The `low-hanging fruit' of small planets
transiting bright stars at short orbital periods will become 
``picked over'' if \tess observes the same sky indefinitely.  
An important
qualitative point of this memo, made in Fig.~\ref{fig:snrf_histogram},
is that after \tesss Primary Mission there will be many objects
remaining for which merely doubling the number of observed transits
will enable their detection.  Eventually though, once \tess is complete
for $R_p<4R_\oplus$ planets orbiting bright stars on $P<20\,\mathrm{day}$ 
orbits, more observations will
only allow us to probe out to longer orbital periods and dimmer stars.
We have not yet quantified when \tess will reach this
regime.

\subsection{The Ephemeris Problem}
\label{sec:ephemeris_times}

\paragraph{Analytic motivation}

For follow-up observations, we will often need to predict future times
of transits or occultations, ideally with an accuracy of an hour or
less. After enough time has passed that the uncertainty has grown to an
operationally significant value, we say that the ephemeris
has gone ``stale,'' \textit{i.e.,} it presents a major obstacle to many 
follow-up programs. For \tesss ground-based follow-up campaign, following up a 
planet
with a stale ephemeris would require much more observing time
for a successful result.  Likewise, for planning
space-based observations, for which observing time is always scarce,
it is extremely important to have a reliable and precise ephemeris.
For mass determination through the Doppler method, a stale transit
ephemeris adds uncertainty to the planetary mass measurements, by
increasing the number of effectively free parameters.

Consider then the problem of estimating $\sigma_{t_c}(T_x)$, the
uncertainty of the mid-transit time $\sigma_{t_c}$ for a given planet
at some time $T_x$ following its last-observed transit.  We begin
analytically: assume that the planet has $N_\mathrm{tra}=2$ observed
transits, spaced an orbital period $P=14$ days apart. Because that
period is one half the nominal \tess dwell time of a given pointing,
it represents the shortest period for which typically
$N_\mathrm{tra}=2$, and as such is the worst-case scenario for predicting
the times of future transits, amongst cases with $N_\mathrm{tra}>1$.
Given two mid-transit times, each measured with the time's uncertainty
$\sigma_0$, separated by $P$, the uncertainty of a future mid-transit
time can be derived by standard least-squares fitting and propagation
of errors (\textit{e.g.}~\citet{lyons_practical_1991}, Equation 2.18):

\begin{equation}
	\sigma_{t_c}(T_x) = \sigma_0 \sqrt{1 + 2 T_x / P + 2 (T_x / P)^2 }
\end{equation}
Note that for observing future transits, $E \equiv T_x / P$ is a positive 
integer, and the above equation can be re-expressed:
\begin{equation}
	\sigma_{t_c}(E) = \sigma_0 \sqrt{1 + 2 E + 2 E^2 },
\end{equation}
which is bounded by the simpler approximation: 
\begin{equation}
	\sigma_{t_c}(E) \lesssim \sigma_0 \left(1+\sqrt{2} E\right), 
\end{equation}
which is exact at $E=0$, has a maximum 8\% fractional error at $E=1$, and 
becomes increasingly accurate as $E$ increases. By $E=20$, the fractional error 
of the latter approximation is less than 1\%.

At 2-minute cadence, a typical value for the per-transit timing uncertainty is 
$\sigma_0 = 4$ minutes.
For example, if $T_x = 2$ years and $P = 2$ weeks, then $E \approx 50$, so 
$\sigma_{t_c}\approx 75\sigma_0$.
Hence the predicted $1\sigma$ uncertainty on its mid-transit time 
is 5 hours (two years later) or 10 hours (four years later).

This leads to a simple rule of thumb: 
\begin{quotation}
	For a two-transit super-Earth, the $1\sigma$ uncertainty in the predicted 
	transit time, in hours, is numerically equal to about
	2$Y$, where $Y$ is the number of years after the Primary Mission.
\end{quotation}
If the transits are observed only at 30-minute cadence, then uncertainty will 
be roughly 4 times greater: $\sigma_0 \sim 16$ minutes. 
This claim (``$4\times$ greater'') is based on Figure 9 
of~\citep{price_transit_2014}, a plot of the effects of finite cadence on 
timing precision. % note that they're using kepler data here 
We compared the precisions of 2-min and 30-min cadences for their specific 
example of $P=10$ days and a dwell time of 1 month.

On the other hand, Fig.~\ref{fig:Ntra_hist_primary} shows that 7 in 8 of the 
planets detected by \tesss Primary Mission will have $N_\mathrm{tra}>3$ and so 
their ephemerides should be better than the worst case example derived 
analytically above.
Rather than generalize the analytic equations, we resort to numerical 
simulations in order to predict the uncertainties of mid-transit times for 
planets expected to be discovered by \tesss Primary Mission.

\begin{figure}[!t]
	\centering
	\includegraphics[scale=1.]{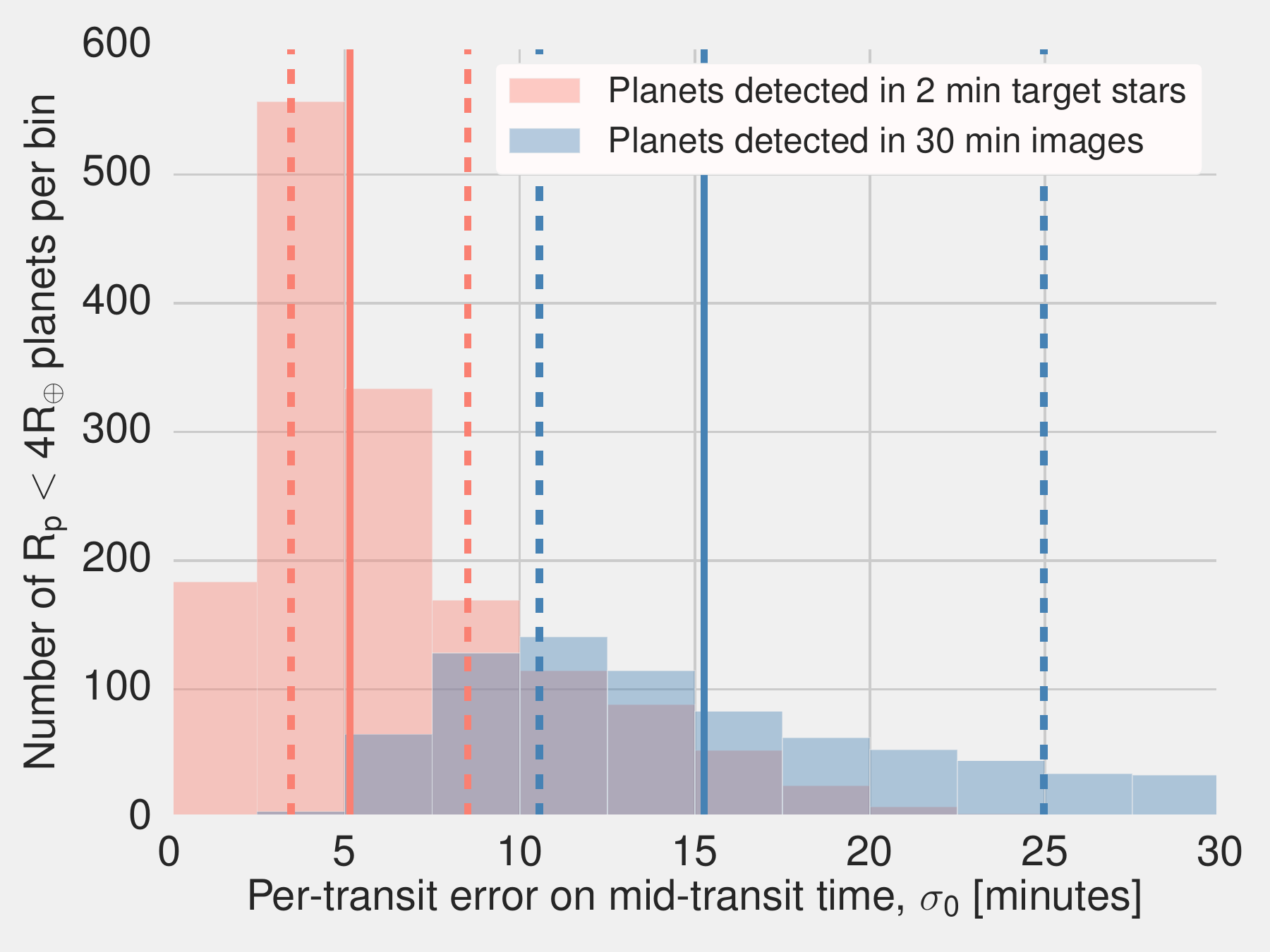}
	\caption{Uncertainty of mid-transit time on a single transit, 
	$\sigma_{t_c}$, for all detected $R_p<4R_\oplus$ planets from the Primary 
	Mission as computed from Eq.~\protect\ref{eq:price_rogers}.
		Solid lines are medians, dashed lines are 25$^\mathrm{th}$ and 
		75$^\mathrm{th}$ percentiles.
	}
	\label{fig:uncertainty_tc_hist}
\end{figure}

\paragraph{Numerical analysis}
We start with the analytic form~\citet{price_transit_2014} derive for the 
per-transit uncertainty on the mid-transit time $\sigma_0$:
\begin{equation}
	\sigma_{0} = \frac{1}{Q} \sqrt{ \frac{\tau T}{2} } \left( 1 - 
	\frac{t}{3\tau} 
	\right)^{-1/2}
	\label{eq:price_rogers_0}
\end{equation}
when $\tau\geq t$ and 
\begin{equation}
	\sigma_{0} = \frac{1}{Q} \sqrt{\frac{t T }{2}} \left( 1 - \frac{\tau}{3t} 
	\right)^{-1/2}
	\label{eq:price_rogers}
\end{equation}
when $t > \tau$,
where $Q$ is the SNR per transit, $t$ is the cadence, $T$ is the transit 
duration, and $\tau$ is the ingress (or egress) time.
We have all the latter terms from our yield simulation, and show the resulting 
distribution of $\sigma_0$ in Fig.~\ref{fig:uncertainty_tc_hist}.
Indeed, our suggested $\sigma_0$ of about 4 minutes for postage stamps and 16 
minutes for full frame images is reasonable, which is good because we computed 
the former from the $t\rightarrow0$ limit of Eq.~\ref{eq:price_rogers_0} 
originally derived by~\citet{carter_analytic_2008}.
\begin{figure}[!t]
	\centering
	\includegraphics[scale=1.]{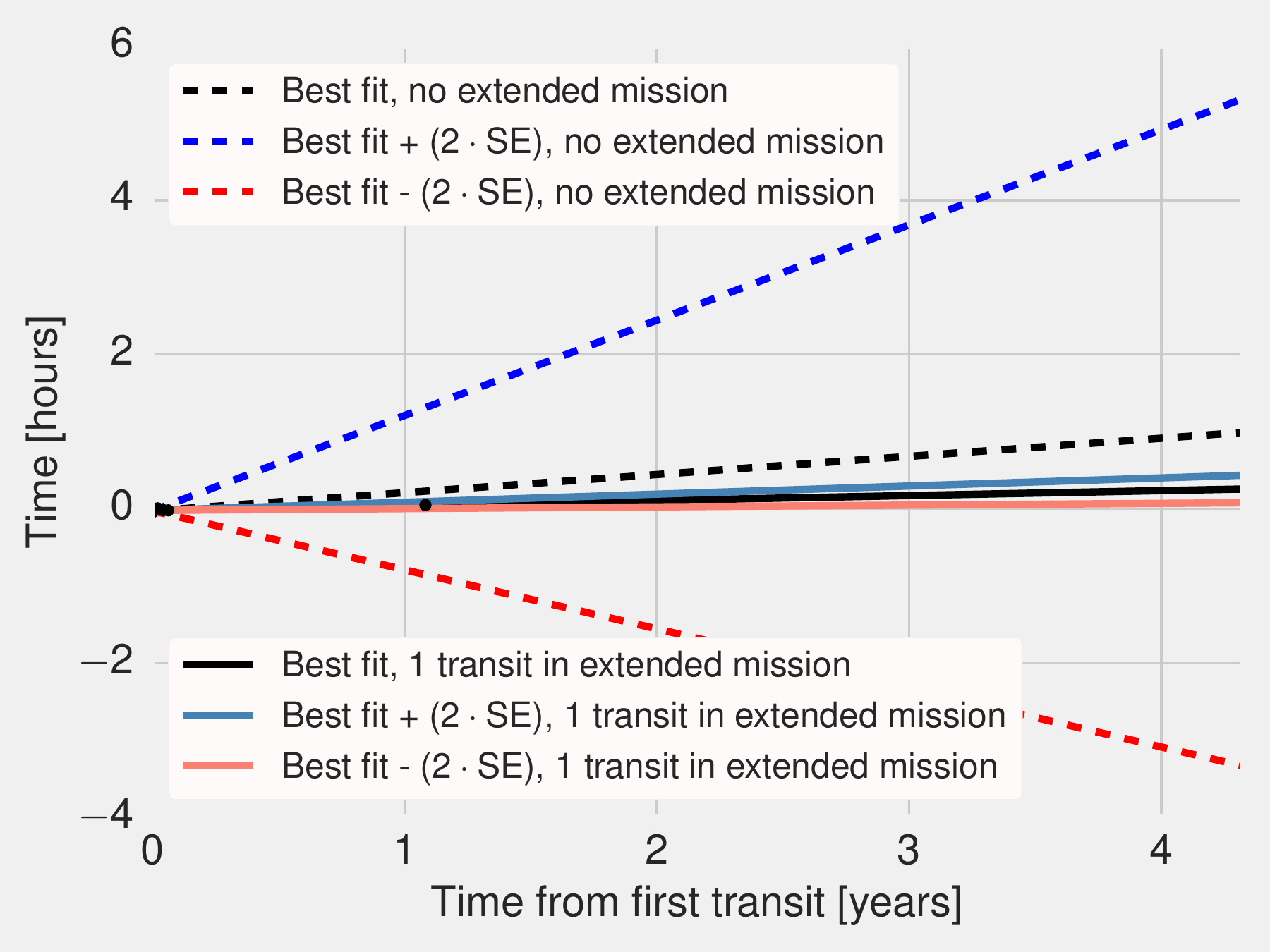}
	\caption{	Observed mid-transit times (dots) and best fits to a linear 
	ephemeris (lines).
		The dashed lines fit 4 data points from a nominal planet.
		The solid lines do the same, but with an additional transit observed 
		one year
		later.
		`SE' is the standard error on the slope which multiplied by 1.96 
		(rounded to 2 in the legend) gives a 95\% confidence interval between 
		the blue and orange lines.
	}
	\label{fig:lowering_uncertainty_tc}
\end{figure}

Given the distributions on per-transit uncertainty of $t_c$, we then took an 
example planet with 4 transits.
We drew ``observed'' mid-transit times from a Gaussian with zero mean and 
standard deviation $\sigma_{0}$, and then ran a linear least squares 
regression. 
We then added just one data point 1 year after the final observed transit, and 
repeated the regression.
This produces a cartoon-plot, Fig.~\ref{fig:lowering_uncertainty_tc}, which 
confirms two expected points:
\begin{enumerate}
	\item Years after the initial discovery, the uncertainty of mid-transit 
	time is of order hours.
	\item If we detect an additional transit 1 year after the final observed 
	transit from the Primary Mission, the uncertainty on the mid-transit time 
	decreases by an order of magnitude.
\end{enumerate}

\begin{figure*}[!t]
	\centering
	\includegraphics[scale=1.]{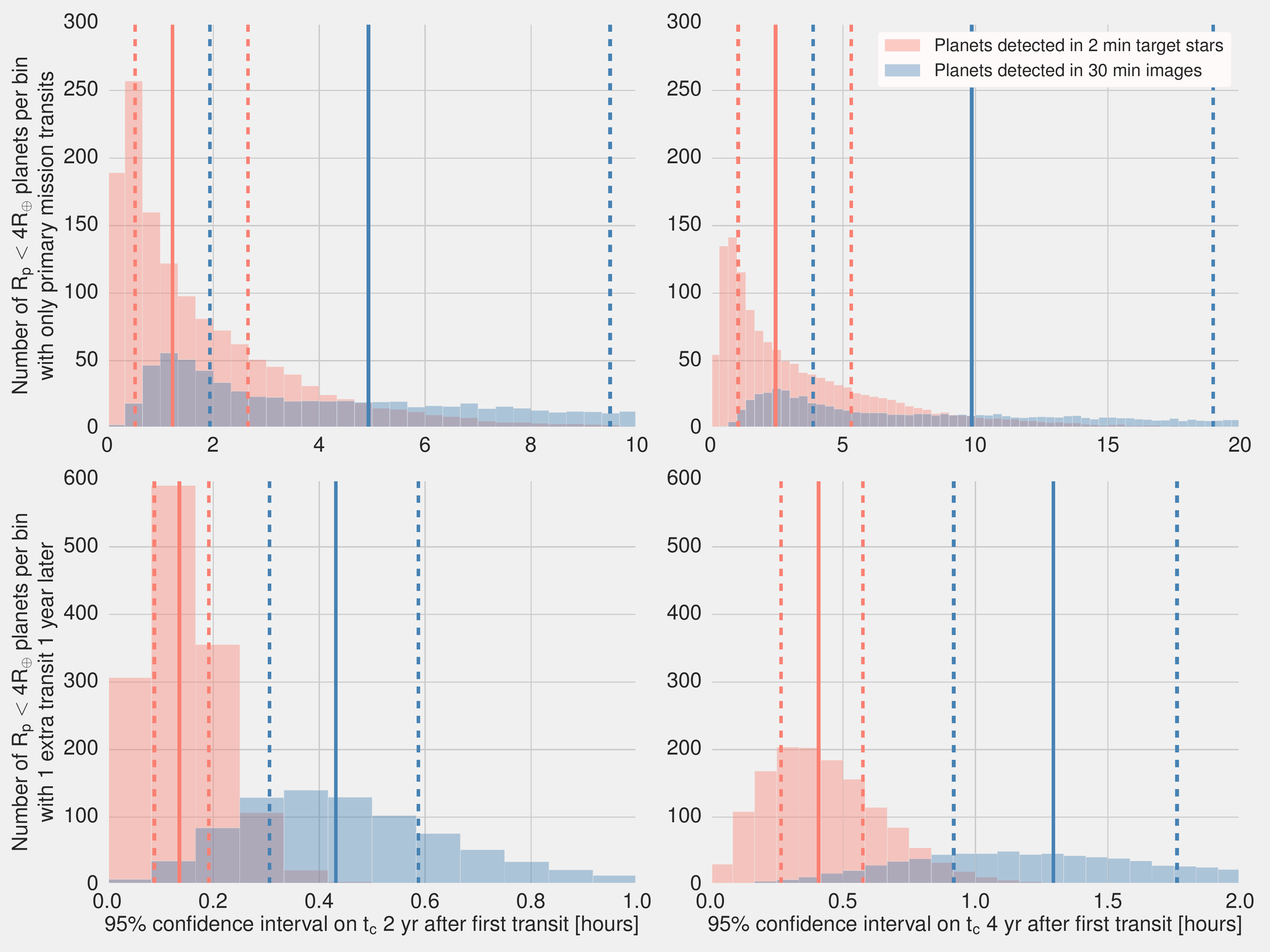}
	\caption{\textit{Top row}: histogram of 95\% confidence intervals 2 
	(\textit{left}) and 4 (\textit{right}) years following the first detected 
	transit in the Primary Mission.
		20 minute bins.
		\textit{Bottom row}: histogram of 95\% confidence intervals 2 
		(\textit{left}) and 4 (\textit{right}) years following the first 
		detected transit in the Primary Mission, but with an additional data 
		point added to the analog of 
		Fig.~\protect\ref{fig:lowering_uncertainty_tc} one year after the 
		transit in the initial time series (5 minute bins).
		Note that the top row's timescale is an order of magnitude greater than 
		the bottom row.
		Solid lines are medians, dashed lines are 25$^\mathrm{th}$ and 
		75$^\mathrm{th}$ percentiles.
	} %all in ephemeris_uncertainties.ipynb
	\label{fig:conf_interval_gets_better}
\end{figure*}
We proceed by repeating the above procedure for every planet, and evaluate 
typical 95\% confidence intervals (``uncertainties'', loosely) for $t_c$, at 
typical times after the first transits, for all of our detected planets.
Specifically, we take them at $T_x=2$ and $4$ years, and get 
Fig.~\ref{fig:conf_interval_gets_better}.
This figure confirms (top left panel) our analytic expectation that the 
uncertainty of mid-transit times in hours should be somewhat less than twice 
the number of years after \tess first observes the planet at 2-min cadence, 
since most such planets have $N_\mathrm{tra}\geq4$.
It also confirms our rough expectation that the uncertainty on FFI mid-transit 
times is roughly 4 times that of postage stamps, although the uncertainties on 
FFIs have a much broader tail than PSs.

More importantly, Fig.~\ref{fig:conf_interval_gets_better} emphasizes the 
importance of refining \tesss ephemerides: if we do not, the typical \tess 
planet will have many hours of uncertainty on its mid-transit time a few years 
following its detection.
If we do follow-up with an Extended Mission, we will be able to predict when 
the planet transits to $\lesssim1$ hour for many more years.
This argues strongly for an Extended Mission which, whether over 1 or 2 years, 
re-observes many if not all of the targets that \tess detects in its Primary 
Mission. 
The smallest-radius Earths and super-Earths may otherwise be much more 
difficult to follow up.

\subsection{Risks and Caveats}
\label{sec:risks_caveats}
\paragraph{Risk that planet detection simulation gets yield wrong:}
What is the risk that we over or under-estimated \tesss planet yield, either in 
the Primary Mission, or in any given Extended Mission?
We summarized the assumptions that went into our yield calculations in 
Sec.~\ref{sec:input_assumptions}.
We made them believing that they were all good enough for useful relative 
comparisons of different sky-scanning scenarios,
even if they are not correct in absolute terms to better than $\approx$50\%.

Highlighting a few of these assumptions, in order of what we feel is decreasing
concern:
\begin{itemize}
	\item [1.) We assume no knowledge of the outcomes of prior transit 
	searches.]
	As indicated in the text, this assumption is worst for the
	\elong\:and \eshort\:scenarios, for which \ktwo and \tesss
	overlap will be important.  Estimating the magnitude of our
	error, assume \ktwo will have observed $70\%$ of the sky in
	the $|\beta|<6^\circ$ band about the ecliptic by 2019.  Of the
	1169 `new' $R_p<4R_\oplus$ planets that \tess detects in
	\elong, 239 of them are within $|\beta|<6^\circ$ of the
	ecliptic, and thus roughly 167 will already have been observed by
	\ktwo (assuming the same stars were selected in \ktwo 
	observations).  
	This simple estimate quantifies our global error in reporting `new'
	planets near the ecliptic ($\sim 15\%$ will not be truly `new'), but 
	neglects the issue of merging the
	two datasets to discover long period planets.  This latter
	opportunity could be an important reason to actually do
	\elong\:or \eshort\:and thus demands detailed study, which we
	recommend below.
	
	\item [2.) We use a SNR threshold of 7.3.] This number was computed 
	by~\citetalias{Sullivan_2015} based on the argument that it would be a 
	sufficient threshold to give one statistical false positive per 
	$2\times10^5$ light curves.
	Applying the same criterion to full frame images would lead to more than 
	one false positive, since full frame images come from a much larger sample 
	of stars.
	Any pipeline that is written to work with \tess data will confront this 
	problem:
	processing $10^8$ vs. $10^6$ stars requires different false positive 
	thresholds.
	Extrapolating from~\citetalias{Sullivan_2015}'s Fig 15, a threshold 
	sufficient to give 0.05 false positives per $2\times10^5$ light curves, or 
	1 false positive per $4\times10^6$ light curves (as from our full frame 
	images), is roughly 7.5.
	Making the same ad-hoc estimate that~\citetalias{Sullivan_2015} did and 
	multiplying by 1.03 for the expected drop in SNR from cosmic ray noise 
	gives a SNR threshold of 7.7 for full frame images.
	Considering our Fig.~\ref{fig:snrf_histogram} (and noting the black line is 
	for the sum of postage stamp and full frame image detections), adopting a 
	SNR threshold of 7.7 for FFIs would mean a loss of $\sim30\times4=120$ 
	planets over two years, or 60 of what we claim are `detected planets' from 
	full frame images in 1 year's Extended Mission.	 We also note that our 
	model lacks any accounting for time-correlated noise and the probabilistic 
	nature of the signal recovery process, which are likely to be more 
	important factors than the purely white-noise statistical fluctuations.

	\item [3.) We use synthetic stars from a single galactic model (TRILEGAL).]
	One check on the robustness of this model would be to compare with other 
	galactic models like GALAXIA~\citep{sharma_galaxia_2011} or the Besa\c con 
	model~\citep{robin2003synthetic}.
	The simplest analytic model, applicable given that \tesss limiting distance 
	for $R_p<4R_\oplus$ detections is $\lesssim 1\ \mathrm{kpc}$, is a 
	distribution of stars uniform in the radial direction and with an 
	exponential profile in the vertical direction. 
	Independent of our numerical simulation,~\citet{winn_searchable_2013} used 
	such a galactic profile, with occurrence rates inferred from \kepler Q1-6. 
	Our results are in order-of-magnitude agreement, with ours being slightly 
	higher, likely due to our incorporation 
	of~\citet{dressing_occurrence_2015}'s M dwarf occurrence rates.
	
	It might eventually be best to use the real star catalog that \tesss Target 
	Selection Working Group is assembling. However, until \gaia DR2 (Q4 2017), 
	this will likely come with large uncertainties on stellar radii, given the 
	difficulty of distinguishing giants and sub-giants from M dwarfs without 
	proper motions.
	An additional challenge is in companion fractions.
	
	While we recommend that these broader checks be performed in future \tess 
	yield calculations, they are excessive for the purposes of this report 
	given that we only used the nearest, brightest, stars 
	($d\lesssim\mathrm{1kpc}, I_c\lesssim16$), with a simple prescription for 
	background contamination, in evaluating \tesss $R_p<4R_\oplus$ detections.
	The uncertainties become much greater if we try to estimate detections of 
	giant planets and false positives throughout the galactic disk.
	That said, we take~\citetalias{Sullivan_2015}'s cross-checks (cf Fig. 5 of 
	that paper) against actual surveys of the local stellar population as 
	indicative that we probably have the number of nearby stars, as well as 
	their properties, correct for the accuracy required in this work.
	%!WINN! #3, 4, 5, 7 not specific to Extended Missions, right?
	%!BOUMA! right. But they should still be here?
	
	\item [4.) At least 2 transits for detection:] recall 
	Figs.~\ref{fig:Ntra_hist} and~\ref{fig:Ntra_hist_primary}.
	If we were to use a more stringent criteria, for instance at least 3 
	transits for detection as the \kepler pipeline currently does, we would 
	lose $\mathcal{O}(200)$ of the planets detected in the Primary Mission, and 
	$\mathcal{O}(100)$ from the Extended Missions.
	This assumption disproportionately affects long-period planets.
	
	%	\item[6.) Multiple planet system distributions:] we assumed that we 
	%could approximate the occurrence rates for transiting multiple planet 
	%systems as repeated draws from the single-planet occurrence distributions 
	%(with independent probability) with added impositions of co-planarity and 
	%orbital stability.

	\item [5.) We neglect instrument aging, spacecraft systematics, etc.]
	To our knowledge detailed models do not currently exist for how \tesss 
	optics and CCDs will degrade with time.
	%	\todo[inline]{talk with Goddard people about this}
	We are also assuming that, as with \ktwo\!, time-correlated spacecraft 
	level systematics can be removed in post-processing.
	These assumptions are reasonable at our current state of pre-flight 
	knowledge, and will need to be changed accordingly as the mission 
	progresses.
	If we were to assume a gradual decline in photometric performance, for 
	instance from an increased number of dead or `hot' pixels, the relative 
	Extended-Mission-to-Extended-Mission comparisons would remain identical.
	The absolute Extended-Mission-to-primary-Mission yields would decrease.
	
	\item [6.) We do not consider the efficacy of the processing pipeline.]
	For instance, \hemis\:will come with period ambiguities and aliasing 
	problems imposed by its 14 day sampling at the `continuous' viewing zones.
	Similar issues are generic across Extended Missions for which we detect a 
	small number of transits in the Primary Mission, and then a small number of 
	transits in the Extended Mission.
	
	A robust way to approach this problem would be to generate a synthetic 
	simulated \tess dataset, \textit{i.e.}, at the image-level, rather than at 
	the idealized phase-folded SNR level from this work, for each Extended 
	Mission's observations.
	Then actually perform astrometry on injected stars, extract light curves, 
	de-trend them, and find their transits.
	This exercise would likely also be a useful way to prepare the SPOC and 
	broader community for what we expect the era of \tess photometry to entail 
	in terms of data quality.	
\end{itemize}

\section{Concluding remarks and recommendations}
\label{sec:conclusions}

Although the science requirements for \tesss Primary Mission have already been
written~\citep{ricker_transiting_2014}, an Extended Mission offers the
entire astronomical community a chance to rethink and reprioritize the use of 
the spacecraft.
A \tess Extended Mission would have wide-ranging applications, and should be 
planned and proposed with much broader goals than exoplanet science alone. 
The purpose of this trade study was to inform the exoplanet-related portion of 
that discussion. By simulating different scenarios for a third year of 
operations, we can anticipate
the numbers and types of planets that would be discovered. Our study has also 
highlighted other issues,
such as the 'stale ephemerides' problem, and the opportunity to change the
selection of the PSs and the cadence of the FFIs for diverse reasons.
This study was not designed to provide any definitive answers; but rather,
to provide materials to use in further discussions.  Our study has also
prompted us to make
some recommendations for future work that would be useful in this regard.

\subsection{Recommendations}
\label{sec:recommendations}
\begin{description}
	
	\item[Deeper analysis of the target prioritization scheme.] How should the
	results of the Primary Mission, or other sources of new data, be used to
	choose target stars for finer time sampling during an Extended Mission?
	
	\item[Optimize cadence.] What is the optimal time sampling for transit 
	detection? For example, observing $2\times10^5$ target stars at 
	4-minute cadence, rather than at 2-minute cadence,
	would allow the FFIs to be returned more frequently.
	This could in turn improve prospects for transit detection.  
	(Preliminary numerical experiments indicate that this is indeed the 
	case.)
	
	In addition, a metric should be devised that
	quantifies, for each star, how much {\it improvement} would
	result from observing at a shorter cadence.
	Stars could then be prioritized according to this improvement
	statistic, rather than an overall planet-detectability statistic.
	For instance if the number of short-cadence stars could be 
	greatly reduced with little effect on the planet detection
	statistics, then this might allow the FFIs to be returned at a
	higher cadence.
	
	\item[Take steps to address the `upgrading cadence' problem:]
	if there is a likely transiting planet in full frame image
	data, upgrading the planet to short cadence in future
	observing sectors improves the probability and fidelity of
	detection. How is this being addressed for the Primary Mission?
	Could this be an argument to observe the southern
	sky in year 3, in order to have extra time to prepare?
	
	\item[Solicit expert advice] from experts in asteroseismology, transient 
	detection,
	and other relevant areas to understand how the parameters of an Extended 
	Mission
	would affect their scientific prospects. Comments on this report sent to
	the authors will be gratefully received. A more formal and comprehensive 
	process would
	be a call for White Papers organized
	by the Guest Investigator program or the \tess Science Office. As
	exemplified in NASA’s 2016 Astrophysics Senior Review,
	everyone benefits from the discussions generated by such
	community feedback~\citep{donahue_senior_2016}.
	
	\item[Consider the relative importance of our proposed exoplanet-detection 
	metrics.]
	Sec.~\ref{sec:comparing_pointing_strategies} summarizes these; the reader 
	may have others to suggest. The crude metric of "total number of new planet 
	detections" is not
	likely to be the most important, and we have found that the scenarios we 
	considered
	differ by $\lesssim30\%$ in
	this regard. 
	
	\item[Simulate combining \tess and \ktwo data from
	\rm{\elong\:} \textit{and} \rm{\eshort}.]  This is perhaps
	the most important qualitative difference between observing
	towards and away from the ecliptic.  Would \tess\!+\ktwo
	enable more discoveries out at long periods than
	alternatives?  How many of the new planets that \tess
	detects on the ecliptic will actually be detected by \ktwo?
	Is the value-added of combining datasets a compelling case
	compared with discovery?
	
	\item[Ensure {\rm \npole} adequately mitigates scattered sunlight.]
	See description in Sec.~\ref{sec:proposed_pointings}.
	
	\item[Point of diminishing returns.] Suppose, for example, the Primary 
	Mission were repeated indefinitely.
	At what point would the planet discovery rate start falling significantly 
	below that of Years 1-3?
	How much fainter would the host stars of new planets be, in each year?
	
\end{description}

\newpage
\section*{Acknowledgements}
We thank Peter Sullivan, for his support in the early phases of this 
work and for helpful comments on our results;
Roland Vanderspek, for fielding numerous questions concerning \tess hardware; 
Jack Lissauer, for input regarding the value of observing \keplers field with 
\textit{TESS};
and Tim Morton, for emphasizing the importance of the false positive problem.
We thank the many other participants in informal discussions on this subject 
over the last few years.
More broadly we are grateful to the hardworking teams 
at MIT, NASA Goddard, NASA Ames, CfA, STScI, Orbital Sciences, and other 
\textit{TESS} partner institutions.
We thank the \tess project for providing the computing resources used in this 
work, and Ed 
Morgan, Isaac Meister, and Kenton Philips for keeping those machines running.

\vspace{0.5cm}
\textit{Facilities}: \tess\!, \kepler

\textit{Software}: matplotlib~\citep{hunter_matplotlib_2007}, 
NumPy~\citep{walt_numpy_2011}, SciPy~\citep{jones_scipy_2001}, 
pandas~\citep{mckinneypandas}, JPL NAIF's SPICE 
library~\citep{acton_SPICE_1996}, and the IDL Astronomy User's 
Library~\citep{landsman_idl_1995}.

\textit{Resources}: This research has made use of the NASA Astrophysics Data 
System and the NASA Exoplanet Archive. The NASA Exoplanet Archive is operated 
by the California Institute of Technology, under contract with the National 
Aeronautics and Space Administration under the Exoplanet Exploration Program.
This paper also makes use of data collected by the Kepler mission. Funding for 
the Kepler mission is provided by the NASA Science Mission directorate.

\pagebreak
\begin{appendices}
	
\addtocontents{toc}{\cftpagenumberson{section}}

%\pagebreak
%\input{models_relevant_to_earth_moon_crossings}
\section{Models Relevant to Earth and Moon Crossings}
\label{sec:appendix}

\begin{figure*}[!h] %[!thb]
	\centering
	\includegraphics{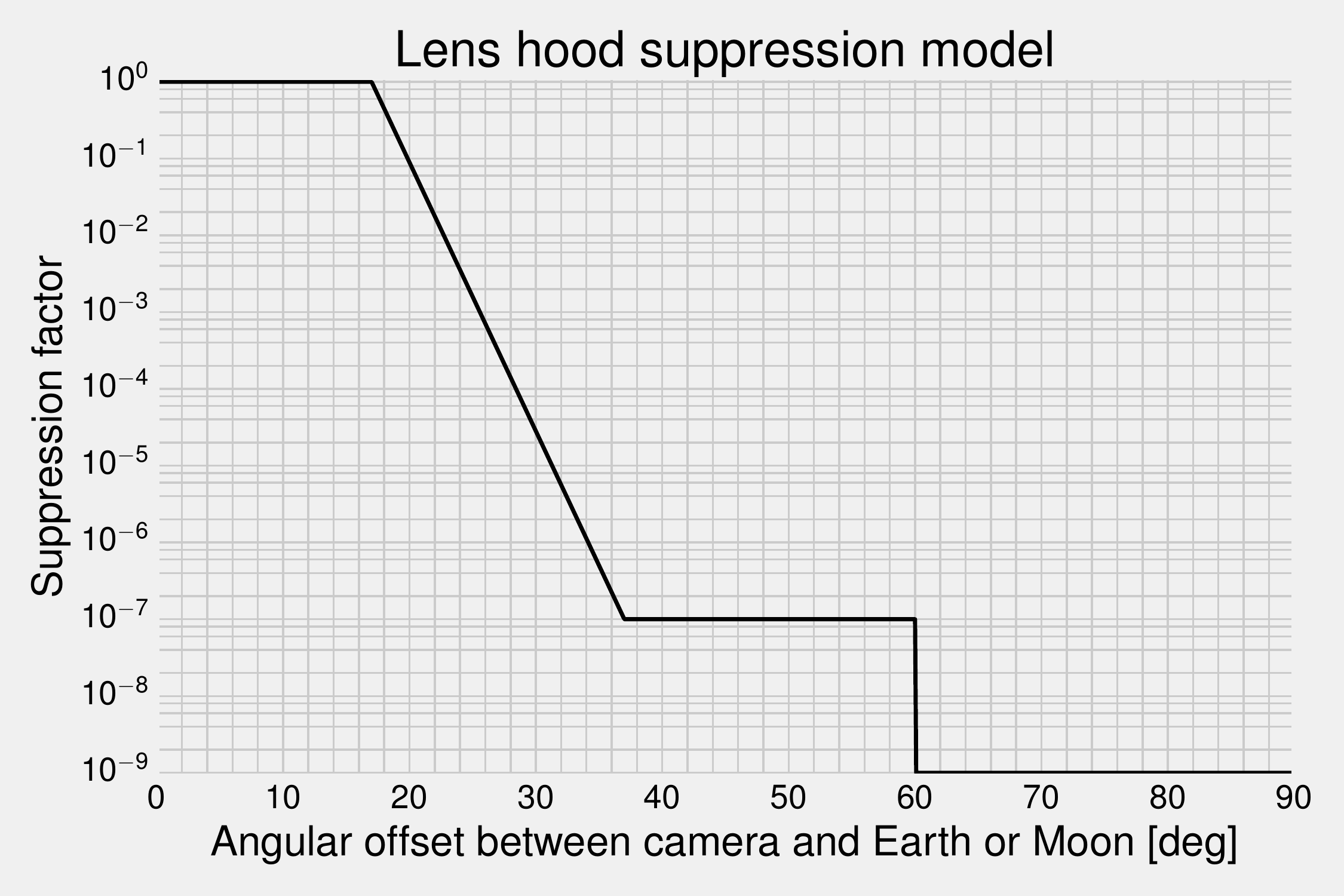}
	\caption{Lens hood suppression plotted against the angular offset to a 
		``point-source'' like the 
		Earth or Moon. This suppression factor is defined as the fraction 
		of 
		incident flux that is blocked by the spacecraft, sunshade, lens 
		hood, 
		or 
		combinations thereof. Source: R. Vanderspek, priv. comm.}
	\label{fig:lens_hood_suppression}
\end{figure*}
\newpage

\begin{figure*}[!th] %[!thb]
	\centering
	\includegraphics{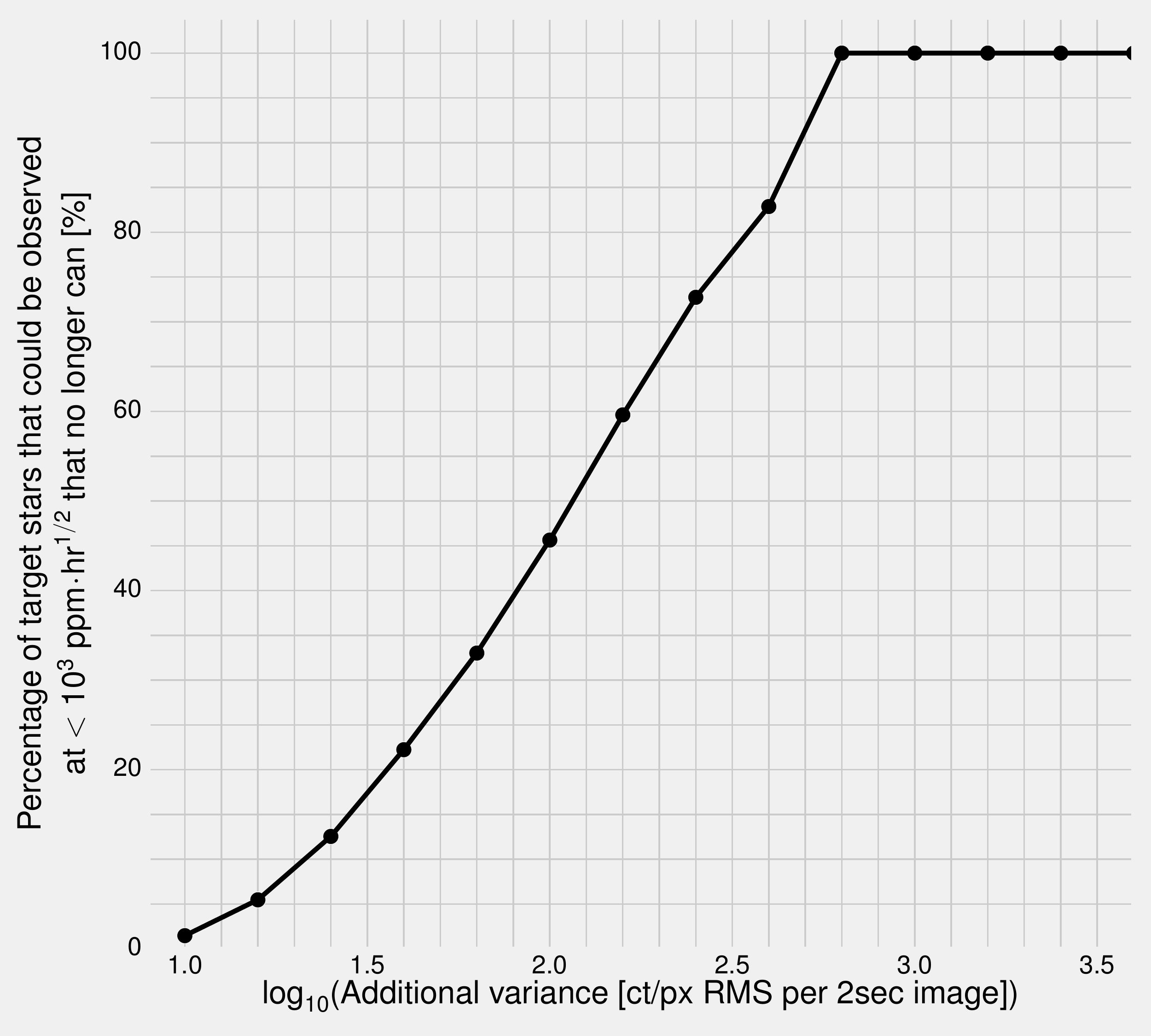}
	\caption{The procedure described in 
		Sec.~\protect\ref{sec:earth_moon_crossings} flags fields to be 
		dropped 
		when 
		there is a mean additional $300\,\mathrm{ct/px/s}$ from the Earth 
		or 
		Moon, 
		corresponding to $\mathrm{log_{10}}(Additional\ variance) = 1.24$ 
		on 
		this 
		plot.
		This reads that $<$10\% of the target stars which could be observed 
		at 		
		$<$1mmag precision over 1 hour cannot when at $F_\mathrm{tresh}$. 
		While this corresponds to the cases with the Earth or Moon 
		$\approx 28^\circ - 34^\circ$ from boresight, once closer, the 
		effects 
		become more serious. 
		Saturation is assumed to occur at a mean flux $> 
		2\times10^{5}\,\mathrm{ct/px}$ per 
		image. 
		Lines between points intended to guide the eye.}
	\label{fig:outage_vs_background}
\end{figure*}
\newpage

\newpage
\clearpage
\pagebreak
\section{Lens Hood Model: Additional Count Rate from Arbitrary Source}
Let $s(\theta)$ be the lens hood suppression as a function of angle 
$\theta$ 
from the camera boresight to a given point-source on the sky. By definition,
\begin{equation}
	s(\theta) \equiv \frac{F_\mathrm{obs}}{F_\mathrm{ns}},
\end{equation}
for $F_\mathrm{obs}$ the observed flux of the source (that which reaches 
the 
CCD), and $F_\mathrm{ns}$ the flux that would be observed with no 
suppression and $\theta=0$.
The observed flux of the source can then be written as a function of 
$\theta$
\begin{align}
	F_\mathrm{obs}(\theta) &= s(\theta) F_\mathrm{ns}  \\
	&= s(\theta) F_0 10^{-0.4(m_\mathrm{ns} - m_0)} ,
\end{align}
for $F_0$ the (non-suppressed) flux corresponding to a source with 
zero-point 
apparent magnitude $m_0$, and $m_\mathrm{ns}$ the apparent magnitude of 
the source with no suppression.
\citet{winn_searchable_2013} tabulates $F_0 = 
1.6\times10^6\,\mathrm{ph/s/cm^2}$ 
for an $I=0$, G2V star.
Thus
\begin{equation}
	F_\mathrm{obs}(\theta) = (1.6\times 10^6) s(\theta) 10^{-0.4 
	m_\mathrm{ns}} 
	\quad \mathrm{[ph/s/cm^2]}.
\end{equation}

We can then write the following expression for $\mu\equiv F_\mathrm{obs} A 
\eta 
/ N$, the mean incident count rate on the pixel of interest:
\begin{equation}
	\mu =  (1.6\times 10^6) s(\theta) 10^{-0.4 m_\mathrm{ns}} \frac{A 
		\eta}{N} \quad \mathrm{[ct/px/s]},
\end{equation}
for $A$ the effective observing area in $\mathrm{cm^2}$, $N$ the 
number of pixels per camera, and $\eta$ the quantum efficiency.
For TESS, $A=69.1\,\mathrm{cm^2}$, $N=4096^2$, and $\eta\approx 1$. Using
these numbers gives
\begin{equation}
	\mu = 6.59  s(\theta) 10^{-0.4 m_\mathrm{ns}}\quad 
	\mathrm{[ct/px/s]}.
	\label{eq:mean_added_flux}
\end{equation}
Assuming a Poisson arrival rate, the standard deviation in the number of 
counts 
per pixel per 2 second readout is then
\begin{equation}
	\sigma \approx \mu^{1/2} = \left[ 13.2 s(\theta) 10^{-0.4 
	m_\mathrm{ns}} 
	\right]^{1/2}\quad \mathrm{[ct/px\ RMS\ per\ 2\ sec\ image]}.
	\label{eq:added_RMS}
\end{equation}

The above expression assumes that the scattered flux from the source 
is uniformly spread across the CCD. The reality may be quite different,
but our purpose in this case is to simply get order-of-magnitude accuracy.
Another caveat is that our zero-point relies on an $I$ magnitude 
calibration -- 
a more accurate approach would use separate bandpass-dependent zero-points.

\begin{figure}[!h]
	\centering
	\includegraphics{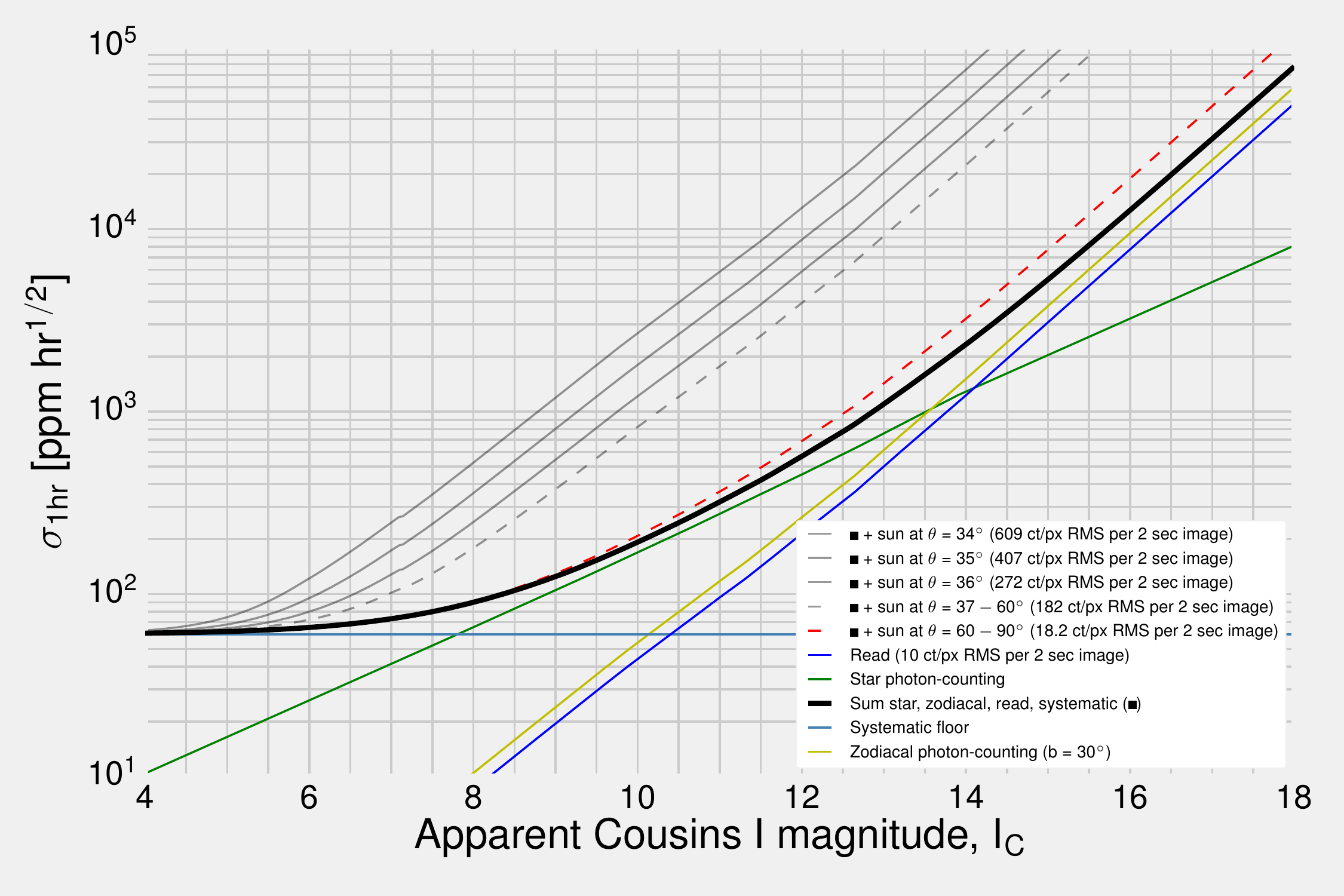}
	\caption{Noise budget including Solar scattered light at different 
	angles 
		$\theta$ from camera boresight. 
		Note that the suppression model 
		(Fig.~\protect\ref{fig:lens_hood_suppression}) 
		accounts solely for lens hood suppression, omitting sunshade 
		suppression (which substantially lowers the dashed red line for most
		orientations of the Sun).
		The number given in brackets is $\sigma$ in 
		Eq.~\protect\ref{eq:added_RMS} --
		the extra standard deviation about the mean.
		Squaring, lines where $\mu \gtrsim 10^5 \mathrm{ct/px/s}$ will 
		saturate 
		the CCD pixels (and thus are not plotted).} 
	\label{fig:sun_scat}
\end{figure}

\begin{figure}[!h]
	\centering
	\includegraphics{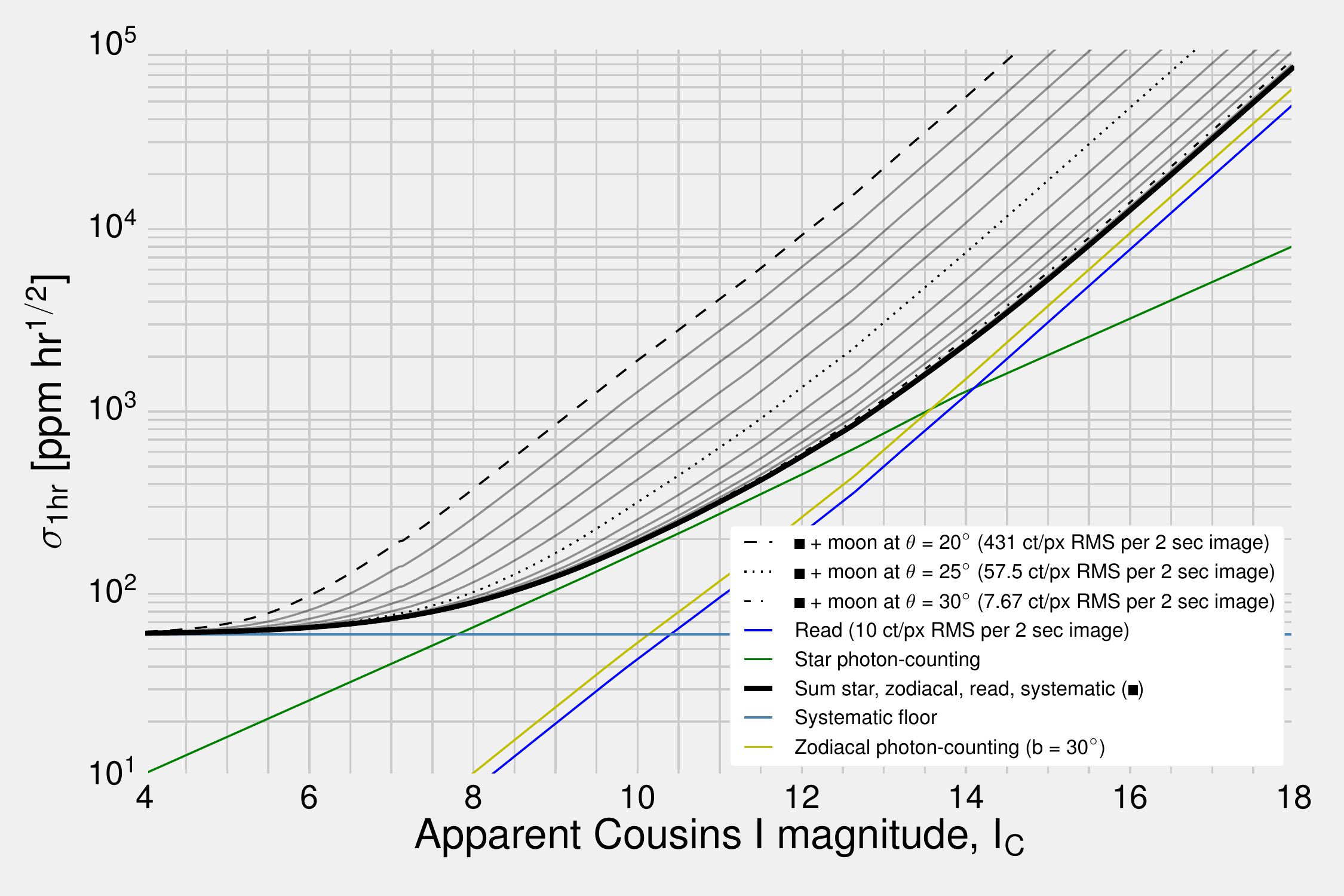}
	\caption{Same as Fig.~\protect\ref{fig:sun_scat}, for scattered 
	Moonlight.
		Gray lines are spaced by $1^\circ$.} 
	\label{fig:moon_scat}
\end{figure}

\begin{figure}[!h]
	\centering
	\includegraphics{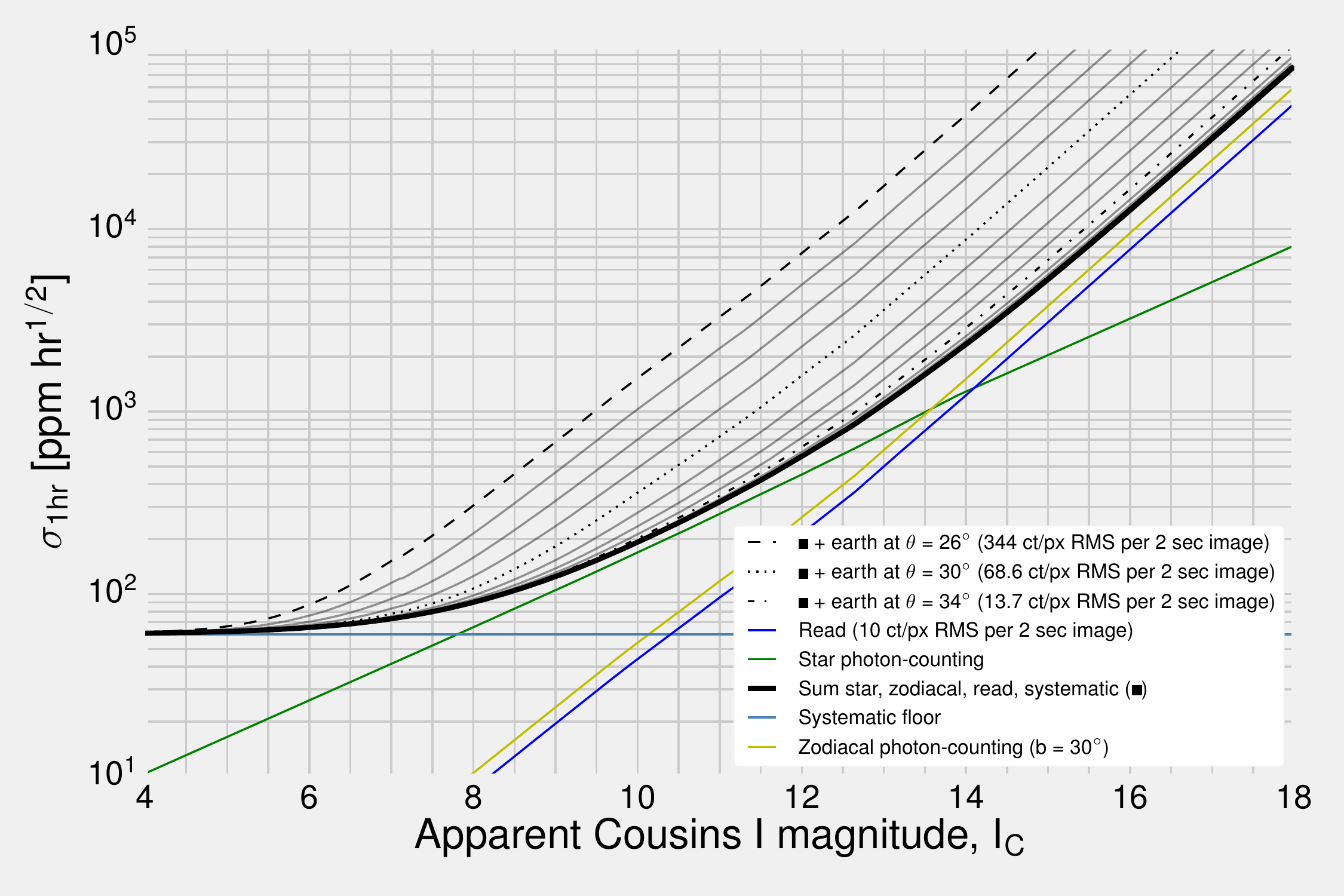}
	\caption{Same as Fig.~\protect\ref{fig:sun_scat}, for scattered 
		Earthlight. Gray lines are spaced by $1^\circ$.} 
	\label{fig:earth_scat}
\end{figure}

To compute the net effect on TESS's noise budget, we add $\sigma$ from 
Eq.~\ref{eq:added_RMS} in quadrature with Eq.~\ref{eq:snr}.
We assume $m_\mathrm{ns}$ values of $-26,-13,\,\mathrm{and}\ -17.76$ for 
the 
Sun, Moon, and Earth respectively.
This gives Figs.~\ref{fig:sun_scat}-\ref{fig:earth_scat}.


\newpage
\clearpage

\section{Changes from~\protect\citet{Sullivan_2015}}
\label{sec:changes_from_S15}
\paragraph{Dilution bug.}
Described in Sec.~\ref{sec:results_from_primary_missions}.

\paragraph{Correction of coordinate assignment for multiple planet systems.}
In the code corresponding to \citetalias{Sullivan_2015}'s released catalog,
``transiting objects'' did not correctly inherit the coordinates of their 
parent star -- they in fact received randomized coordinates within that 
star's 
HealPix tile.
While this is not a problem for the statistical results presented in 
\citetalias{Sullivan_2015}, it makes reverse-engineering multiple planet 
distributions from \citetalias{Sullivan_2015}'s Table 6 nearly impossible.
The updated catalog associated with this white paper has coordinates which 
are 
unique system identifiers. An associated new column is a number 
($\mathtt{hostID}$), unique to each Monte Carlo realization of the updated 
code, with the same function.

\paragraph{On the angular dependence of \tesss pixel response function:} 
In our SNR calculation, we do not keep track of individual times for every 
transit. 
How then do we assign PSFs to different transits from the same object that 
fall 
on different regions of the CCDs, and thus should have slightly better or 
degraded PSFs? (Largest cumulative flux fraction at the CCD's center, 
smallest 
at the corners). 

~\citetalias{Sullivan_2015} dealt with this by computing the mean of all 
the 
field angles (distance from the center of the CCD axis), and then passing 
this 
mean into a look-up table for PSFs based on four PSFs that had been 
computed 
from a ray-tracing model at four different field angles.
~\citetalias{Sullivan_2015} then `observes' each eclipsing object with a 
single 
class of PSF.
This leads to the implausible phenomenon that extra observations can 
actually 
\textit{lower} the SNR of an eclipsing object if they are taken with an 
unfavorable field angle/PSF. 
This effect is largely off-set by the extra pointing increasing the SNR, 
but 
for Extended Missions (coming back to the same objects at potentially very 
different field angles) it winds up reducing observed SNR for $\sim$3\% of 
detected objects.

In Extended Missions (as well as in the Primary Mission) we expect stars to 
land on very different regions of the CCD over the course of being observed.
We simplify this in our work by assuming that all stars land on the center 
of 
the \tess CCDs (the green curve of~\citetalias{Sullivan_2015} Fig. 13). 
This assumption is justified because our chief point of quantitative 
comparison 
is the ability of different pointing strategies to impact \tesss planet 
yield 
in Extended Missions, and there is little \textit{a priori} reason to 
assume 
that any one pointing scenario should be biased for an extra amount of 
stars to 
land on the `bad regions' of \tesss CCDs.
%We also note that the PRF performance only decreases noticeably for $1 - 
%\pi(12^\circ)^2 / (24^\circ)^2 \approx 21\%$ of the pixel area, and this 
%modification has the benefit of omitting the rare ``extra-observations 
%lead to 
%reduced SNR'' effect described in the above paragraph.
	
\end{appendices}

\newpage
\addcontentsline{toc}{section}{References}
\bibliography{biblio.bib}

\end{document}